\documentclass[journal]{IEEEtran}

\usepackage{fancyhdr}
\usepackage{setspace}
\usepackage[normalem]{ulem}
\usepackage[hyphens]{url}
\usepackage[caption=false]{subfig}
\usepackage{amsmath}
\usepackage{bm}
\usepackage{booktabs}
\usepackage{url}
\usepackage{multirow}
\usepackage{threeparttable}
\usepackage{array}
\usepackage{graphicx}
\usepackage{makecell}
\usepackage[square,sort&compress,comma,numbers]{natbib}
\usepackage[usenames,dvipsnames]{xcolor}
\title
{
    Eyeriss v2: A Flexible Accelerator for Emerging Deep Neural Networks on Mobile Devices
}
\author
{
    Yu-Hsin Chen, \IEEEmembership{Student Member,~IEEE,}
    Tien-Ju Yang, \IEEEmembership{Student Member,~IEEE,}
    Joel Emer, \IEEEmembership{Fellow,~IEEE,}
    and Vivienne Sze, \IEEEmembership{Senior Member,~IEEE}
}
\begin{document}
\maketitle
\begin{abstract}

A recent trend in deep neural network (DNN) development is to extend the reach of deep learning applications to platforms that are more resource and energy constrained, e.g., mobile devices. These endeavors aim to reduce the DNN model size and improve the hardware processing efficiency, and have resulted in DNNs that are much more \emph{compact} in their structures and/or have high data \emph{sparsity}. These compact or sparse models are different from the traditional large ones in that there is much more variation in their layer shapes and sizes, and often require specialized hardware to exploit sparsity for performance improvement. Therefore, many DNN accelerators designed for large DNNs do not perform well on these models. In this work, we present Eyeriss v2, a DNN accelerator architecture designed for running compact and sparse DNNs. To deal with the widely varying layer shapes and sizes, it introduces a highly flexible on-chip network, called hierarchical mesh, that can adapt to the different amounts of data reuse and bandwidth requirements of different data types, which improves the utilization of the computation resources. Furthermore, Eyeriss v2 can process sparse data directly in the compressed domain for both weights and activations, and therefore is able to improve both processing speed and energy efficiency with sparse models. Overall, with sparse MobileNet, Eyeriss v2 in a 65nm CMOS process achieves a throughput of 1470.6 inferences/sec and 2560.3 inferences/J at a batch size of 1, which is 12.6$\times$ faster and 2.5$\times$ more energy efficient than the original Eyeriss running MobileNet.

\end{abstract}
\section{Introduction}
\label{sec:introduction}

The development of deep neural networks (DNNs) has shown tremendous progress in improving accuracy over the past few years~\cite{ijcv2015-russakovsky}. In addition, there has been an increasing effort to reduce the computational complexity of DNNs, particularly for those targeted at mobile devices~\cite{sze_pieee_2017}. Various different techniques have been widely explored in the design of DNN models including reduced precision of weights and activations~\cite{nips2015-courbariaux-binaryconnect,eccv2016-rastegari-xnor_net,lee2017lognet, li2016ternary,moons2016energy,judd2016proteus}, compact network architectures~\cite{cvpr2015-szegedy,arxiv2017-howard,arxiv2016-iandola} (i.e., \emph{compact} DNNs), and increasing sparsity in the filter weights~\cite{nips1990-lecun-opt_brain_damage,nips2015-han,yang2016} (i.e., \emph{sparse} DNNs). While these approaches provide theoretical reductions in the size and number of operations and storage cost, specialized hardware is often necessary to translate these theoretical benefits into measurable improvements in energy efficiency and processing speed.

Support for reduced precision has been demonstrated in recent hardware implementations, including Envision~\cite{isscc2017-moons}, Thinker~\cite{thinker_vlsi_2017}, UNPU~\cite{isscc2018-unpu}, Loom~\cite{dac2018-loom}, and Stripes~\cite{micro2016-stripes}. These works have shown various methods that efficiently translate reduced bitwidth from 16-bits down to 1-bit into both energy savings and increase in processing speed. Specialized hardware for binary networks have also been widely explored~\cite{andri2016yodann,BRein2017,2018xnorsram,bankman2018always,Valavi2018machine}. In this work, we focus on complementary approaches that have been less explored, specifically the support for diverse filter shapes for compact DNNs, as well as support for processing in the compressed domain for sparse DNNs. While compact and sparse DNNs have fewer operations and weights, they also introduce new challenges in hardware design for DNN acceleration.  

\subsection{Challenges for Compact DNNs}

\begin{figure}
    \centering
    \subfloat[]
    {
        \includegraphics[width=0.8\linewidth]{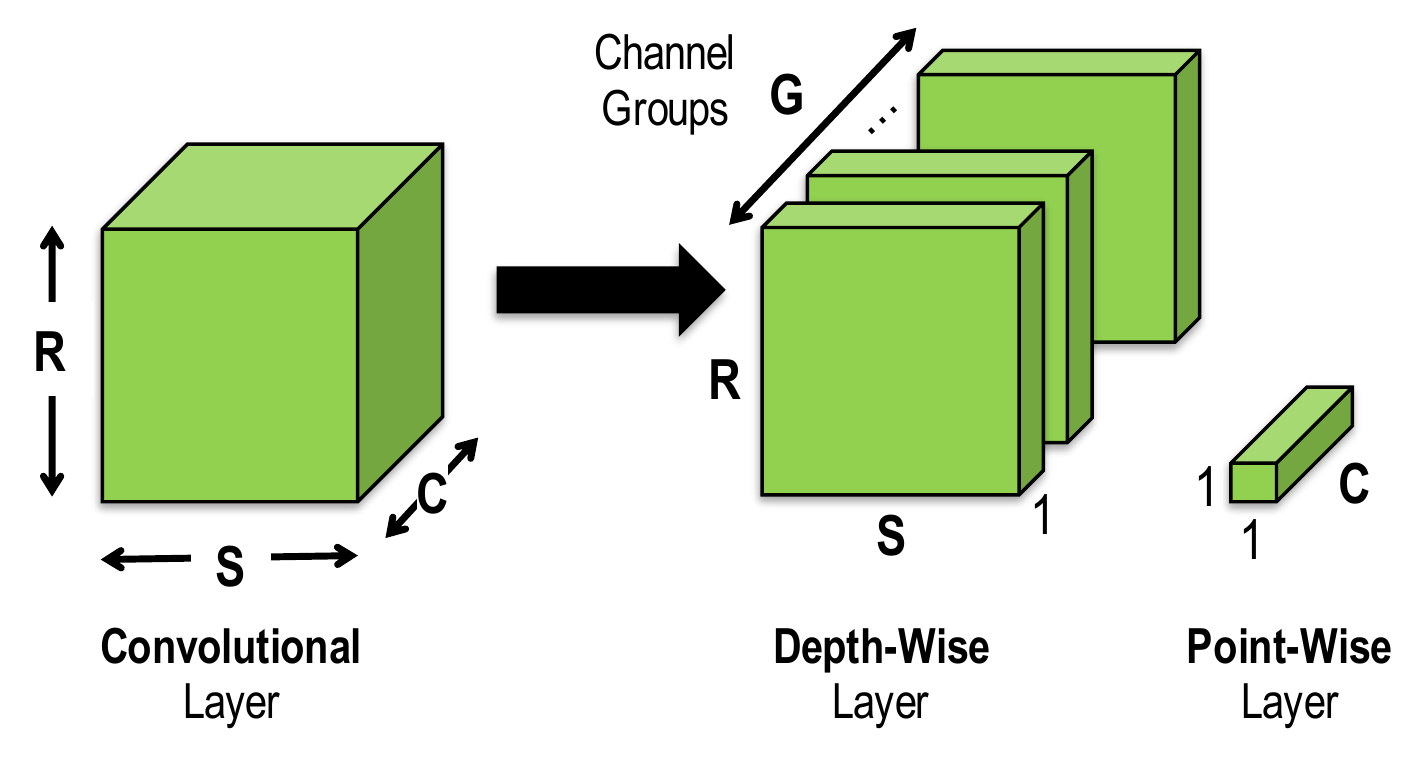}
        \label{fig:mobilenet_decompose}
    }\\
    \subfloat[]
    {
        \includegraphics[width=0.45\linewidth]{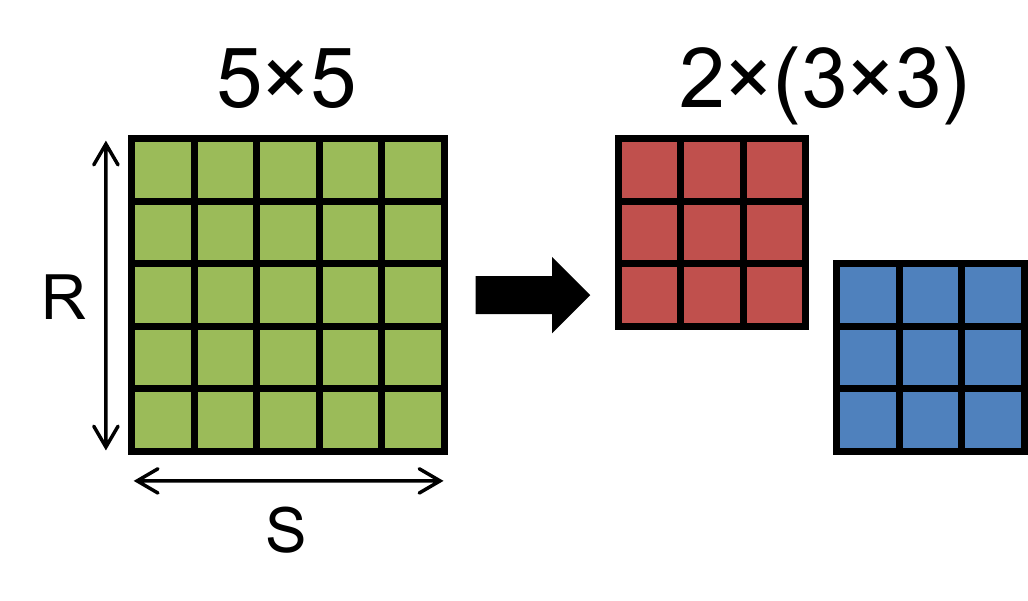}
        \label{fig:2D_decompose}
    }
    \subfloat[]
    {
        \includegraphics[width=0.45\linewidth]{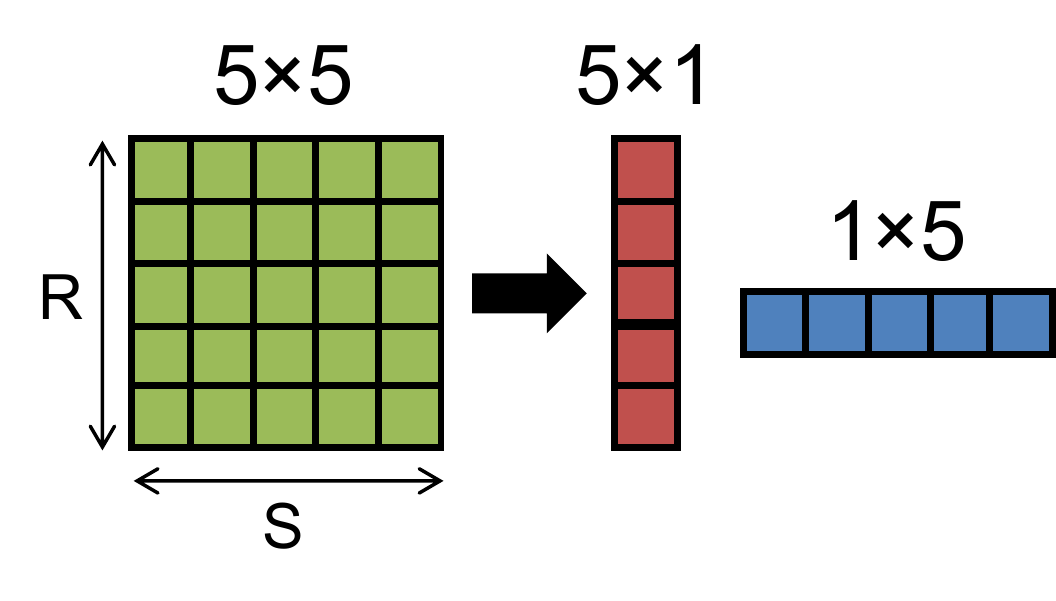}
        \label{fig:2D_decompose_1D}
    }
    \caption{   Various filter decomposition approaches~\cite{arxiv2017-howard, iclr2015-simonyan, szegedy2016rethinking}.
            }
    \vspace{-10pt}
    \label{fig:fiter_decomposition}
\end{figure}

The trend for compact networks is evident in how the iconic DNNs have evolved over time. Early models, such as AlexNet~\cite{nips2012-krizhevsky} and VGG~\cite{iclr2015-simonyan}, are now considered \textit{large} and over-parameterized. Techniques such as using deeper but narrower network structures and bottleneck layers were proposed to pursue higher accuracy while restricting the size of the DNN (e.g., GoogLeNet~\cite{cvpr2015-szegedy} and ResNet~\cite{cvpr2016-he}). This quest further continued with a focus on drastically reducing the amount of computation, specifically the number of multiply-and-accumulates (MACs), and the storage cost, specifically the number of weights. Techniques such as filter decomposition as shown in Fig.~\ref{fig:fiter_decomposition} have since become popular for building \textit{compact} DNNs targeted at mobile devices (e.g., SqueezeNet~\cite{arxiv2016-iandola} and MobileNet~\cite{arxiv2017-howard}). This evolution has resulted in a more diverse set of DNNs with widely varying shapes and sizes. 

One effect of compact DNNs is that \textit{any data dimension in a DNN layer can diminish}. In addition, due to latency constraints, it is increasingly desirable to run DNNs at smaller batch sizes (i.e., smaller $N$). Table~\ref{table:diminishing_dnn_data_dimensions} summarizes the data dimensions that are used to describe a DNN layer and the common reasons for each dimension to diminish. This suggests that less assumptions can be made on the dimensions of a DNN layer.  

\begin{table}
    \centering
    \resizebox{\columnwidth}{!}{%
        \begin{tabular}{|c|l|l|}
        \hline
        \multicolumn{2}{|c|}{\multirow{2}{*}{\textbf{Data Dimension}}}  & \multicolumn{1}{c|}{\textbf{Common Reasons for}} \\
        \multicolumn{2}{|c|}{}                                          & \multicolumn{1}{c|}{\textbf{Diminishing Dimension}} \\
        \hline
        \multirow{2}{*}{$G$}        & number of                         & \multirow{2}{*}{non-depth-wise layers} \\
                                    & channel groups                    &                                         \\
        \hline        
        $N$                         & batch size                        & low latency requirements \\
        \hline
        \multirow{2}{*}{$M$}        & number of                         & (1) bottleneck layers \\
                                    & output channels                   & (2) depth-wise layers  \\
        \hline
        \multirow{3}{*}{$C$}        & \multirow{2}{*}{number of}        & (1) layers after bottleneck layers \\
                                    & \multirow{2}{*}{input channels}   & (2) depth-wise layers \\
                                    &                                   & (3) first layer (e.g., 3 in visual inputs) \\
        \hline
        \multirow{2}{*}{$H$ / $W$}  & input feature map                 & \multirow{2}{*}{deeper layers in a DNN} \\
                                    & height/width                      &                                         \\
        \hline
        \multirow{2}{*}{$R$ / $S$}   & \multirow{2}{*}{filter height/width}  & (1) point-wise layers (i.e., 1$\times$1) \\
                                    &                                   & (2) decomposed layers (i.e., R$\times$1, 1$\times$S) \\
        \hline
        \multirow{2}{*}{$E$ / $F$}  & output feature map                & (1) deeper layers in a DNN \\
                                    & height/width                      & (2) fully-connected (FC) layers \\
        \hline
    \end{tabular}
    }
    \vspace{+5pt}
    \caption{   Reasons for diminishing data dimensions in a DNN layer.
            }
    \vspace{-10pt}
    \label{table:diminishing_dnn_data_dimensions}
\end{table}

For hardware designers, widely varying DNN layer shapes, especially diminishing dimensions, is challenging as they result in changes in a key property of DNNs: \textit{data reuse}, which is the number of MACs that use the same piece of data, i.e., MACs/data. Most DNN accelerators rely on data reuse as a means to improve efficiency. The amount of data reuse for each of the three data types in a DNN layer, i.e., input activations (iacts), weights and partial sums (psums), is a function of the layer shape and size. For example, the amount of iact reuse is proportional to the number of output channels as well as the filter size in a layer. Therefore, diminished data dimensions suggest that it is more difficult to exploit data reuse from any specific dimension. 

\begin{figure}
    \centering
    \subfloat[Input activations (iacts)]
    {
        \includegraphics[width=0.47\linewidth]{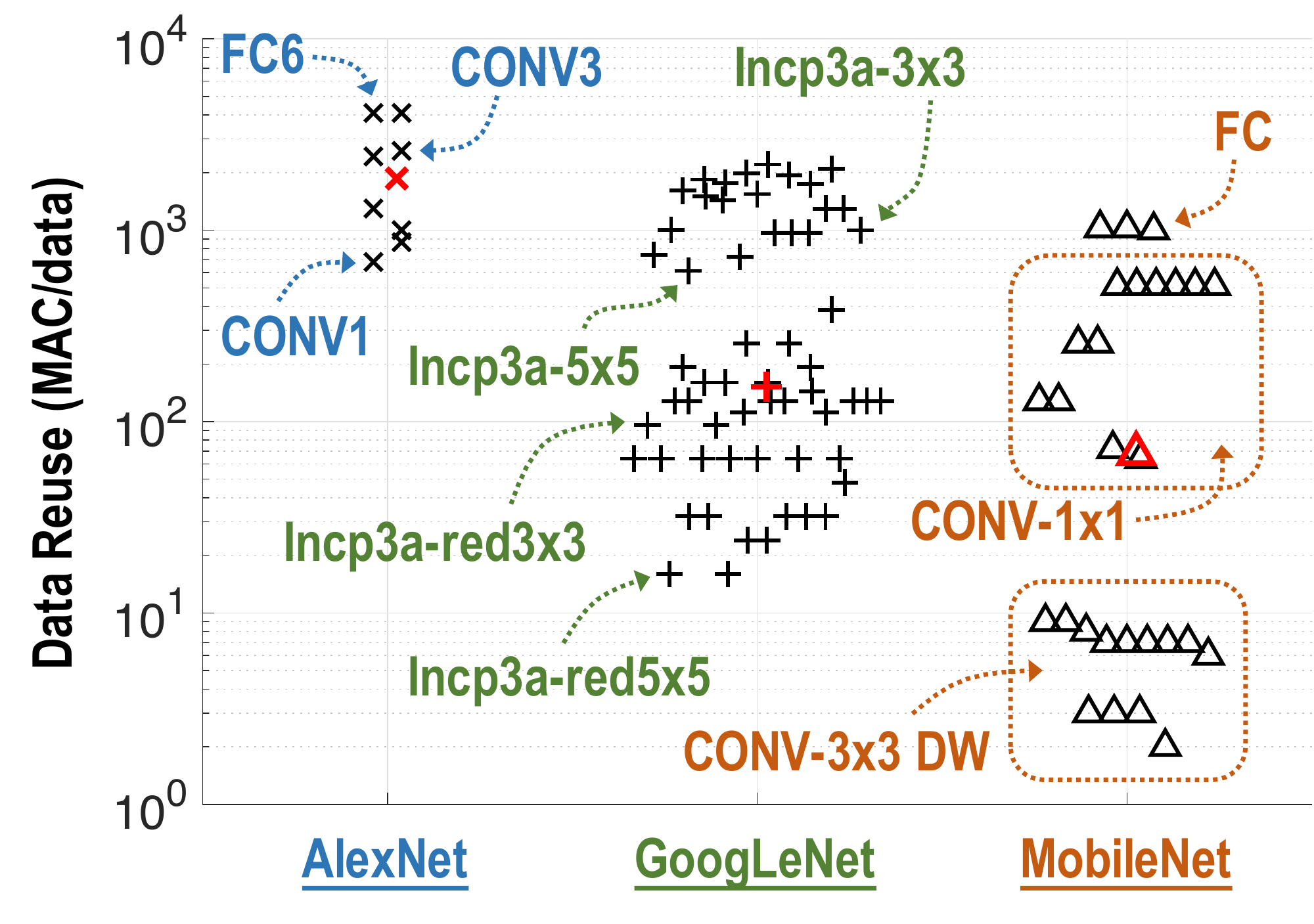}
        \label{fig:nominal_data_reuse_iacts}
    }
    \subfloat[Weights (batch size = 1)]
    {
        \includegraphics[width=0.47\linewidth]{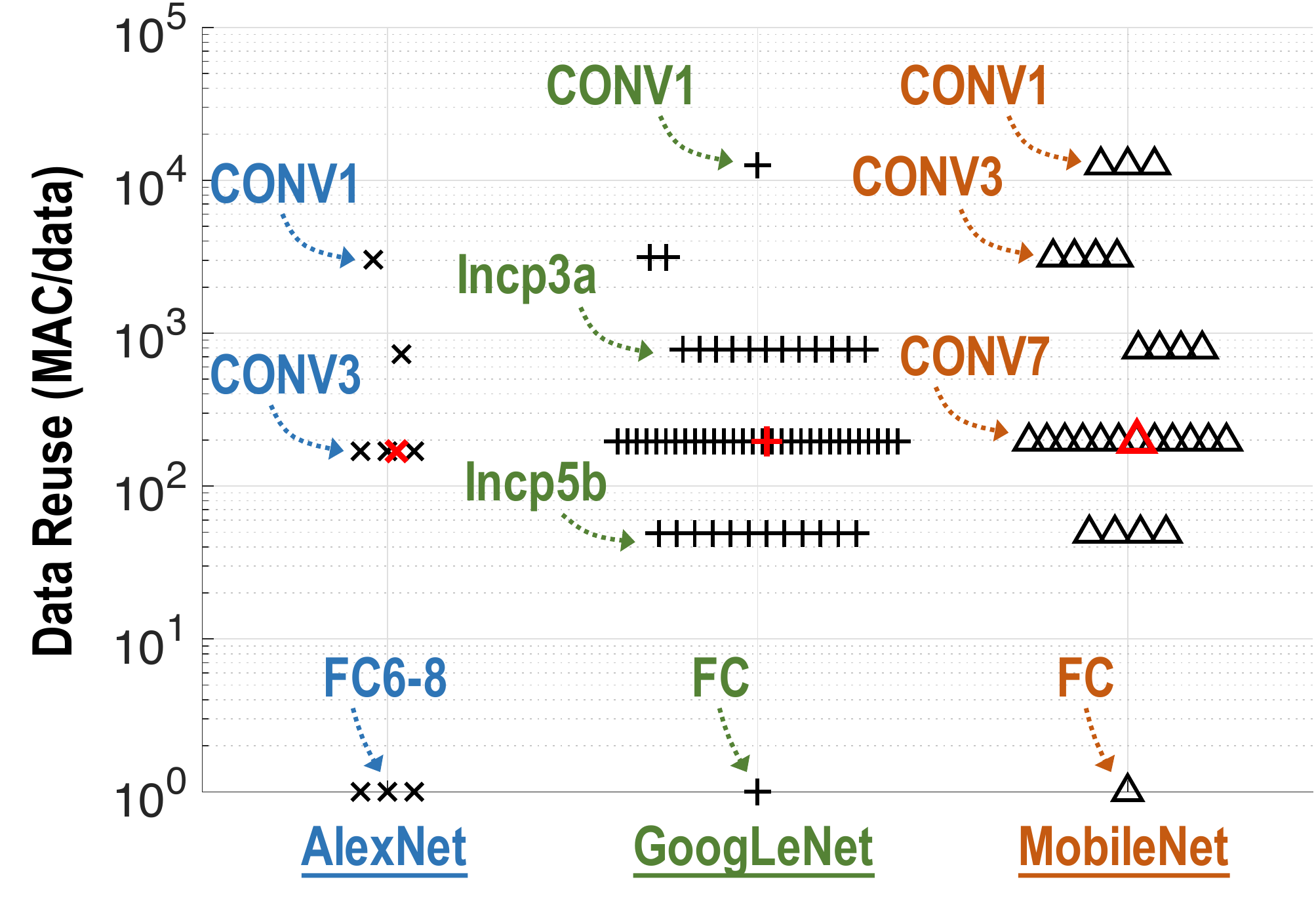}
        \label{fig:nominal_data_reuse_weights}
    }\\
    \subfloat[Partial sums (psums)]
    {
        \includegraphics[width=0.47\linewidth]{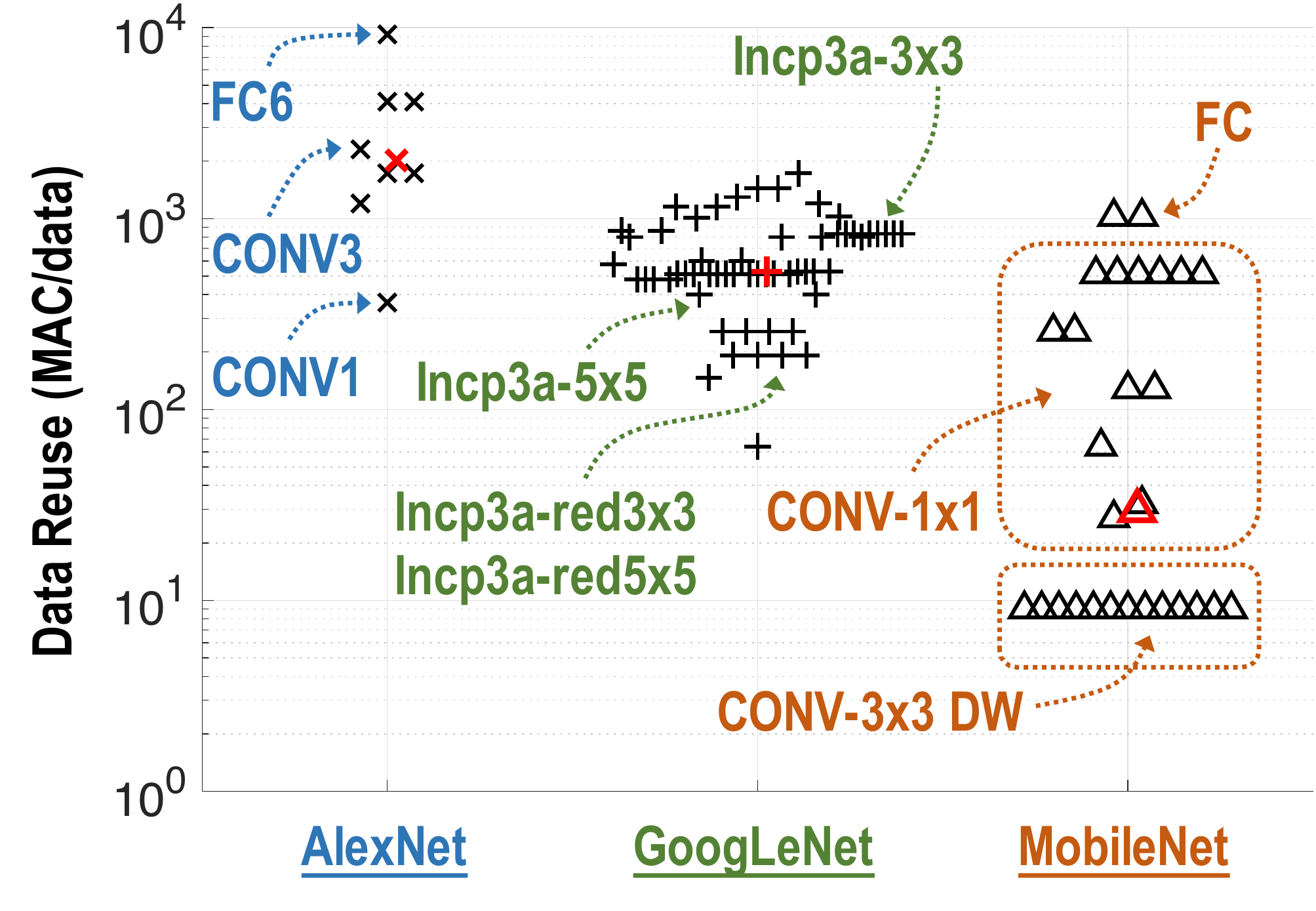}
        \label{fig:nominal_data_reuse_psums}
    }
    \vspace{-5pt}
    \caption{   Data reuse of the three data types in each layer of the three DNNs. Each data point represents a layer, and the red point indicates the median amount of data reuse among all the layers in a DNN. For example, \emph{incp3a-red5$\times$5} means the reduction layer with 5$\times$5 filters in Inception module 3a in GoogLeNet.
            }
    \vspace{-10pt}
    \label{fig:dnn_data_reuse}
\end{figure}

Fig.~\ref{fig:dnn_data_reuse} shows that the variation in data reuse increases in all data types in more recent DNNs, and the amount of reuse also decreases in iacts and psums. This variation and overall reduction in data reuse makes the design of DNN accelerators more challenging in two ways.

\subsubsection{Array Utilization}
\begin{figure}
    \centering
    \subfloat[Spatial accumulation]
    {
        \includegraphics[width=0.40\linewidth]{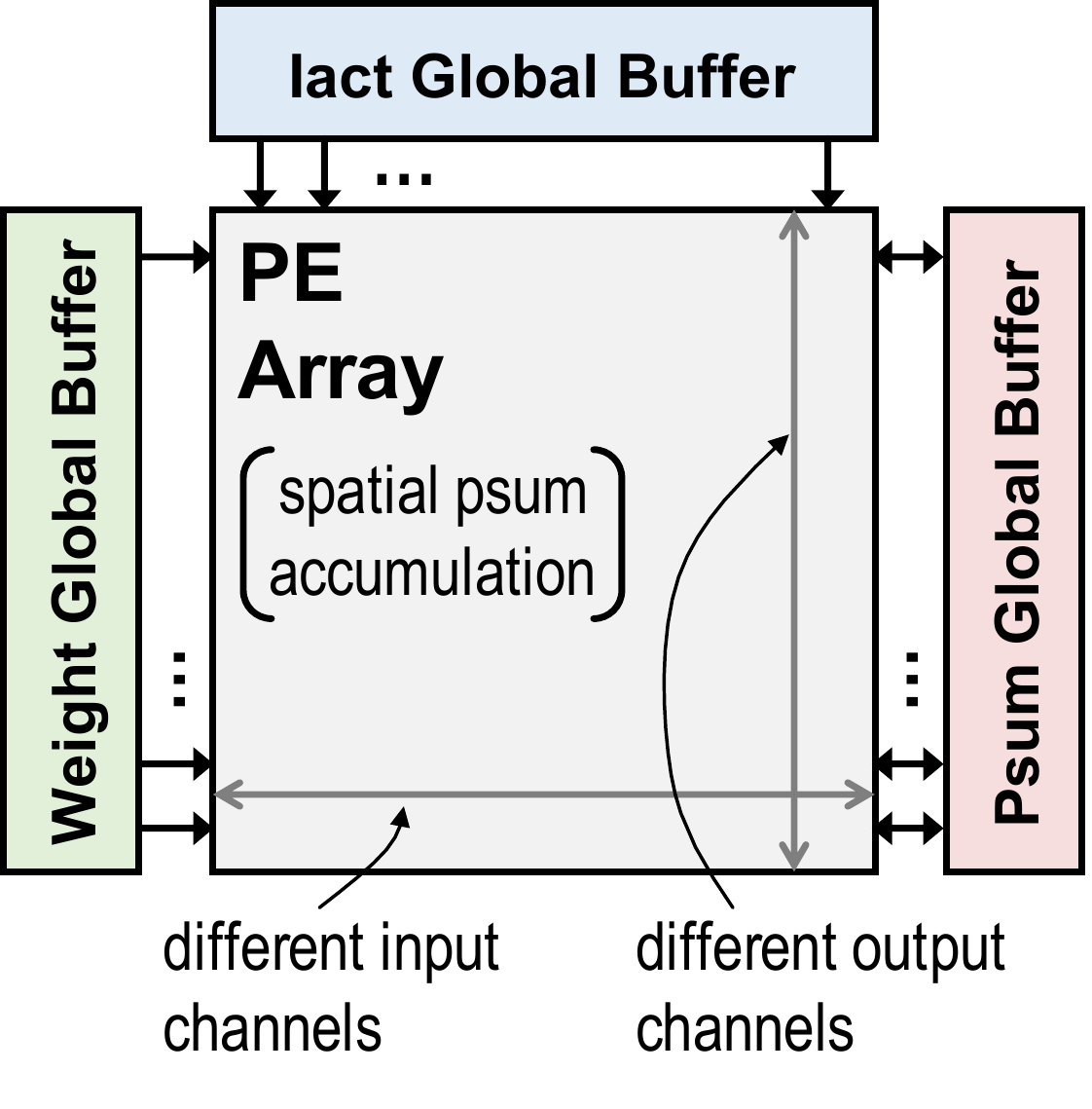}
        \label{fig:spatial_accumulation_array}
    }
    \subfloat[Temporal accumulation]
    {
        \includegraphics[width=0.50\linewidth]{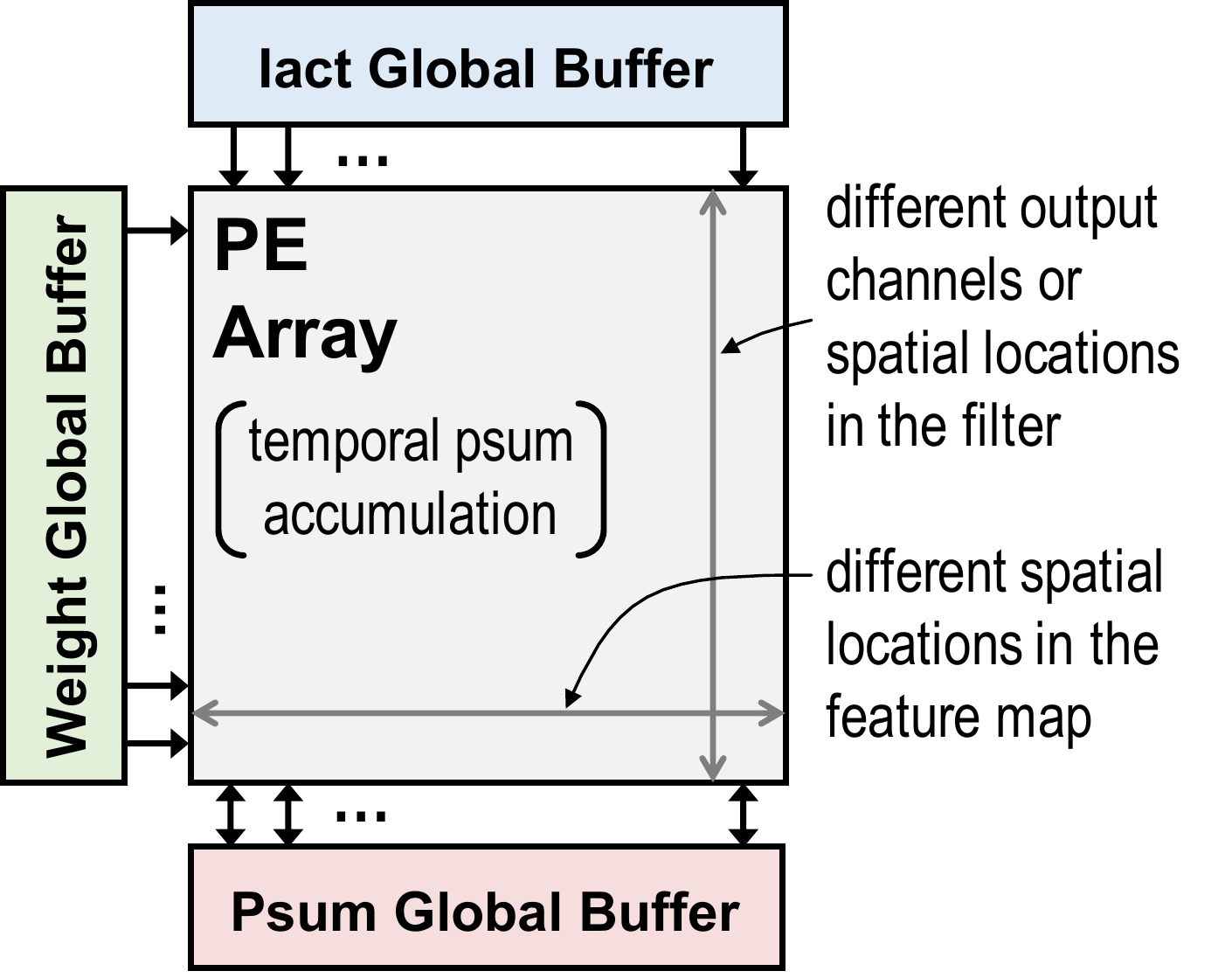}
        \label{fig:temporal_accumulation_array}
    }
    \caption{   Two common DNN accelerator designs: (a) Spatial accumulation array~\cite{arxiv2015-gupta,isca2015-du,nvdla,jouppi2017datacenter}: iacts are reused vertically and psums are accumulated horizontally. (b) Temporal accumulation array~\cite{isscc2017-moons,thinker_vlsi_2017,isscc2018-unpu}: iacts are reused vertically and weights are reused horizontally.
            }
    \vspace{-10pt}
    \label{fig:existing_dnn_accelerators}
\end{figure}

Many existing DNN accelerators~\cite{isscc2017-moons, thinker_vlsi_2017, isscc2018-unpu,  arxiv2015-gupta, isca2015-du, nvdla, jouppi2017datacenter} rely on a set of pre-selected data dimensions to exploit both high parallelism across an array of processing elements (PEs) for high performance and data reuse for high energy efficiency. For instance, Fig.~\ref{fig:existing_dnn_accelerators} shows two designs that are commonly used. A spatial accumulation array architecture (Fig.~\ref{fig:spatial_accumulation_array}), which is often used for a weight-stationary dataflow, relies on both output and input channels to map the operations spatially onto the PE array to exploit parallelism. At the same time, each iact can be reused across the PE array vertically with weights from different output channels, while psums from the PEs in the same row can be further accumulated spatially together before written back to the global buffer. Similarly, a temporal accumulation array architecture (Fig.~\ref{fig:temporal_accumulation_array}), which is often used for a output-stationary dataflow, relies on another set of data dimensions to achieve high compute parallelism. In this case, each iact is still reused vertically across different PEs in the same column, while each weight is reused horizontally across PEs in the same row.

\begin{figure}[t]
    \begin{center}
        \includegraphics[width=0.97\linewidth]{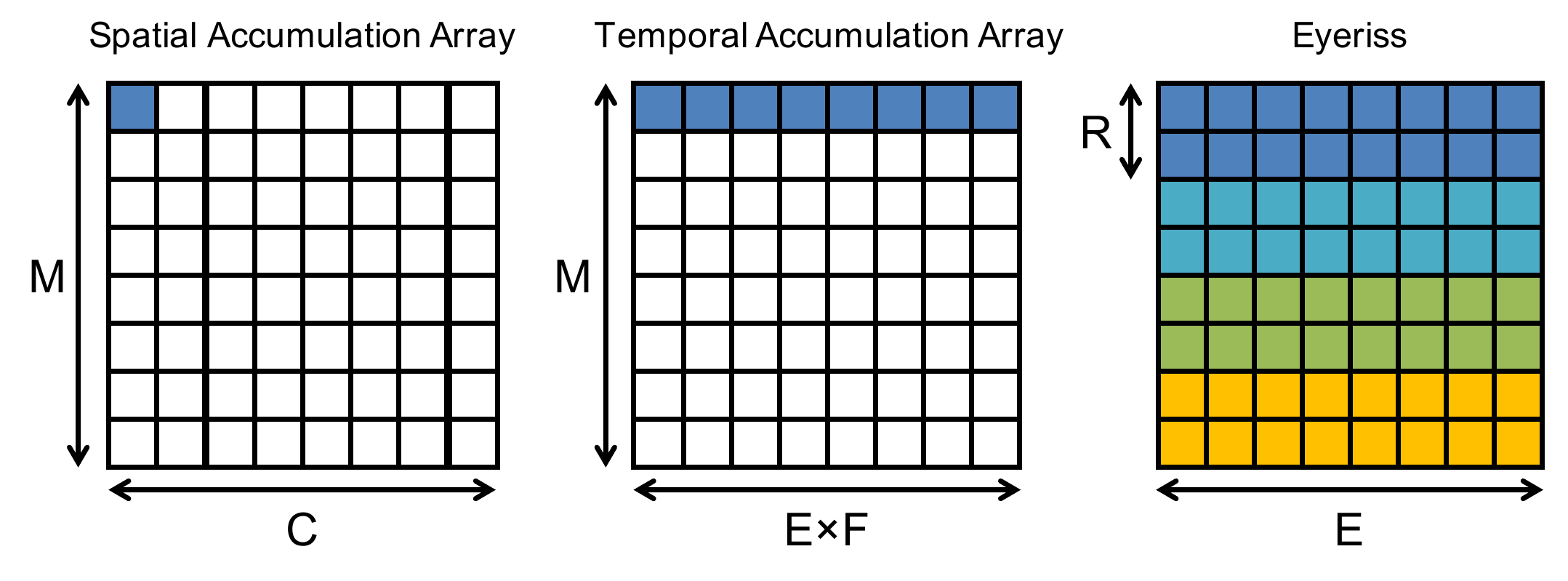}
        \caption{   Array utilization of different architectures for depth-wise (DW) layers in MobileNet. The colored blocks are the utilized part of the PE array. For Eyeriss~\cite{jssc2016-chen}, the different colors denote the parts that run different channel groups ($G$). Please refer to Table~\ref{table:diminishing_dnn_data_dimensions} for the meaning of the variables.
                }
        \label{fig:dw_utilization}
    \end{center}
\end{figure}

When the set of pre-selected data dimensions diminish due to a change in DNN shapes and sizes, e.g., the number of output channels in a layer ($M$) is less than the height of the PE array, efficiency decreases. Specifically, these spatial mapping constraints result in both reduced array utilization (i.e., fewer PEs are used) as well as lower energy efficiency. Furthermore, these inefficiencies are magnified as the size of the PE array is scaled up, because the diminished dimension is even more likely to be unable to fill the array. For example, as shown in Fig.~\ref{fig:dw_utilization}, the aforementioned spatial and temporal accumulation arrays will find it difficult to fully utilize the array due to the lack of input and output channels in the depth-wise (DW) layers in MobileNet. In contrast, Eyeriss~\cite{jssc2016-chen} can still achieve high array utilization under such circumstances by mapping the independent channel groups onto different part of the PE array due to the flexibility of its Row-Stationary (RS) dataflow.

\subsubsection{PE Utilization}
A lower data reuse also implies that a higher data bandwidth is required to keep the PEs busy. If the on-chip network (NoC) for data delivery to the PEs is designed for high spatial reuse scenarios, e.g., a broadcast network, the insufficient bandwidth can lead to reduced utilization of the PEs (i.e., increased stall cycles), which further reduces accelerator performance. For instance, even though Eyeriss can better utilize the array as shown in Fig.~\ref{fig:dw_utilization}, its broadcast NoC (which supports multicast) is not going to provide adequate bandwidth to support high throughput processing at high parallelism, thus the performance will still suffer. However, if the NoC is optimized for high bandwidth scenarios, e.g., many unicast networks, it may not be able to take advantage of data reuse when available.

An additional challenge lies in the fact that all DNNs that the hardware needs to run will \emph{not be known at design time}~\cite{sysml2018-chen}; as a result, the hardware has to be flexible enough to efficiently support a wide range of DNNs. To build a truly flexible DNN accelerator, the new challenge is to design an architecture that can accommodate a wide range of shapes and sizes of DNN layers. In other words, the data has to be flexibly mapped spatially according to the specific shape and size of the layer, instead of with a set of pre-selected dimensions, in order to maximize the utilization of the PE array. Also, the data delivery NoC has to be able to provide high bandwidth when data reuse is low while still being able to exploit data reuse with high parallelism when the opportunity presents itself. 

\subsection{Challenges for Sparse DNNs}

Sparse activations naturally occur in DNNs for several reasons. One is that many DNNs use the rectified linear unit (ReLU) as the activation function, which sets negative values to zero; this sparsity tends to increase in deeper layers and can go above 90\%. Another increasingly important reason is that many popular DNNs are in the form of autoencoders~\cite{noh2015learning,dosovitskiy2015flownet,laina2016deeper} or generative adversarial networks (GAN)~\cite{goodfellow2014generative}, which contain decoder layers that use zero insertion to up-sample the input feature maps, resulting in over 75\% zeros.

There has also been a significant amount of work to make the weights in a DNN sparse. Various metrics are used to decide which weights to prune (i.e., set to zero), including saliency~\cite{nips1990-lecun-opt_brain_damage}, magnitude~\cite{nips2015-han}, and energy consumption~\cite{yang2016}.  These pruned networks have weight sparsity of up to 90\%.

Sparsity in weights and activations can be translated into improved energy efficiency and processing speed in two ways:
(1) The MAC computation can be either gated or skipped; the former reduces energy while the latter reduces both energy and cycles. (2) The weights and activations can be compressed to reduce the amount of storage and data movement; the former reduces energy while the latter reduces both energy and cycles. However, it is quite challenging to design DNN accelerators that can actually harness these benefits from sparsity due to the following reasons:

\subsubsection{Irregular Accesses Patterns}
Computation gating can effectively translate sparsity in both weights and activations into energy savings, and its implementation can be realized at a low cost by recognizing if either the weight or activation is zero and gating the datapath switching and memory accesses accordingly. For example, Eyeriss has demonstrated gating for sparse activations. 

To improve throughput in addition to saving energy consumption, it is desirable to skip the cycles of processing MACs that have zero weights or iacts. However, this requires more complex read logic as it must find the next non-zero value to read without wasting cycles reading zeros. A natural way to address this issue is to keep the weights and iacts in a compressed format that can indicate the location of the next non-zero relative to the current one. However, compressed formats tend to be of variable length and thus must be accessed sequentially. This makes it difficult to divide up the compressed data for parallel processing across PEs without compromising compression efficiency. Furthermore, this presents a challenge if sparsity in \emph{both} weights and activation must be simultaneously recognized, as it is difficult to `jump ahead' (e.g., skip non-zero weights when the corresponding iact is zero) for many of the most efficient compression formats; the irregularity introduced by jumping ahead also prevents the use of pre-fetching as a means of improving throughput. Thus, the control logic to process the compressed data can be quite complex and adds overhead to the PEs.

Accordingly, there has been limited hardware in this space. Cnvlutin~\cite{albericio2016cnvlutin} only supports skipping cycles for activations and does not compress the weights, while Cambricon-X~\cite{zhang2016cambricon} does not keep activations in compressed form. Due to the complexity of the logic to skip cycles for both weights and activations, existing hardware for sparse processing is typically limited to a specific layer type. For instance, EIE targets fully-connected (FC) layers~\cite{han2016eie}, while SCNN targets convolutional (CONV) layers~\cite{isca2017-parashar}.

\subsubsection{Workload Imbalance and PE Utilization}
With computation skipping for sparse data, the amount of work to be performed at each PE now depends on sparsity. Since the number of non-zero values varies across different layers, data types, or even regions within the same filter or feature map, it creates an imbalanced workload across different PEs and the throughput of the entire DNN accelerator will be bounded by the PE that has the most non-zero MACs. This leads to a decrease in PE utilization. 

\subsection{Contributions of This Work}

To address these challenges, we present Eyeriss v2, a flexible architecture for DNN processing that can adapt to a wide range of filter shapes and sizes used in compact DNNs such as MobileNet. This is achieved through the design of a highly flexible on-chip network (NoC), which is currently the bottleneck for dealing with a more diverse set of DNNs. In addition, Eyeriss v2 also supports sparse DNNs by exploiting the sparsity in the weights and activations across a variety of DNN layers and translates them into improvements in both energy efficiency and processing speed. Finally, similar to the original Eyeriss, Eyeriss v2 does not make any assumption about whether the total storage capacity required by a DNN layer can fit on-chip or not; instead, it optimizes the way to tile data of different types to achieve high on-chip reuse and energy efficiency. In summary, the contributions of this paper include:

\begin{itemize}
    \item A novel NoC, called hierarchical mesh, that is designed to adapt to a wide range of bandwidth requirements. When data reuse is low, it can provide high bandwidth (via unicast) from the memory hierarchy to keep the PEs busy; when data reuse is high, it can still exploit spatial data reuse (via multicast or broadcast) to achieve high energy efficiency. For a compact DNN such as MobileNet, the hierarchical mesh increases the throughput by 5.6$\times$ and energy efficiency by 1.8$\times$. (Section~\ref{sec:noc})
    
    \item A PE that exploits the sparsity in weights and activations to achieve improved throughput and energy efficiency across a variety of DNN layers. Data is kept in compressed sparse column (CSC) format for both on-chip processing and off-chip access to reduce storage and data movement costs. Mapping of the weights to a PE is performed by taking the sparsity into account to increase reuse within PE, and can therefore reduce the impact of workload imbalance. Overall, exploiting sparsity results in an additional 1.2$\times$ and 1.3$\times$ improvement in throughput and energy efficiency, respectively, for MobileNet. (Section~\ref{sec:sparse_processing})
    
    \item A flexible accelerator, Eyeriss v2, that combines the above contributions to efficiently support both compact and sparse DNNs. Eyeriss v2 running sparse MobileNet is 12.6$\times$ faster and 2.5$\times$ more energy efficient than the original Eyeriss (scaled to the same number of PEs and storage capacity as Eyeriss v2), i.e., Eyeriss v1, running MobileNet (49.2M MACs). Eyeriss v2 is also 42.5$\times$ faster and 11.3$\times$ more energy efficient with sparse AlexNet compared to Eyeriss v1 running AlexNet (724.4M MACs). Finally, Eyeriss v2 running sparse MobileNet is 225.1$\times$ faster and 42.0 $\times$ more energy efficient than Eyeriss v1 running AlexNet. It is evident that supporting sparse and compact DNNs have a significant impact on speed and energy consumption.
    (Section~\ref{sec:results})
\end{itemize}
\section{Architecture Overview}
\label{sec:arch_overview}

\begin{figure}[t]
    \centering
    \includegraphics[width=0.97\linewidth]{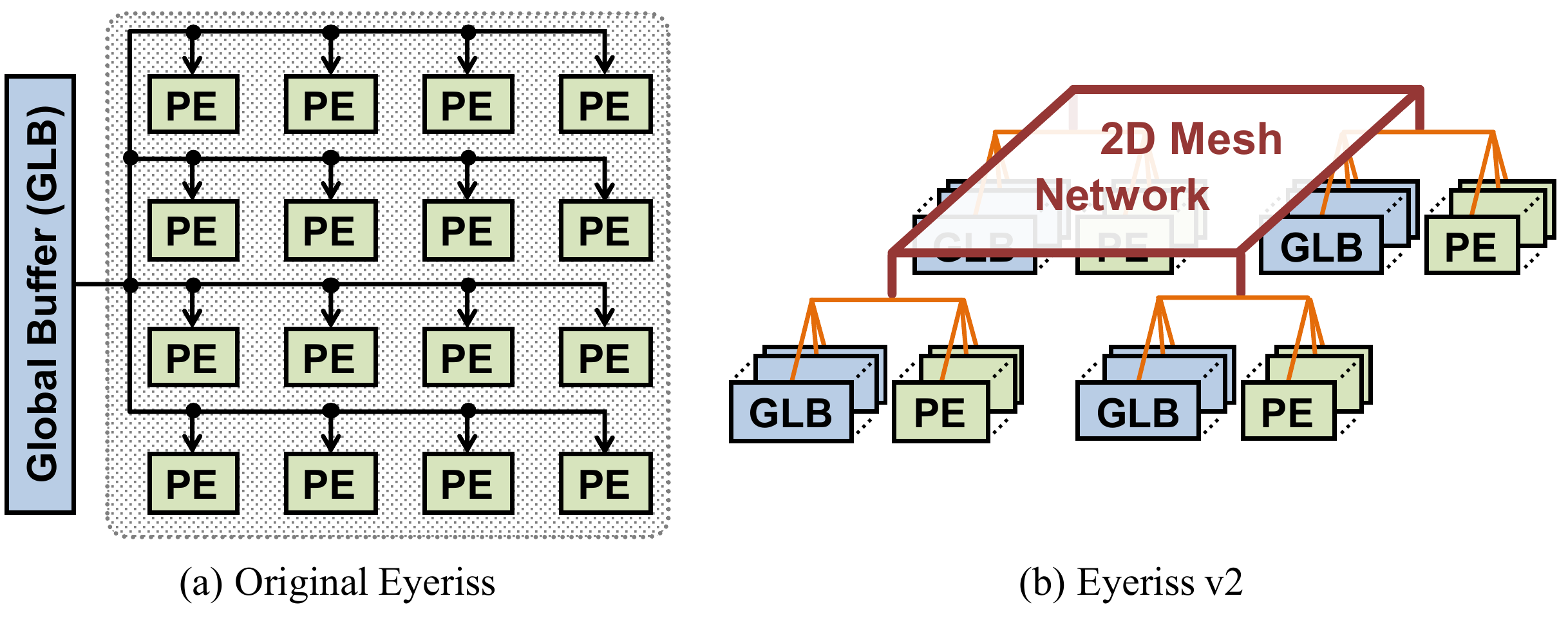}
    \caption{   Comparison of the architecture of original Eyeriss and Eyeriss v2.
            }
    \label{fig:eyeriss_v1_vs_v2}
\end{figure}

Fig.~\ref{fig:eyeriss_v1_vs_v2} shows a comparison between the original Eyeriss~\cite{jssc2016-chen} and the Eyeriss v2 architecture. Similar to the original Eyeriss architecture, Eyeriss v2 is composed of an array of processing elements (PE), where each PE contains logic to compute multiply-and-accumulate (MAC) and local scratch pad (SPad) memory to exploit data reuse, and global buffers (GLB), which serve an additional level of memory hierarchy between the PEs and the off-chip DRAM. Therefore, both the original Eyeriss and Eyeriss v2 have a two-level memory hierarchy. The main difference is that Eyeriss v2 uses a hierarchical structure, where the PEs and GLBs are grouped into clusters in order to support a flexible on-chip network (NoC) that connects the GLBs to the PEs at low cost; in contrast, the original Eyeriss used a flat multicast NoC between the GLB and PEs. As with the original Eyeriss, Eyeriss v2 uses separate NoCs to transfer each of the three data types, i.e., input activation (iact), weight, and partial sums (psums), between the GLBs and PEs, with each NoC tailored for the corresponding dataflow of that data type. Details of the NoC are described in Section~\ref{sec:noc}.

\begin{figure}
    \begin{center}
        \includegraphics[width=0.97\linewidth]{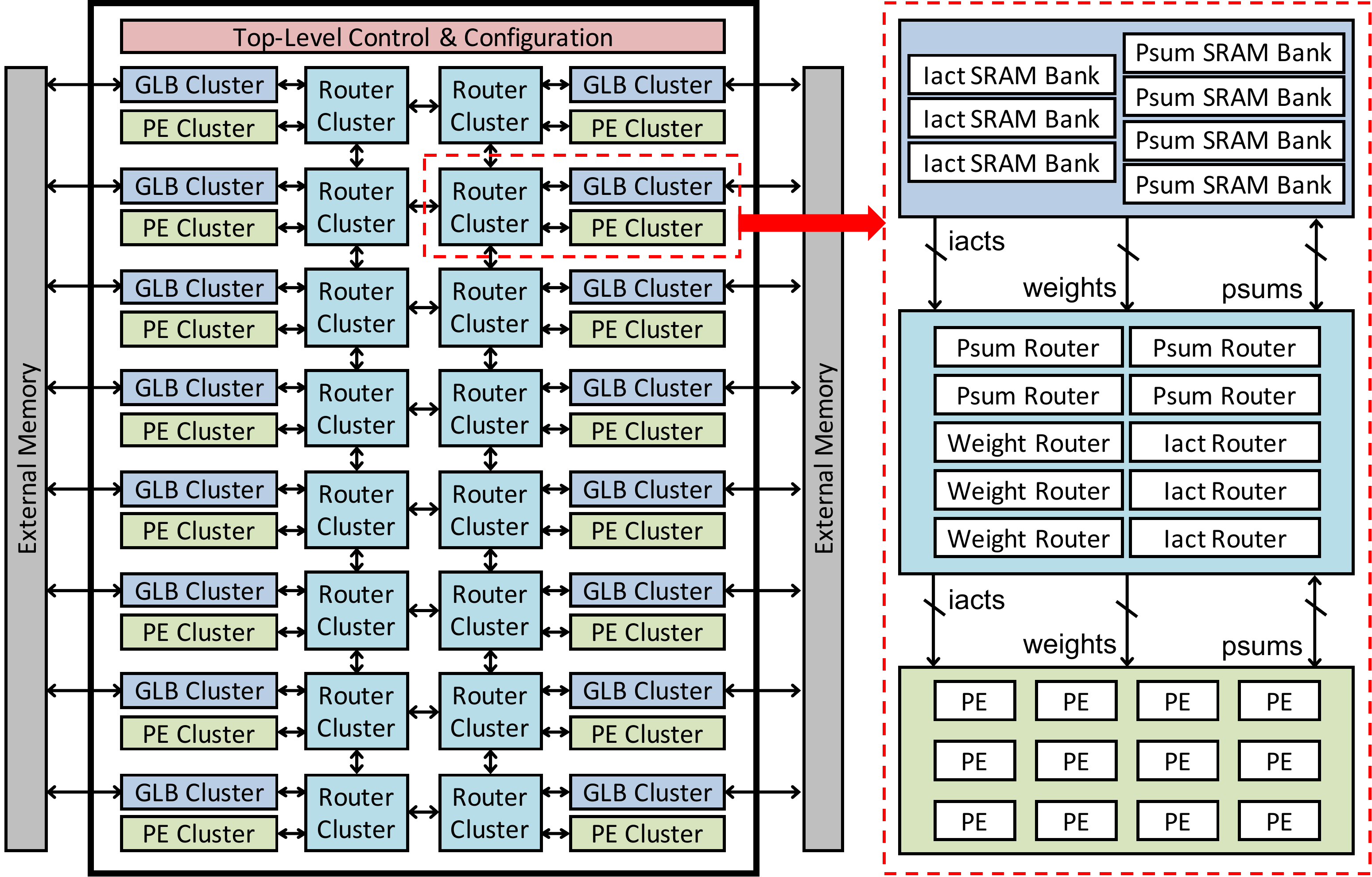}
        \caption{   Eyeriss v2 top-level architecture.
                }
        \label{fig:eyeriss_v2_top}
    \end{center}
\end{figure}

Fig.~\ref{fig:eyeriss_v2_top} shows the top-level architecture of Eyeriss v2 and Table~\ref{table:hierarchy} summarizes the components in the architecture. It consists of 16 PE clusters and 16 GLB clusters arranged in an 8$\times$2 array. Each PE cluster contains 12 PEs arranged in a 3$\times$4 array. Each GLB cluster has a capacity of 12 KB and consists of SRAMs that are banked for different data types: iacts have three banks, each of which is 1.5 kB, and psums have four banks, each of which is 1.875 kB.  

\begin{table}
    \centering
    \footnotesize
    \begin{tabular}{|l|l|}
        \hline
        \textbf{Hierarchy}                          &   \textbf{\# of Components}   \\
        \Xhline{3\arrayrulewidth}
        \multirow{3}{*}{\textbf{Cluster Array}}     &   8$\times$2 PE clusters      \\
                                                    &   8$\times$2 GLB clusters     \\
                                                    &   8$\times$2 router clusters  \\
        \hline
        \textbf{PE cluster}                          &   3$\times$4 PEs              \\
        \hline
        \multirow{2}{*}{\textbf{GLB cluster}}       &   3$\times$1.5 kB SRAM banks for iacts      \\
                                                    &   4$\times$1.875 kB SRAM banks for psums      \\
        \hline
        \multirow{3}{*}{\textbf{router cluster}}    &   3 iact routers (4 src/dst ports, 24b/port)    \\
                                                    &   3 weight routers (2 src/dst ports, 24b/port)    \\
                                                    &   4 psum routers (3 src/dst ports, 40b/port)    \\
        \hline
    \end{tabular}
    \vspace{+5pt}
    \caption{   Eyeriss v2 architecture hierarchy.
            }
    \label{table:hierarchy}
\end{table}

A hierarchical NoC is used to connect the PEs and GLBs: the PE and GLB clusters are connected through 2D mesh on-chip networks that consist of router clusters. Within each router cluster, there are 3, 3, and 4 routers for iact, weight and psum, respectively. Between the PE cluster and the router cluster, an all-to-all NoC is used to connect all the PEs to the routers for each data type. Between the GLB cluster and the router cluster, each router is paired with a specific port of the GLB cluster, which can read from and write to one SRAM bank or off-chip I/O. Therefore, data from either off-chip or a GLB cluster first goes into the router cluster, and then can be unicast to the local PE cluster, multicast to PE clusters on the same row or column in the mesh network, or broadcast to all PE clusters. The decision is based on the shape and size of the DNN layer and the processing dataflow. The design motivation and implementation details of this hierarchical mesh network and the dataflow are described in Section~\ref{sec:noc}.

The data movement through the two-level memory hierarchy on Eyeriss v2 is as follows:
\begin{itemize}
    \item \emph{iacts} are read from off-chip into the GLB cluster, where they can be stored into the GLB memory or get passed directly to the router cluster depending on the configuration. 
    \item \emph{psums} are always stored in the GLB memory once they get out of the PE cluster. The final output activations skip the GLB cluster and go directly off-chip. 
    \item \emph{weights} are not stored in GLB and get passed to the router clusters and eventually stored in the SPads in each PE directly.
\end{itemize}
Eyeriss v2 adopts the Row-Stationary (RS) dataflow~\cite{isca2016-chen} used in the original Eyeriss, and further explores tiling the MAC operations spatially across PEs through any layer dimension, including the channel group dimension ($G$ in Table~\ref{table:diminishing_dnn_data_dimensions}). This is especially important for layers such as the depth-wise (DW) CONV layers in MobileNet, which lacks the input and output channels that are commonly used for spatial tiling and therefore greatly improves the array utilization.

Each PE contains multiply-and-accumulate (MAC) datapaths designed to process 8-bit fixed-point iacts and weights, which is the commonly accepted bitwidth for inference. Since many layers receive iacts after ReLU, the iacts can be set to either signed or unsigned, which further extends the scale of iact representation. Psums are accumulated at 20-bit precision, which has shown no accuracy impact in our experiments. When the accumulation is done, the 20-bit psums are converted back to 8-bit output activations and sent off-chip. The PE contains separate SPads for iact, psum and weights. Details of the PE architecture are described in Section~\ref{sec:sparse_processing}.

Finally, Eyeriss v2 has a two-level control logic similar to the original Eyeriss. The system-level control coordinates the off-chip data accesses and data traffic between the GLB and PEs, and the lower-level control is within each PE and controls the progress of processing of each PE independently. The chip can be reconfigured to run the dataflow that maximizes the energy efficiency and throughput for the processing of each DNN layer. This includes setting up the specific data traffic pattern of the NoCs, data accesses to the GLB and SPads, and workload distribution for each PE. For each layer, a 2134-bit command that describes the optimized configuration is sent to the chip and accessed statically throughout the processing of this layer. Only one layer is processed at a time. When the processing for a layer is done, the chip is reconfigured for the processing of the next layer.
\section{Flexible Hierarchical Mesh On-Chip Network}
\label{sec:noc}

One of the key features required to support compact DNNs is a flexible and efficient on-chip network (NoC). This section will provide details on the implementation of the NoC in Eyeriss v2 as well as describe how the NoC is configured for various use cases.

\subsection{Motivation}
\label{sec:noc_motivation}

The NoC is an indispensable part of modern DNN accelerators, and its design has to take the following factors into consideration: (1) support processing with high parallelism by efficiently delivering data between storage and datapaths, (2) exploit data reuse to reduce the bandwidth requirement and improve energy efficiency, and (3) can be scaled at a reasonable implementation cost.

\begin{figure}[t]
    \begin{center}
        \includegraphics[width=0.97\linewidth]{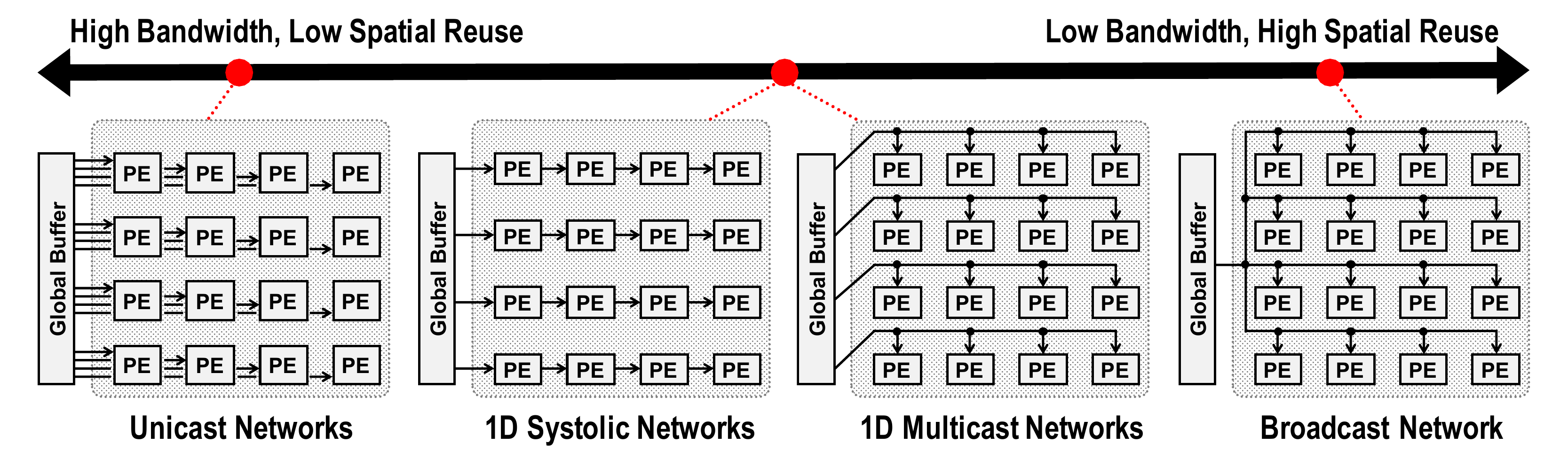}
        \caption{   Common NoC Designs.
                }
        \label{fig:common_noc_designs}
    \end{center}
\end{figure}

Fig.~\ref{fig:common_noc_designs} shows several NoC designs commonly used in DNN accelerators. Due to the property of DNN that data reuse for all data types cannot be maximally exploited simultaneously, a mixture of these NoCs is usually adopted for different data types. For example, a DNN accelerator can use a 1D horizontal multicast network to reuse the same weight across PEs in the same row and a 1D vertical multicast network to reuse the same iact across PEs in the same column. This setup will then require an unicast network that gathers the unique output activations from each PE. This combination, however, implies that each weight needs to have the amount of reuse with different iacts at least equal to the width of the PE array, and the number of iact reuse with different weights at least equal to the height of the PE array. If these conditions are not fulfilled, the PE array will not be fully utilized, which will impact both throughput and energy efficiency.

\begin{figure*}
    \centering
    \includegraphics[width=0.97\linewidth]{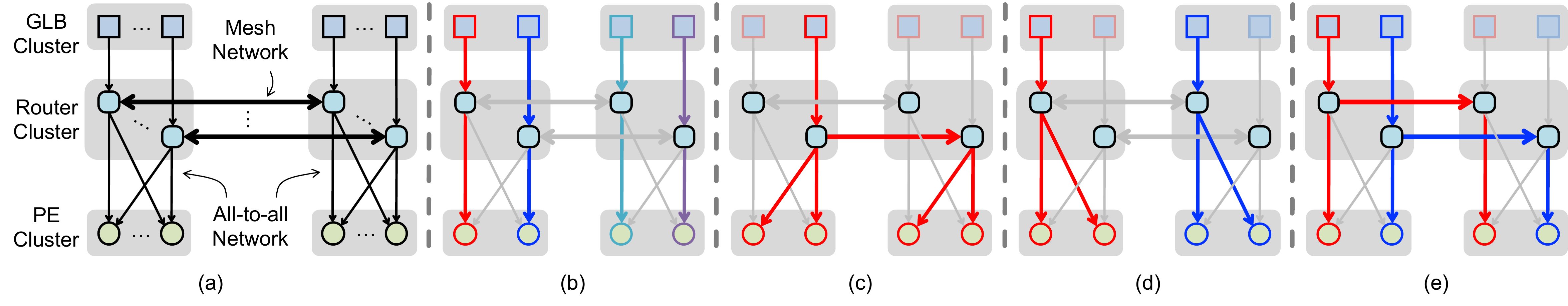}
    \caption{   (a) High-level structure of the hierarchical mesh network (HM-NoC), and its different operating modes: (b) High bandwidth mode, (c) High reuse mode, (d) grouped-multicast mode, and (e) interleaved-multicast mode. In each mode, the colored arrows show the routing path; different colors denote the path for unique data.
            }
    \label{fig:hmn}
\end{figure*}

While it was easy to satisfy such conditions with large DNNs, the rise of compact DNN models has made this design approach less effective. The key reason is that it is much more difficult to assume the amount of data reuse or required data bandwidth for each data type, as it will vary across layers and DNN models. For example, the lack of input or output channels in the depth-wise layers of MobileNet or in the bottleneck layers of ResNet and GoogLeNet has made it very difficult to efficiently utilize the aforementioned example well due to its rigid NoC design. In layers such as fully-connected layers, commonly used in RNNs and CNNs, it will also require a large batch size to improve the amount of reuse for weights, which can be challenging in real-time applications that are sensitive to the processing latency.

The varying amount of data reuse for each DNN data type across different layers or models pose a great challenge to the NoC design. The broadcast network can exploit the most data reuse, but its low source bandwidth can limit the throughput when data reuse is low. The unicast network can provide the most source bandwidth but misses out on the data reuse opportunity when available. Taking the best from both worlds, an all-to-all network that connects any data sources to any destinations can adapt to the varying amount of data reuse and bandwidth requirements. However, the cost of its design increases quadratically with the number of nodes, e.g., PEs, and therefore is difficult to scale up to the amount of parallelism required for DNN accelerators.

\subsection{High-Level Concept and Use Cases}
\label{ssec:noc_use}

To deal with this problem, we propose the hierarchical mesh network (HM-NoC) in Eyeriss v2 as shown in Fig.~\ref{fig:hmn}a. HM-NoC takes advantage of the all-to-all network, but solves the scaling problem by creating a two-level hierarchy. The all-to-all network is limited within the scope of a cluster at the lower level. In Eyeriss v2, there are only 12 PEs in each cluster, which effectively reduce the cost of the all-to-all network. At the top level, the clusters are further connected with a mesh network. While this example shows a 2$\times$1 mesh, Eyeriss v2 uses a 8$\times$2 mesh. Scaling up the architecture at the cluster level with the mesh network is much easier than with the all-to-all network since the implementation cost increases linearly instead of quadratically.

Fig~\ref{fig:hmn}b to~\ref{fig:hmn}e shows how the HM-NoC can be configured into four different modes depending on the data reuse opportunity and bandwidth requirements.
\begin{itemize}
    \item In the \emph{high bandwidth mode} (Fig.~\ref{fig:hmn}b), each GLB bank or off-chip data I/O can deliver data independently to the PEs in the cluster, which achieves \emph{unicast}. 
    \item In the \emph{high reuse mode} (Fig.~\ref{fig:hmn}c), data from the same source can be routed to all PEs in different clusters, which achieves \emph{broadcast}.
    \item For situations where the data reuse cannot fully utilize the entire PE array with broadcast, different multicast modes, specifically \emph{grouped-multicast} (Fig.~\ref{fig:hmn}d) and \emph{interleaved-multicast} (Fig.~\ref{fig:hmn}e), can be adopted according to the desired multicast patterns.
\end{itemize}

\begin{figure}
    \centering
    \subfloat[CONV layers]
    {
        \includegraphics[width=0.3\linewidth]{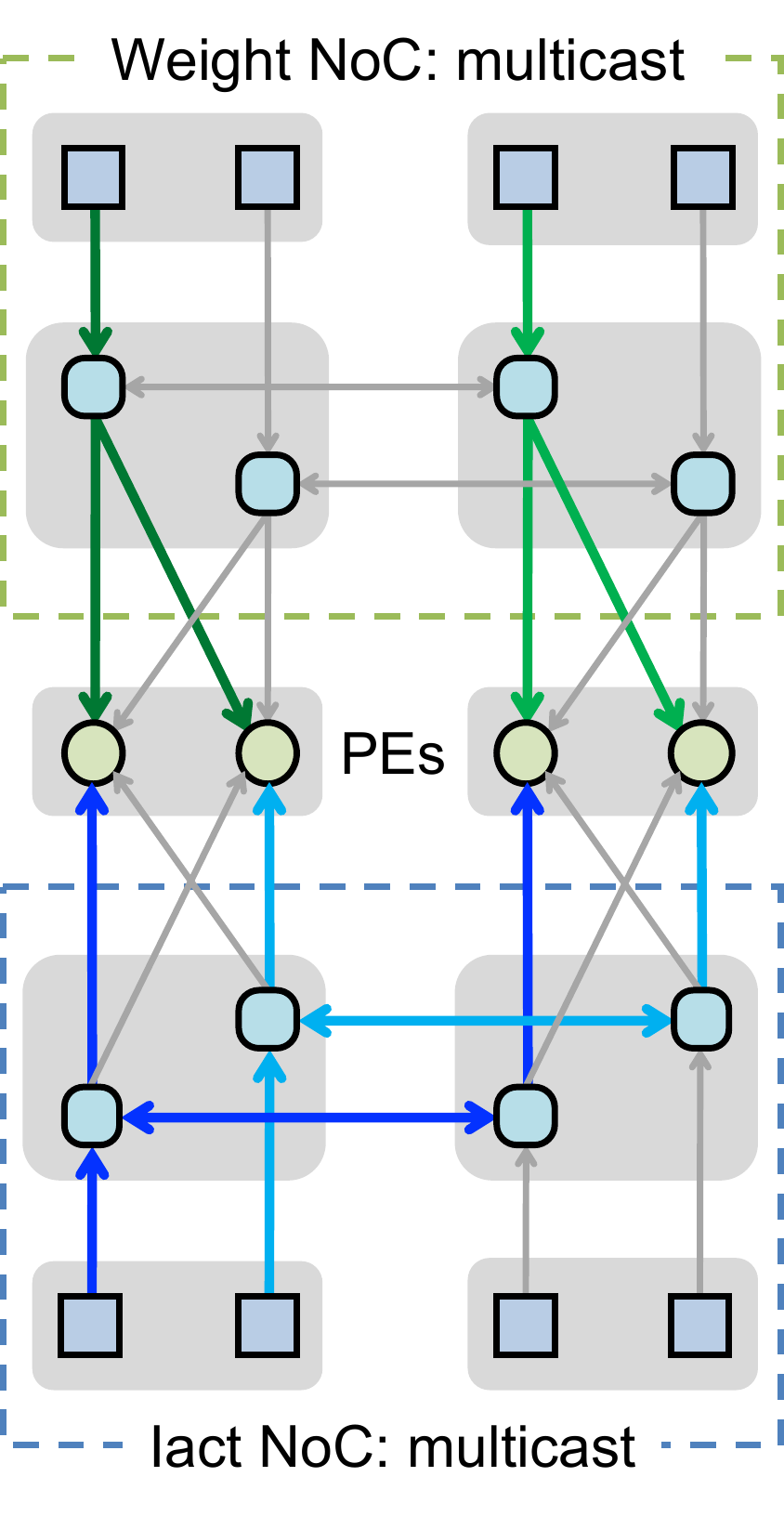}
        \label{fig:hmn_conv_layers}
    }
    \subfloat[DW CONV layers]
    {
        \includegraphics[width=0.3\linewidth]{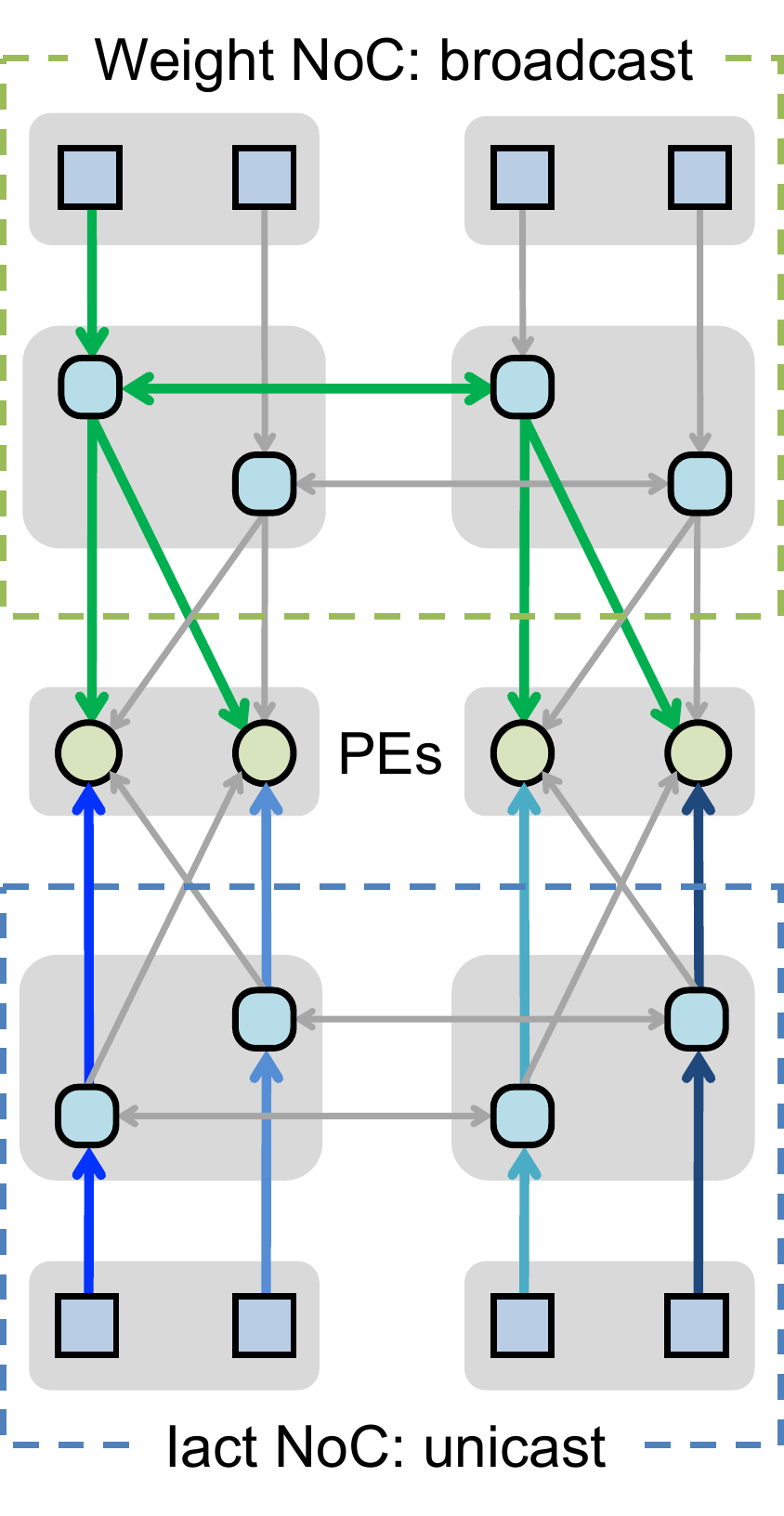}
        \label{fig:hmn_dw_layers}
    }
    \subfloat[FC layers]
    {
        \includegraphics[width=0.3\linewidth]{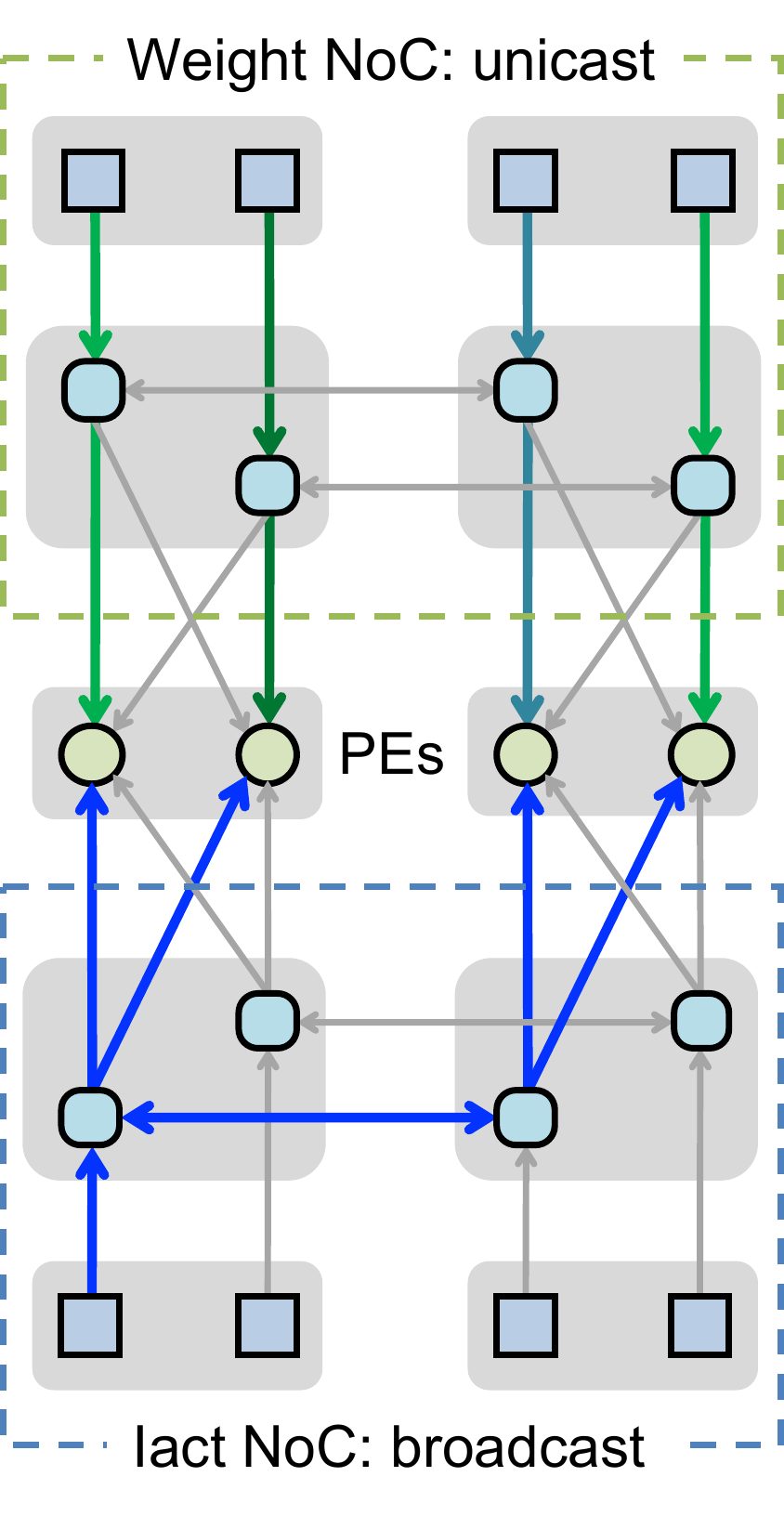}
        \label{fig:hmn_fc_layers}
    }
    \caption{   Examples of weight and iact hierarchical mesh networks configured in different modes for different types of DNN layers: (a) CONV layers; (b) depth-wise (DW) CONV layers; (c) fully-connected (FC) layers. Green arrows and blue arrows show the routing paths in the weight and iact NoC, respectively.
            }
    \label{fig:hmn_examples}
\end{figure}

Fig.~\ref{fig:hmn_examples} shows several example use cases of how HM-NoC adapts different modes for different types of layers. For simplicity, we are only showing a simplified case with 2 PE clusters with 2 PEs in each cluster, and it omits the NoC for psums. However, the same principles apply to NoC for all data types and at larger scales.
\begin{itemize}
    \item Conventional CONV layers (Fig.~\ref{fig:hmn_conv_layers}): In normal CONV layers, there is plenty of data reuse for both iacts and weights. To keep all 4 PEs busy at the lowest bandwidth requirement, we need 2 iacts and 2 weights from the data source (ignoring the reuse from SPad). In this case, either the HM-NoC for iact or weight has to be configured into the grouped-multicast mode, while the other one configured into the interleaved-multicast mode. 
    \item Depth-wise (DP) CONV layers (Fig.~\ref{fig:hmn_dw_layers}): For DP CONV layers, there can be nearly no reuse for iacts due to the lack of output channels. Therefore, we can only exploit the reuse of weights by broadcasting the weights to all PEs while fetching unique iacts for each PE.
    \item Fully-connected (FC) layers (Fig.~\ref{fig:hmn_fc_layers}): Contrary to the DP CONV layers, FC layers usually see little reuse for weights, especially when the batch size is limited. In this case, the modes of iact and weight NoCs are swapped from the previous one: the weights are now unicast to the PEs while the iacts are broadcast to all PEs.
\end{itemize}

\subsection{Implementation Details}
\label{ssec:noc_imp}
To support the various uses cases described in Section~\ref{ssec:noc_use}, each of the HM-NoC employs circuit-switched routing, which mainly consists of muxes and is statically configured by the configuration bits as described in Section~\ref{sec:arch_overview}. Therefore, the implementation cost of each router is very low. A separate HM-NoC is implemented for each data type (iact, psum, and weights) that is tailored for their given dataflow. The specifications of the routers for each data type are summarized in Table~\ref{table:hierarchy}. For iacts and weights, each port has a bitwidth of 24 bits such that it can send and receive three 8b uncompressed iact values or two 12b compressed iact run-data pairs per cycle. Section~\ref{sec:sparse_processing} describes the compression format in more detail.  For psum, each port has a bitwidth of 40-bit to send and receive two psums per cycle. We will now describe how the routers, GLB and PEs are connected in the HM-NoC for each data type.

\subsubsection{HM-NoC for input activations}
\begin{figure}
    \begin{center}
        \includegraphics[width=0.70\linewidth]{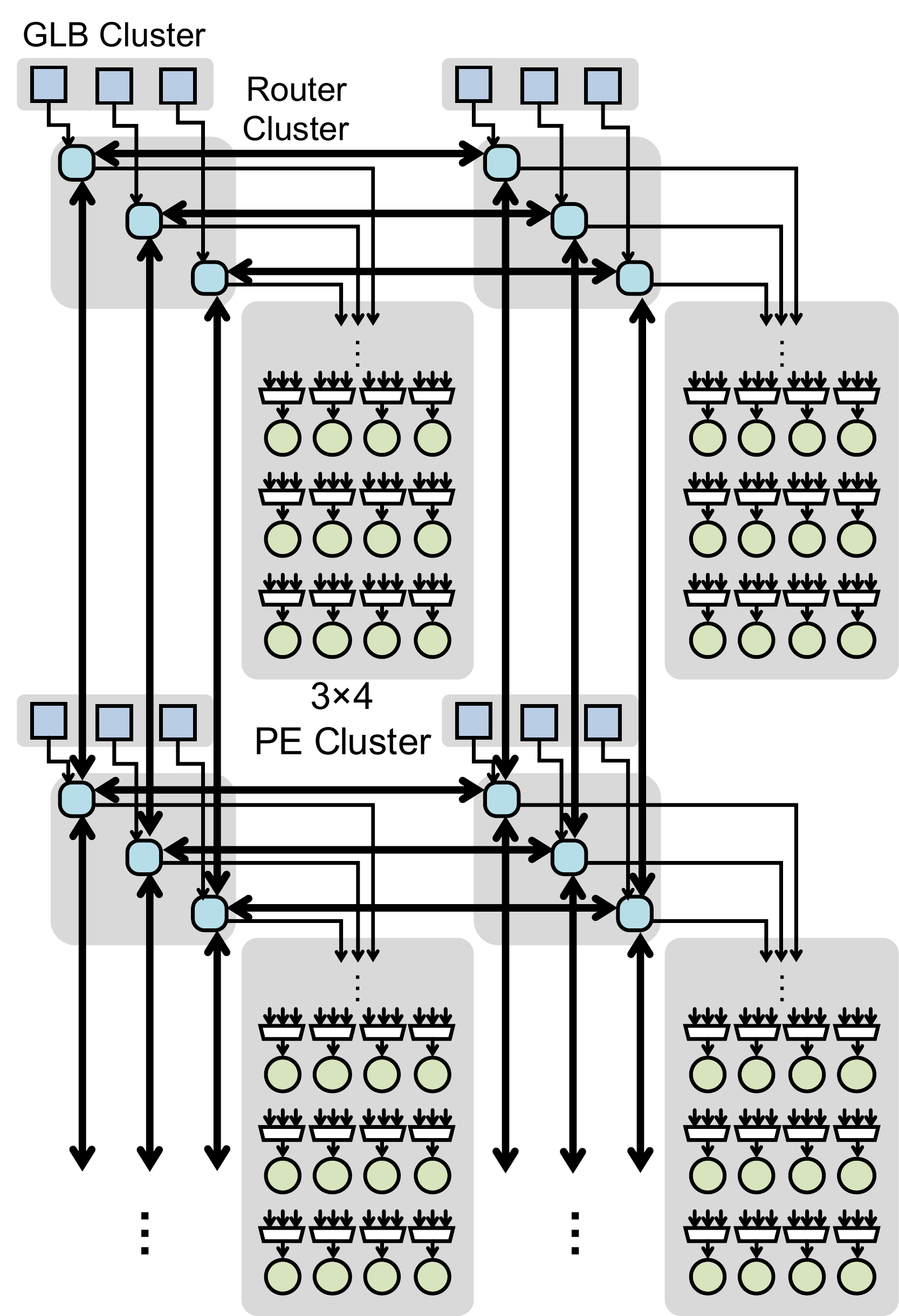}
        \caption{   Hierarchical mesh network for input activations. This only shows the top 2$\times$2 of the entire 8$\times$2 cluster array.
                }
        \label{fig:hm-noc_iacts}
    \end{center}
\end{figure}

The HM-NoC implementation for \emph{iacts} is shown in Fig.~\ref{fig:hm-noc_iacts}. There are three iact routers per router cluster, one for each iact SRAM bank in the GLB cluster. Each router for iact has four source ports (to receive data) and four destination ports (to transmit data). Three of the source and destination ports are used to receive and transmit data from the other clusters in the mesh, which are highlighted with bold arrows in Fig.~\ref{fig:hm-noc_iacts}; while a mesh network typically requires four pairs of source and destination ports, we only require three pairs since we only have 8$\times$2 clusters and thus either the east or west port can be omitted. The fourth source port connects to the GLB cluster to receive data either from the memory bank or off-chip, and the fourth destination port connects to all the 3$\times$4 PEs in the cluster.  Thanks to the all-to-all network in the PE cluster, data from any router can go to any PE in the same cluster.

\begin{figure}
    \begin{center}
        \includegraphics[width=0.85\linewidth]{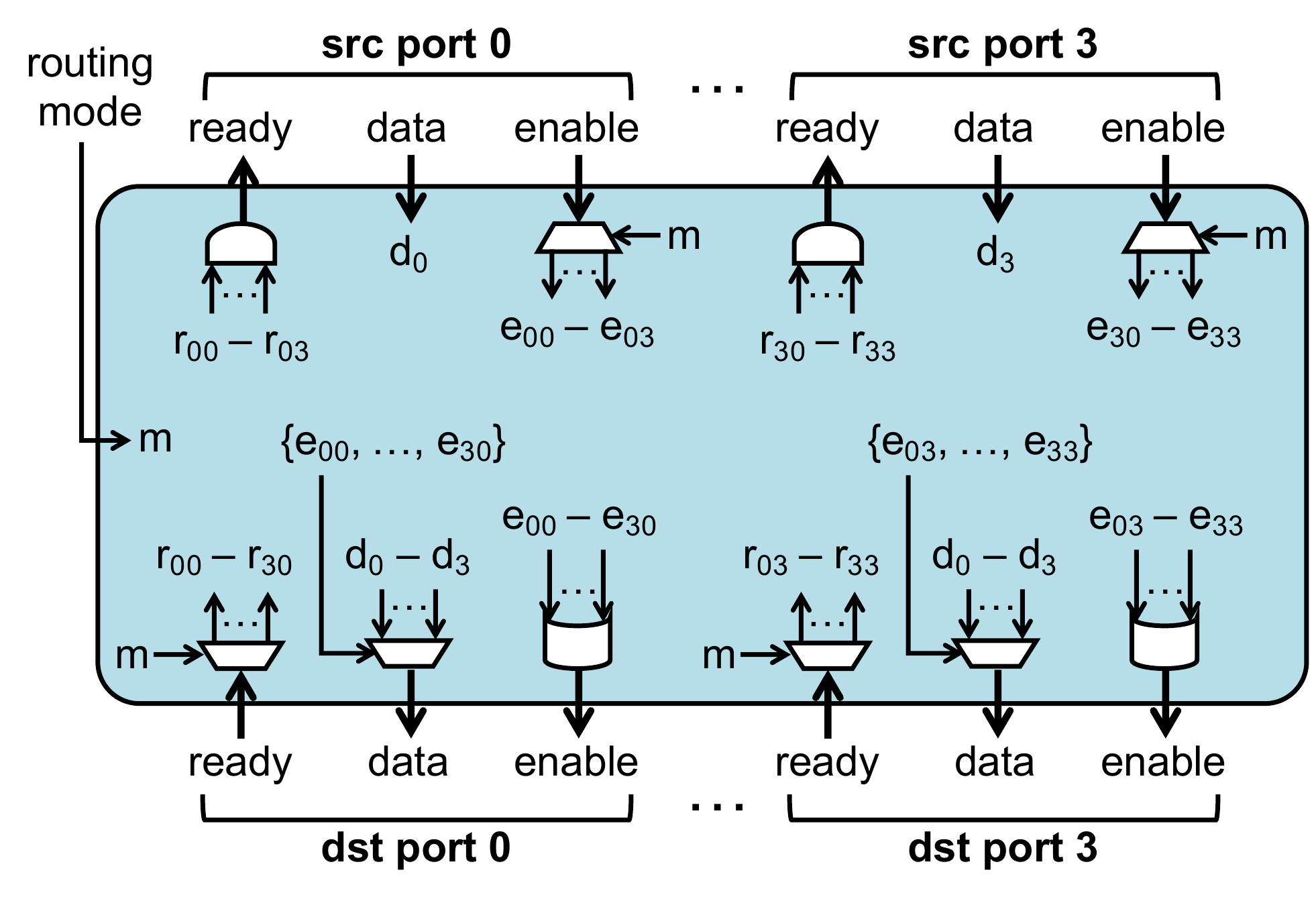}
        \caption{   Implementation details of the mesh network router for input activations. Routers for the other data types use similar logic but with different numbers of ports. $d$, $e$, $r$ and $m$ are data, enable, ready and routing mode signals, respectively.
                }
        \label{fig:iact_mesh_router}
    \end{center}
\end{figure}

Fig.~\ref{fig:iact_mesh_router} shows the implementation details of the mesh network router for iacts. It has four source (src) and four destination (dst) ports. In addition to data ($d$), each port also has two additional signals, ready ($r$) and enable ($e$), for hand-shaking. Each source port generates four enable signals (e.g., $e_{00}$-$e_{03}$), each for one destination port, based on its own enable signal and the statically configured routing mode ($m$). The routing mode can be one of the following: unicast, horizontal multicast, vertical multicast, or broadcast. It determines which ports can be enabled for passing data. For example, in the horizontal multicast mode, ports that connect to other routers in the vertical direction of the mesh network will not be enabled. At each destination port, the destination-specific enable signals from all source ports (e.g., $e_{00}$-$e_{30}$ for destination port 0) go through an OR gate to generate the final enable output. The ready signal from the destination ports to the source ports are generated in a similar fashion with the difference that the source-specific ready signals (e.g., $r_{00}$-$r_{03}$ for source port 0) go through an AND gate to generated the final ready output at each source port. The output data from all source ports ($d_0$-$d_3$) is MUXed at each destination port, and is chosen based on the enable signals from the source ports, i.e., the data from the enabled source port will be passed through.

\subsubsection{HM-NoC for weights}
\begin{figure}
    \begin{center}
        \includegraphics[width=0.75\linewidth]{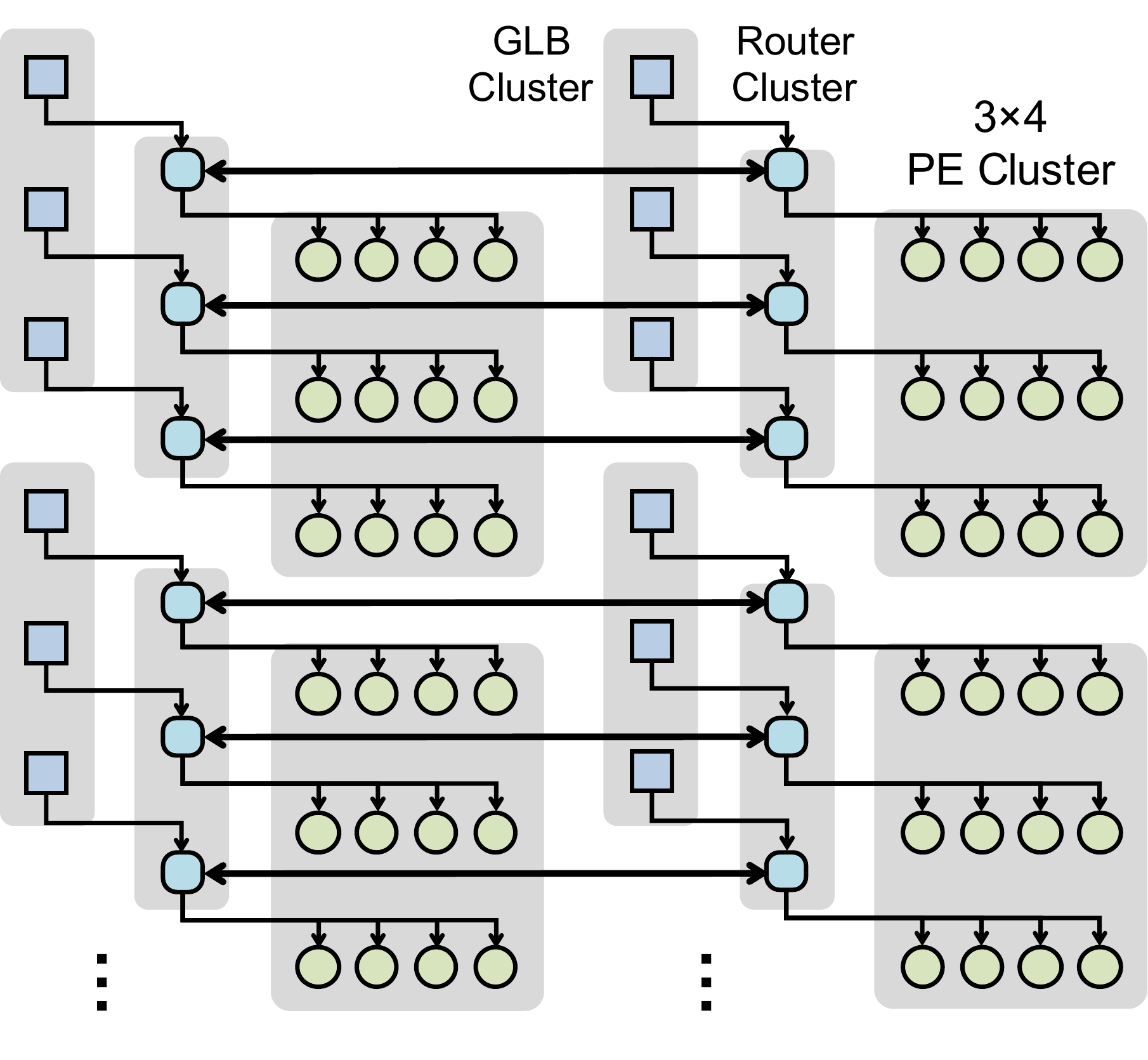}
        \caption{   Hierarchical mesh network for weights. This only shows the top 2$\times$2 of the entire 8$\times$2 cluster array.
                }
        \label{fig:hm-noc_weights}
    \end{center}
\end{figure}

The HM-NoC implementation for \emph{weights} is shown Fig.~\ref{fig:hm-noc_weights}.  There are three weight routers per router cluster, one for each row of PEs within a cluster. Since Eyeriss v2 uses the RS dataflow, a significant amount of weight reuse can be exploited using the SPad within the PE, and only spatial reuse across horizontal PEs needs to be further exploited. Therefore, the implementation of the NoC for the weights can be simplified at both levels of the HM-NoC to reduce cost but still satisfy the flexibility requirements. Specifically, the vertical connections of the 2D mesh between the clusters can be removed. Furthermore, within each cluster, each router only needs to connect to one row of PEs. Accordingly, each weight router has two source ports and two destination ports. A source port and a destination port are used to receive and transmit weights coming from neighboring cluster; again, we only need one pair of ports here since we only have 8$\times$2 clusters and thus either the east or west port can be omitted. The second source port connects to the GLB cluster to receive data from off-chip, while the second destination port connects to one row of PEs within the cluster. The implementation of the mesh network router for weights is similar to that in Fig.~\ref{fig:iact_mesh_router}.

\subsubsection{HM-NoC for partial sums}
\begin{figure}
    \begin{center}
        \includegraphics[width=0.75\linewidth]{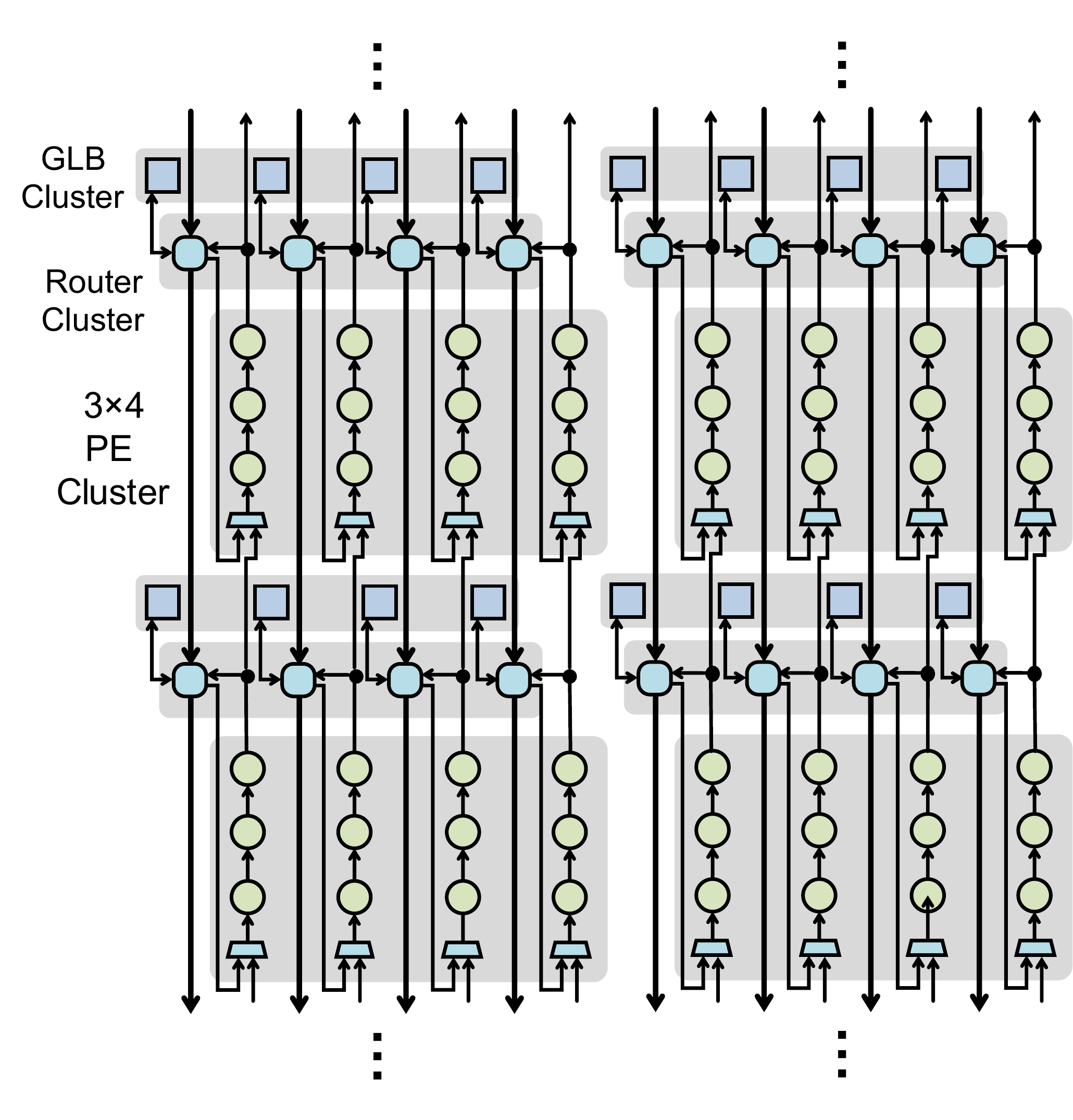}
        \caption{   Hierarchical mesh network for psums. This only shows the a 2$\times$2 portion of the entire 8$\times$2 cluster array.
                }
        \label{fig:hm-noc_psums}
    \end{center}
\end{figure}

The HM-NoC implementation for \emph{psums} is shown in Fig.~\ref{fig:hm-noc_psums}. There are four psum routers per router cluster, one for each psum SRAM bank in the GLB cluster or, equivalently, one for each column of PEs within a cluster. Similar to the weight NoC, the psum NoC is simplified for its given dataflow; specifically, the psums are only allowed to be accumulated across PEs in the vertical direction. This is due to the fact that, in the row-stationary dataflow, weights are reused across PEs horizontally, which makes it impossible to accumulate psums across PEs horizontally. Thus, the horizontal connections of the 2D mesh between the clusters can be removed since psums won't be passed horizontally. Within each cluster, the PEs are vertically connected and each router in the cluster only needs to transmit the psum from the psum bank in the GLB cluster to the bottom of each PE column, and receive the updated psum from the top of the same PE column. Accordingly, each psum router has three source ports and three destination ports. One of the source ports is used to receive data from the neighboring router cluster to the north, while one of the destination ports is used to transmit data to the neighboring router cluster to the south. The second pair of source and destination ports are assigned to the psum bank in the GLB, while the third destination port is assigned to the bottom PE in a column of the PE cluster and the third source port is assigned to the top PE in a column of the PE cluster.

\subsection{Scalability}
\label{ssec:noc_scalability}

A key design focus of the HM-NoC is to enable strong scaling for Eyeriss v2. In other words, as the architecture scales with more PEs, the performance, i.e., throughput, should scale accordingly for the same problem size. Performance, however, is a function of many factors, including the dataflow, NoC design, available on-chip and off-chip data delivery bandwidth, etc. To examine the impact of the HM-NoC, we will assume no limitation on the off-chip bandwidth and no workload imbalance (i.e., no sparsity) in the following scalability experiments.

We profile the performance of Eyeriss v2 at three different scales: 256 PEs, 1024 PEs, and 16384 PEs, where each PE is capable of processing at 1 MAC/cycle. The PE cluster for all scales has a fixed array size of 4$\times$4 PEs, and the number of PE clusters scales at 4$\times$4, 8$\times$8, and 32$\times$32. For comparison, we also examine the scalability of the original Eyeriss, i.e., Eyeriss v1, at the same set of scales. For Eyeriss v1, the PEs are arranged in square arrays, i.e., 16$\times$16, 32$\times$32, and 128$\times$128. Both versions of Eyeriss use the row-stationary dataflow. For rapid evaluation of architectures at large scales, we have built an analytical model that can search for the operation mappings with the best performance at different scales considering the data distribution and bandwidth limitations of different NoC designs in the two versions of Eyeriss. 

\begin{figure}
    \centering
    \subfloat[Eyeriss v2]
    {
        \includegraphics[width=0.47\linewidth]{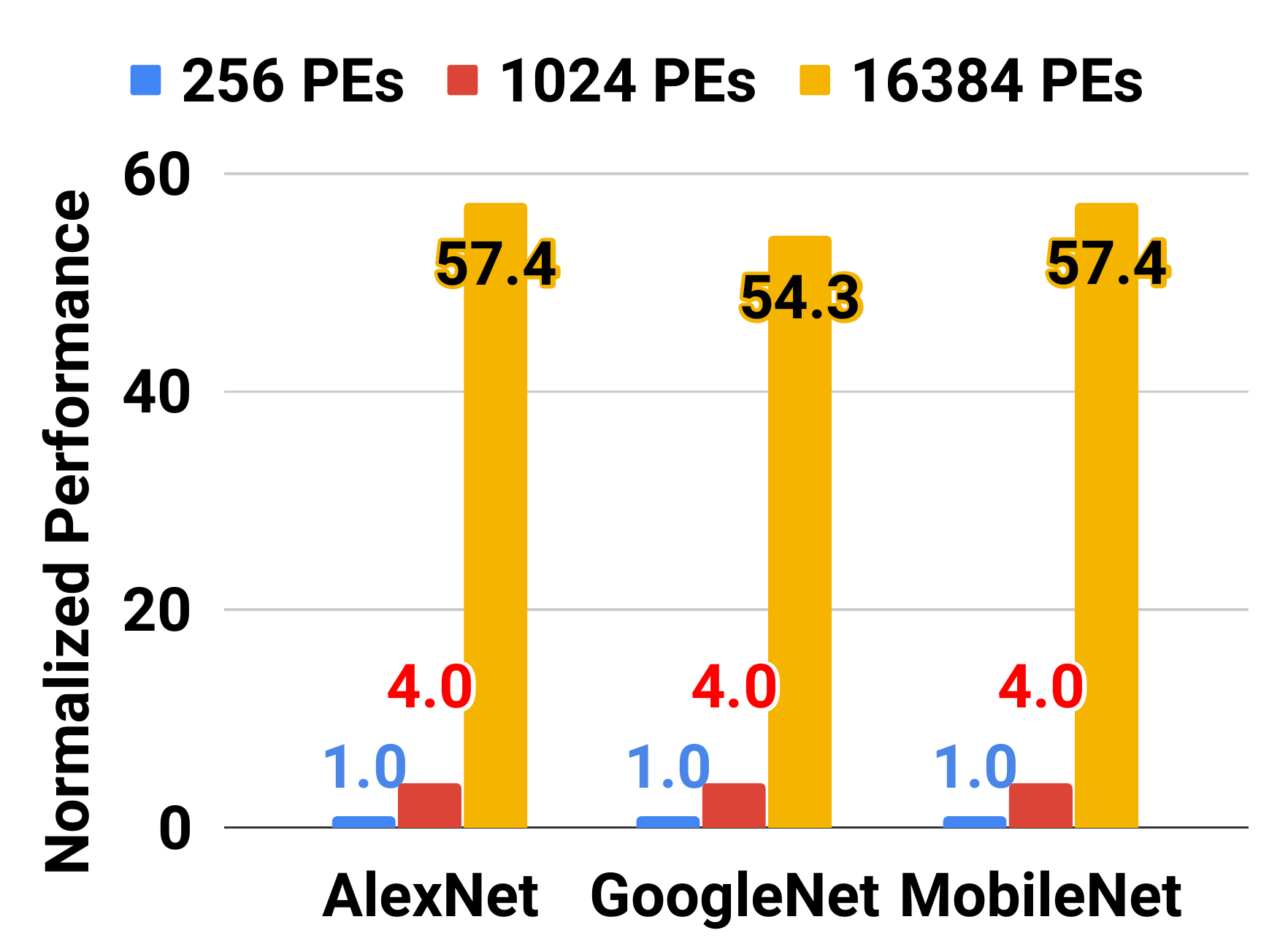}
        \label{fig:eyeriss_v2_scalability}
    }
    \subfloat[Eyeriss v1]
    {
        \includegraphics[width=0.47\linewidth]{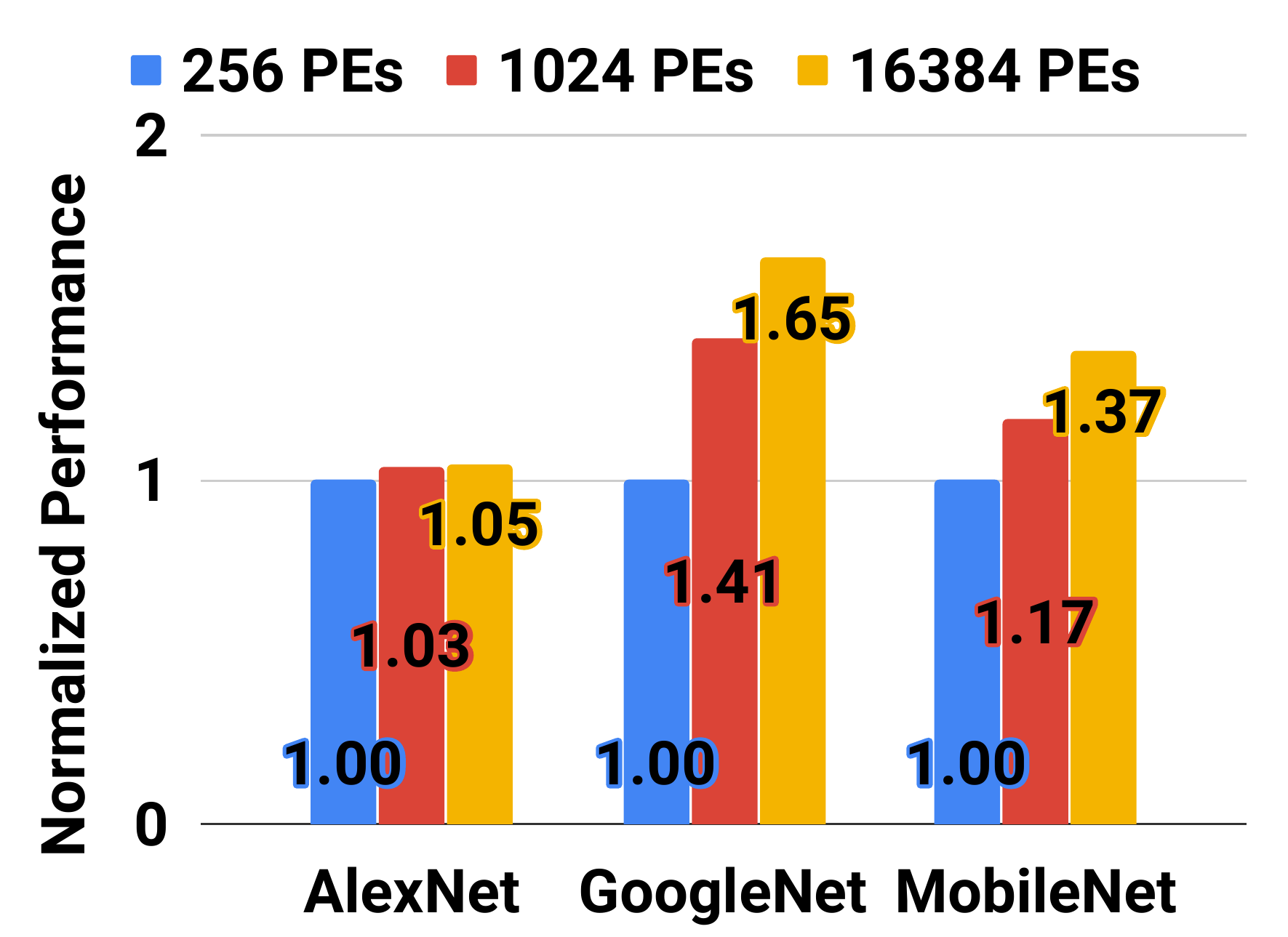}
        \label{fig:eyeriss_v1_scalability}
    }
    \caption{   Normalized performance of (a) Eyeriss v2 and (b) Eyeriss v1 running AlexNet, GoogLeNet, and MobileNet with a batch size of 1 at three different scales. Note that the MobileNet model has a width multiplier of 1.0 and an input size of 224$\times$224, which is different from the MobileNet benchmarked in Section~\ref{sec:results}.
            }
    \label{fig:scalability}
\end{figure}

Fig.~\ref{fig:eyeriss_v2_scalability} and~\ref{fig:eyeriss_v1_scalability} show the normalized performance of Eyeriss v2 and Eyeriss v1, respectively, running three DNNs: AlexNet, GoogLeNet, and MobileNet (with width multiplier of 1.0 and input size of 224$\times$224)\footnote{The large MobileNet model used here is not used for performance and energy efficiency benchmarking in Section~\ref{sec:results} since the post-place-and-route simulation turn-around time is not practical; the smaller MobileNet model also has the same accuracy as AlexNet, which makes it a better comparison.} at the three different scales with a batch size of 1. For all three DNNs, the performance of Eyeriss v2 scales linearly from 256 to 1024 PEs, and achieves more than 85\% of the linearly scaled performance at 16384 PEs. In contrast, the performance of Eyeriss v1 hardly improves when scaled up. This is due to the insufficient bandwidth provided by the broadcast NoC in Eyeriss v1 as discussed in Section~\ref{sec:noc_motivation}. For example, the performance of the FC layers in AlexNet and depth-wise layers in MobileNet do not see any improvement going from 256 PEs to 16384 PEs in Eyeriss v1 due to the insufficient NoC bandwidth for delivering weights and input activation, respectively, to the PEs. The HM-NoC in Eyeriss v2, however, is capable of adapting to the bandwidth requirements, therefore achieving higher performance at large scales. The HM-NoC is doing so while still being able to exploit available data reuse to achieve high energy efficiency, which will be demonstrated in Section~\ref{sec:results-perf_analysis} and~\ref{sec:results-benchmark}. Also note that, at large scales, the external data bandwidth will eventually become the performance bottleneck, and it will require more efforts to integrate the accelerator into the system to harness its full potential.

The implementation of the HM-NoC described in Section~\ref{ssec:noc_imp} targets the size of 8$\times$2 PE clusters, and will require modifications when scaled up. Specifically, the mesh routers for input activations and weights need an extra pair of source and destination ports in order to handle data delivery for more than two columns of PE clusters. As the area and energy cost of the router grows with the number of ports, the overall cost will increase. However, the same routers can then be used for any architectural scales. Also, as will be shown in Section~\ref{sec:results}, the entire NoC only accounts for less than 3\% of the area and 6\%-10\% of the total energy consumption. The additional complexity in the routers is unlikely to add significant cost. In addition, the proportion of cost of different components will stay roughly constant as the system scales thanks to the design of the hierarchical mesh network.
\section{Sparse Processing with SIMD Support}
\label{sec:sparse_processing}

In the original Eyeriss, sparsity of input activations (iacts), i.e., zeros, is exploited to improve energy efficiency by gating the switching of logic and data accesses. In Eyeriss v2, we want to exploit sparsity further in \emph{both} weights and iacts and improve not only energy efficiency but also throughput.  Whereas the original Eyeriss only used compression between the GLB and off-chip DRAM, in Eyeriss v2, we keep the data in compressed form all the way to the PE. Processing in the compressed domain provides benefits in terms of reducing on-chip bandwidth requirements as well as on-chip storage, which can result in energy savings and throughput improvements. However, as compressed data often has variable length, this presents challenges in terms of how to manipulate the data (e.g., distributing data across PEs, and sliding window processing within the PE). In this section, we will introduce a new PE architecture that can process sparse data in the compressed domain for higher throughput. We will also introduce support for SIMD in the PE such that each PE can process two MACs per cycle.

\subsection{Sparse PE Architecture}
\label{sec:sparse_pe_arch}

\begin{figure}[t]
    \begin{center}
        \includegraphics[width=0.85\linewidth]{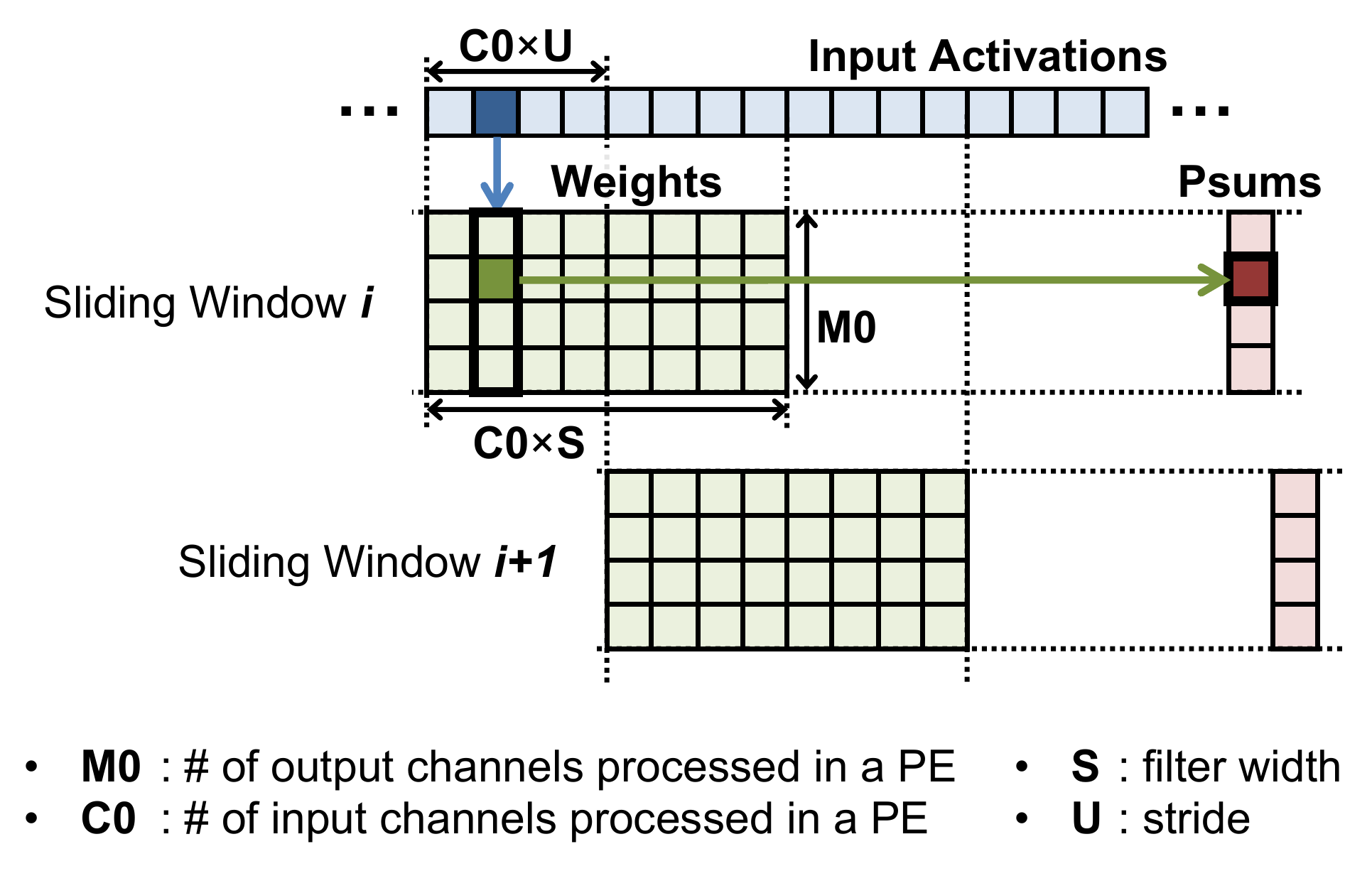}
        \caption{   Processing in the PE.
                }
        \label{fig:rs_pe_processing}
    \end{center}
\end{figure}

Fig.~\ref{fig:rs_pe_processing} illustrates how the PE processes uncompressed weights and iacts in the original Eyeriss, where $M0$ and $C0$ are the output and input channels processed within the PE, $S$ is the filter width, and $U$ is the stride. Recall that for the row-stationary dataflow, multiple 1-D rows of weights and iact are mapped to a given PE and processed in a sliding window fashion; here, the $C0 \times M0$ rows of weights with width $S$ are assigned to the PE, and the weights belong to $M0$ output channels and $C0$ input channels. For each iact, the PE runs through $M0$ MAC operations sequentially in consecutive cycles with the corresponding column of $M0$ weights in the weight matrix, and accumulates to $M0$ partial sums (psums). By going through a window of $C0\times S$ iacts in the stream, the processing goes through all $M0\times C0\times S$ weights in the matrix and accumulates to the same $M0$ psums. It then slides to the next window in the iact stream by replacing $C0\times U$ iacts at the front of the window with new ones, and repeats the processing with the same weight matrix but accumulates to another set of psums. Note that the access pattern of weights goes through the entire weight matrix once sequentially in a column-major fashion for each window of iacts.

To speedup the processing when the iacts and/or weights are sparse, the goal is to read only the non-zero data in the iact stream and the weight matrix for processing. In addition, we only want to perform the read when \emph{both} iact and weights are non-zero. The challenge, however, is to correctly and efficiently address data for all three data types. For example, when jumping between non-zero iacts in a window, the access pattern of weights does not go through the weight matrix sequentially anymore. Instead, additional logic is required to fetch the corresponding column of weights for the non-zero iact, which is not deterministic. Similarly, when jumping between non-zero weights in a weight column, it also has to calculate the address of the corresponding psum instead of just incrementing the address by one. Since the access order is also not deterministic, prefetching from the weight SPad is very challenging.

In order to achieve the processing of sparse data as described above, we take advantage of the compressed sparse column (CSC) compression format similar to what is described in~\cite{han2016eie, dorrance2014scalable}. For each non-zero value in the data, the CSC format records a \textit{count} value that indicates the number of leading zeros from the previous non-zero value in the uncompressed data stream; this is similar to the run length in run length coding (RLC). The count value can then be used to calculate the address change between the non-zero data.  The added advantage of CSC over RLC is that it has an additional \textit{address} value that allows the data to be broken into segments (e.g., columns) for easy handling, which we will discuss next; this, of course, also adds overhead in the compression.

\begin{figure}[t]
    \begin{center}
        \includegraphics[width=0.7\linewidth]{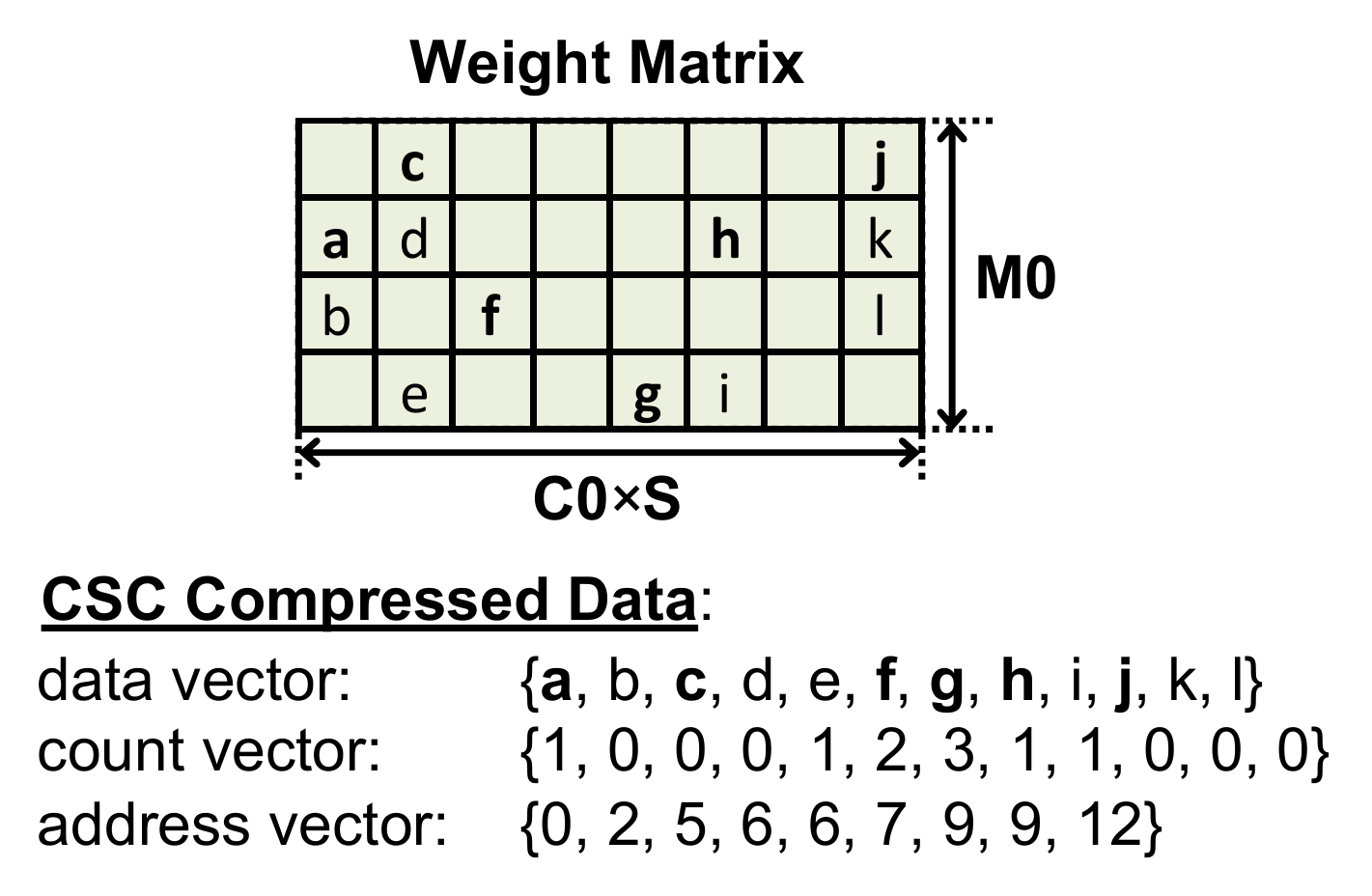}
        \caption{   Example of compressing sparse weights with compressed sparse column (CSC) coding.   The first non-zero weight in each column is highlighted in bold; the address vector points to the location of these weights within the data vector. If there are no non-zero weights in a column next location is repeated (e.g., repeated 6 in address vector reflects the all zero column between \textbf{f} and \textbf{g}).
                }
        \label{fig:csc_compression}
    \end{center}
\end{figure}

Both the iact and weights are encoded in the CSC format. For iacts, the data stream is divided into non-overlapping $C0\times U$ segments, and each segment is CSC encoded separately. Doing so enables sliding window processing, which replaces a segment of data with a new one from the stream when the window slides. Since the data length of each segment will be different after CSC coding, additional information is needed to address each encoded segment. Therefore, for each encoded segment, an \textit{address} value is also recorded in the CSC format that indicates the start address of the encoded segment in the entire encoded stream. The filter weights are also encoded with CSC compression by dividing each column of $M0$ weights as a segment and encoding each segment separately. This helps enable fast access of each column of non-zero weights. 

Fig.~\ref{fig:csc_compression} shows an example of CSC compressed weights. The characters in the weight matrix indicate the locations of non-zero values. To read the non-zero weights from a specific column, e.g., column 1 (assuming indexing starts from 0), the PE first reads address[1] and address[2] from the address vector in the CSC compressed weights, which gives the inclusive lower bound and non-inclusive upper bound of the addresses, i.e., 2 and 5, respectively, for reading the data and count vector. The first address (in this example, address[1]) is the location of the first non-zero weight in each column, highlighted in bold in Fig.~\ref{fig:csc_compression}, within the data vector; it then goes through the three non-zero weights in the column, i.e., $c$, $d$ and $e$, to perform the computation. If there is no non-zero weight in a column, the location of the next first non-zero value is repeated (e.g., since there are no non-zero values in column 3, the value 6 which is the location of \textbf{$g$}, is repeated such that the difference in consecutive address values is zero, which reflects the all zero column.). At the same time, the corresponding addresses of the psums to update can be calculated by accumulating the counts from the count vector.

In the CSC format, the count vector is an overhead in addition to the non-zero data. If the bitwidth of the count is low, it may affect the compression efficiency when sparsity is high since the number of consecutive zeros can exceed the maximum count. If the count bitwidth is high, however, the overhead of the count vector becomes more significant. From our experiments, setting each count at 4b yields the best compression rate for the 8b iact and weights. Therefore, each count-data pair is 12b and is stored in a 12b word of the data SPad for both iact and weight. This is similar to setting the run-length in the RLC, where 5b was allocated to the run-length  in~\cite{jssc2016-chen}.

In summary, both the weights and iacts can be processed directly in the CSC format. The processing can skip the zeros entirely without spending extra cycles, thus improving the processing throughput as well as energy efficiency.

\begin{figure}
    \begin{center}
        \includegraphics[width=0.98\linewidth]{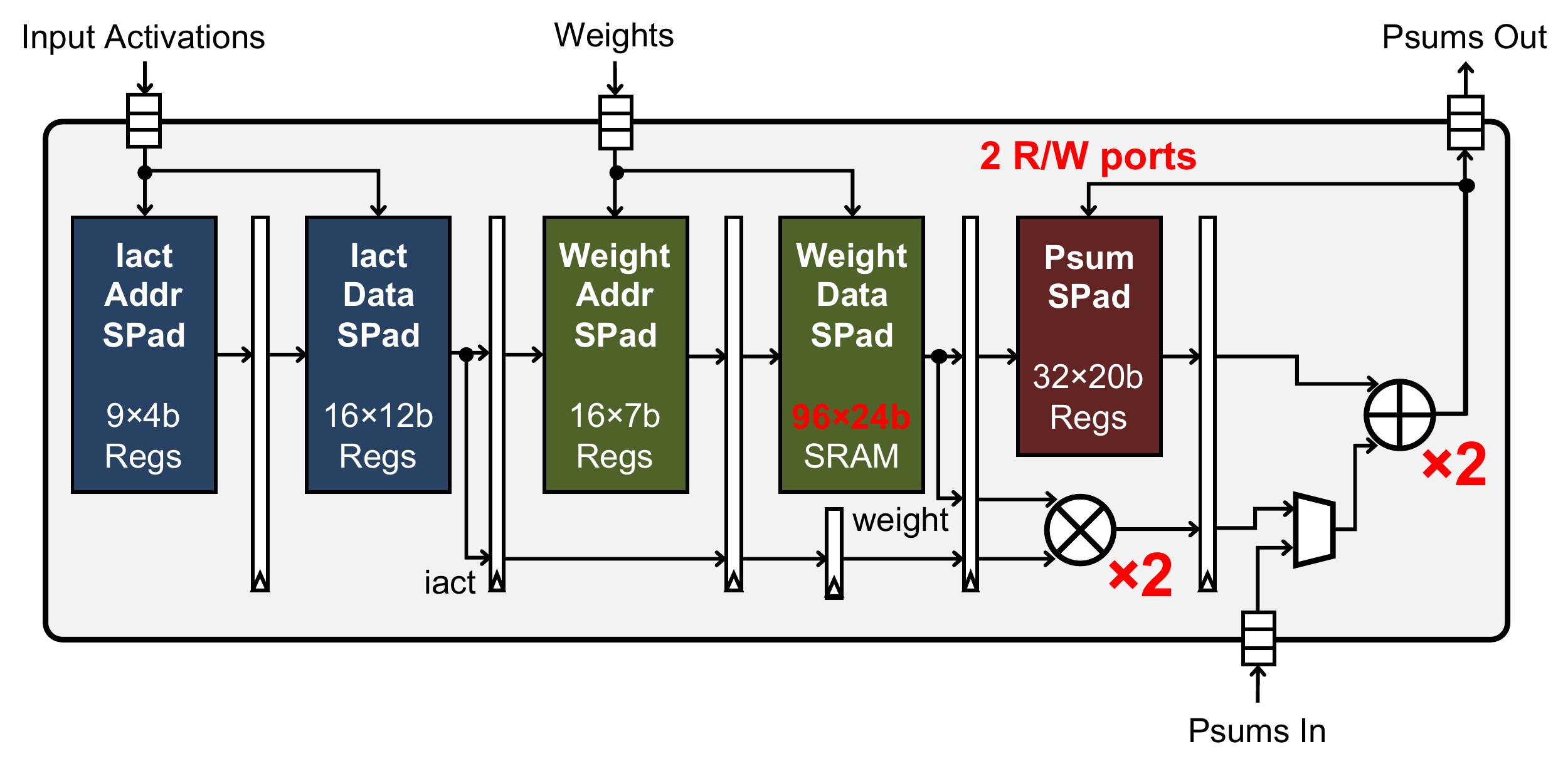}
        \caption{   Eyeriss v2 PE Architecture. The address SPad for both iact and weight are used to store addr vector in the CSC compressed data, while the data SPad stores the data and count vectors. The text in red denote changes for SIMD (Section~\ref{sec:pe_simd}).
                }
        \label{fig:eyeriss_v2_pe}
    \end{center}
\end{figure}

Fig.~\ref{fig:eyeriss_v2_pe} shows the block diagram of the \textit{sparse PE} that can perform the processing of CSC encoded iacts and weights directly as described above. Processing only non-zero data in the compressed format introduces read dependencies. For the compressed format, the address must be read before the data-count pair. To ensure only non-zero values are read, iact is read before the weight such that a non-zero weight is only read when the corresponding iact is non-zero.  To handle these dependencies while still maintaining throughput, the PE is implemented using seven pipeline stages and five SPads. The first two pipeline stages are responsible for fetching non-zero iacts from the SPads. The iact address SPad stores the address vector of the CSC compressed iacts, which is used to address the reads from the iact data SPad that holds the non-zero data vector as well as the count vector. After a non-zero iact is fetched, the next three pipeline stages read the corresponding weights. Similarly, there is a weight address SPad to address the reads from the weight data SPad for the correct column of weights. The final two stages in the pipeline perform the MAC computation on the fetched non-zero iact and weight, and then send the updated psum either back to the psum SPad or out of the PE.

Since either iact and weight can be zero or non-zero, there are three possible scenarios: 
\begin{itemize}
    \item \emph{If the iact is zero}, the CSC format will ensure that it is not read from the spad and therefore no cycles are wasted. 
    \item \emph{If the iact is not zero}, its value will be fetched from the iact data SPad and passed to the next pipeline stage. 
    \begin{itemize}
        \item \emph{If there are non-zero weights corresponding to the non-zero iacts}, they will be passed down the pipeline for computation. The zero weights will be skipped since the weights are also encoded with the CSC format. 
        \item \emph{If there are no non-zero weights corresponding to the non-zero iacts}, the non-zero iacts will not be further passed down in the pipeline. This may not necessarily introduce bubbles in the pipeline since the later stages, i.e., after the weight data Spad stage, can still be working on the computation for the previous non-zero iact if it has multiple corresponding weights.
    \end{itemize}
\end{itemize}

In the Eyeriss v2 PE, the sizes of the iact address and data SPads are 9$\times$4b and 16$\times$12b, respectively, which allow for a maximum iact window size of 16. The sizes of the weight address and data SPads are 16$\times$7b and 96$\times$24b, respectively. This allows for a maximum weight matrix size of 96$\times$(24b/12b)$=$192. The size of the psum SPad is 32$\times$20b, and allows for a maximum weight matrix height of 32 (i.e., maximum number of output channels $M0$). If we fully utilize the iact SPads and psum SPad, it will require a weight matrix size of 32$\times$16$=$512, which is larger than the limit of 192; however, the sparse PE design takes advantage of the fact that the sparse pattern of weights is known at compile time; therefore, it is possible to guarantee that the compressed weights will fit in a smaller SPad. Table~\ref{table:eyeriss_v2_pe_sparse_mapping_alexnet} shows how many weights are stored in the SPad of each PE for sparse AlexNet. While in most layers the nominal number of weights is higher than 192, the number of non-zero weights after compression in the worst case is smaller and fits in the SPad for processing. By mapping more non-zero weights into each PE instead of mapping based on the nominal number of weights, more operations are performed in a PE, which statistically helps to reduce the amount of workload imbalance caused by the sparsity.

\begin{table}[t]
    \centering
    \footnotesize
    \begin{tabular}{|c|c|c|c|c|c|}
        \hline
                        &   \multirow{2}{*}{\boldsymbol{$M0$}}  &   \multirow{2}{*}{\boldsymbol{$C0$}}  &   \multirow{2}{*}{\boldsymbol{$S$}}   &   \multicolumn{2}{c|}{\textbf{Num. of Non-Zero Weights}}   \\
        \cline{5-6}
                        &                                   &                                   &                                   &   \textbf{Nominal}&   \textbf{Compressed}         \\
        \Xhline{3\arrayrulewidth}
        \textbf{CONV1}  &   12                              &   1                               &   11                              &   132             &   64                  \\
        \textbf{CONV2}  &   32                              &   2                               &   5                               &   320             &   86                  \\
        \textbf{CONV3}  &   32                              &   5                               &   3                               &   480             &   126                 \\
        \textbf{CONV4}  &   24                              &   4                               &   3                               &   288             &   100                 \\
        \textbf{CONV5}  &   32                              &   4                               &   3                               &   384             &   174                 \\
        \hline
        \textbf{FC6}    &   32                              &   2                               &   6                               &   384             &   92                  \\
        \textbf{FC7}    &   32                              &   15                              &   1                               &   480             &   84                  \\
        \textbf{FC8}    &   32                              &   15                              &   1                               &   480             &   170                 \\
        \hline
    \end{tabular}
    \vspace{+5pt}
    \caption{   Distribution of weights in sparse AlexNet to the SPad\protect\\in each PE of Eyeriss v2.}
    \label{table:eyeriss_v2_pe_sparse_mapping_alexnet}
\end{table}

Since the degree of sparsity varies across different DNNs and data types, the PE is also designed to adapt to the scenarios when sparsity is low. In such cases, the PE can directly take in uncompressed iacts and weights instead of the CSC compressed versions to reduce the overhead in data traffic. Both iact and weight address SPads are not used and therefore clock-gated to save energy consumption, and the count in the CSC format is fixed to zero to address the data SPads correctly for processing.

\subsection{SIMD Support in PE}
\label{sec:pe_simd}

Profiling results of the PE implementation shows that the area and energy consumption of the MAC unit is insignificant compared to other components in a PE. In Eyeriss, for example, the MAC unit takes less than 5\% of the PE area, and only consumes 2\%--9\% of the PE power. This motivates the exploration of SIMD processing in a PE in order to achieve speedup of at most two times.

SIMD is applied to the PE architecture as shown in Fig.~\ref{fig:eyeriss_v2_pe} by fetching two weights instead of one for computing two MAC operations per cycle with the same iact, i.e., a SIMD width of two. The changes are noted in the red text in the figure. SIMD processing not only improves the throughput but also further reduces the number of iact reads from the SPad. In terms of architectural changes, SIMD requires the word width of the weight data SPad to be two-word wide, which is why the size of the weight data SPad is 96$\times$24b instead of 192$\times$12b. The psum SPad also has to have two read and two write ports for updating two psums per cycle.  In the case where only an odd number of non-zero weights exist in the column of $M0$ weights, the second 12b of the last 24b word in a column of non-zero weights is filled with zero. When the PE logic encounters the all-zero count-data pair, it clock-gates the second MAC datapaths as well as the read and write of the second ports in the psum SPad to avoid unnecessary switching, which reduces power consumption. 
\section{Implementation Results}
\label{sec:results}

Eyeriss v2 was implemented in a 65nm CMOS process and the specifications of the design are summarized in Table~\ref{table:eyeriss_v2_spec}. The design was placed-and-routed and the results reported in this section are from post-layout cycle-accurate gate-level simulations with (1) technology library from the worst PVT corner, (2) switching activities profiled from running the actual weights of the DNNs and data from the ImageNet dataset~\cite{ijcv2015-russakovsky}, and (3) a batch size of one, which represents a more challenging setup for energy efficiency and throughput, but captures the low latency use case.

\begin{figure}[t]
    \centering
    \subfloat[Overall Area Breakdown]
    {
        \includegraphics[width=0.43\linewidth]{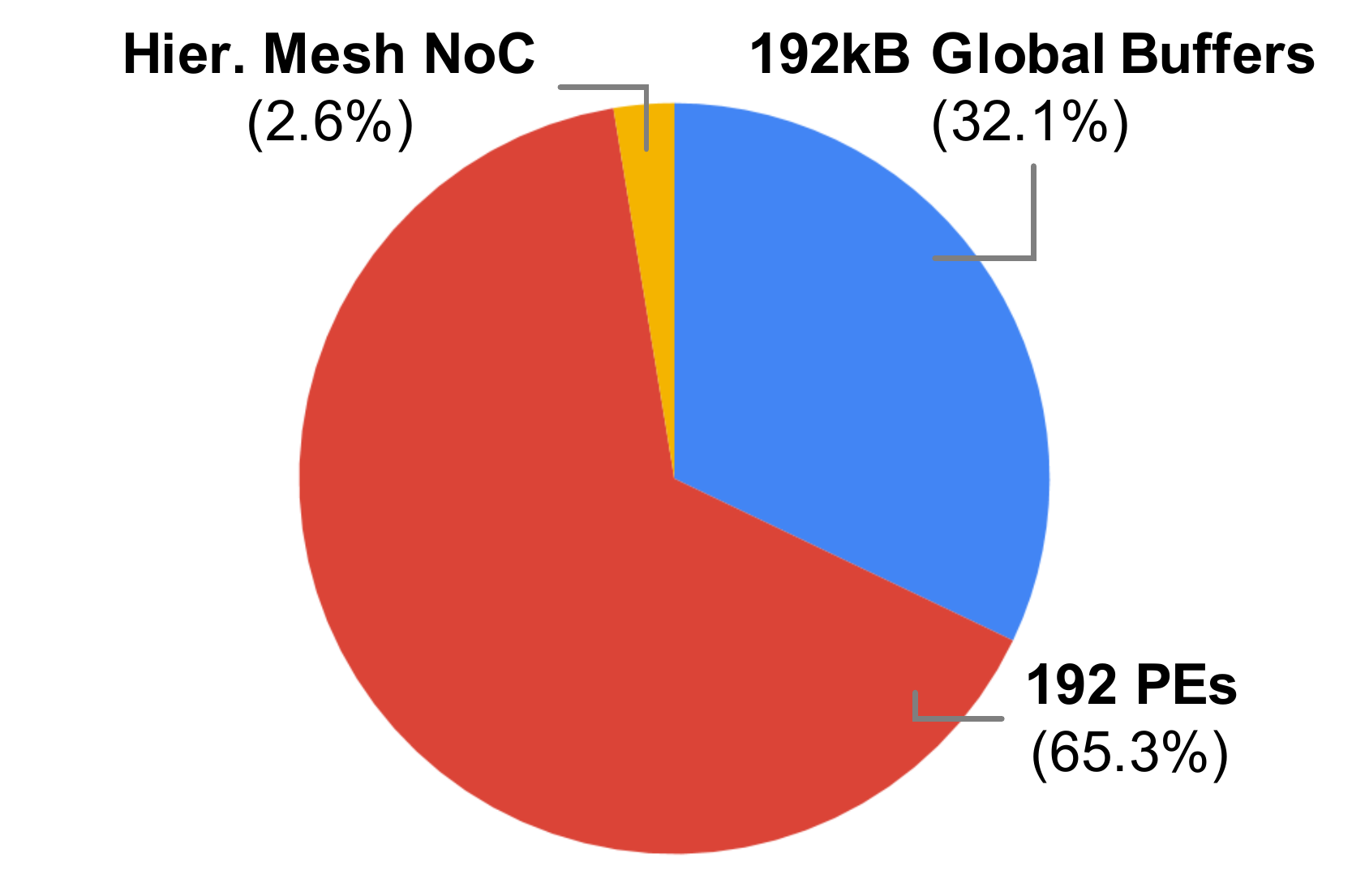}
        \label{fig:eyeriss_v2_overall_area_breakdown}
    }
    \subfloat[PE Area Breakdown]
    {
        \includegraphics[width=0.55\linewidth]{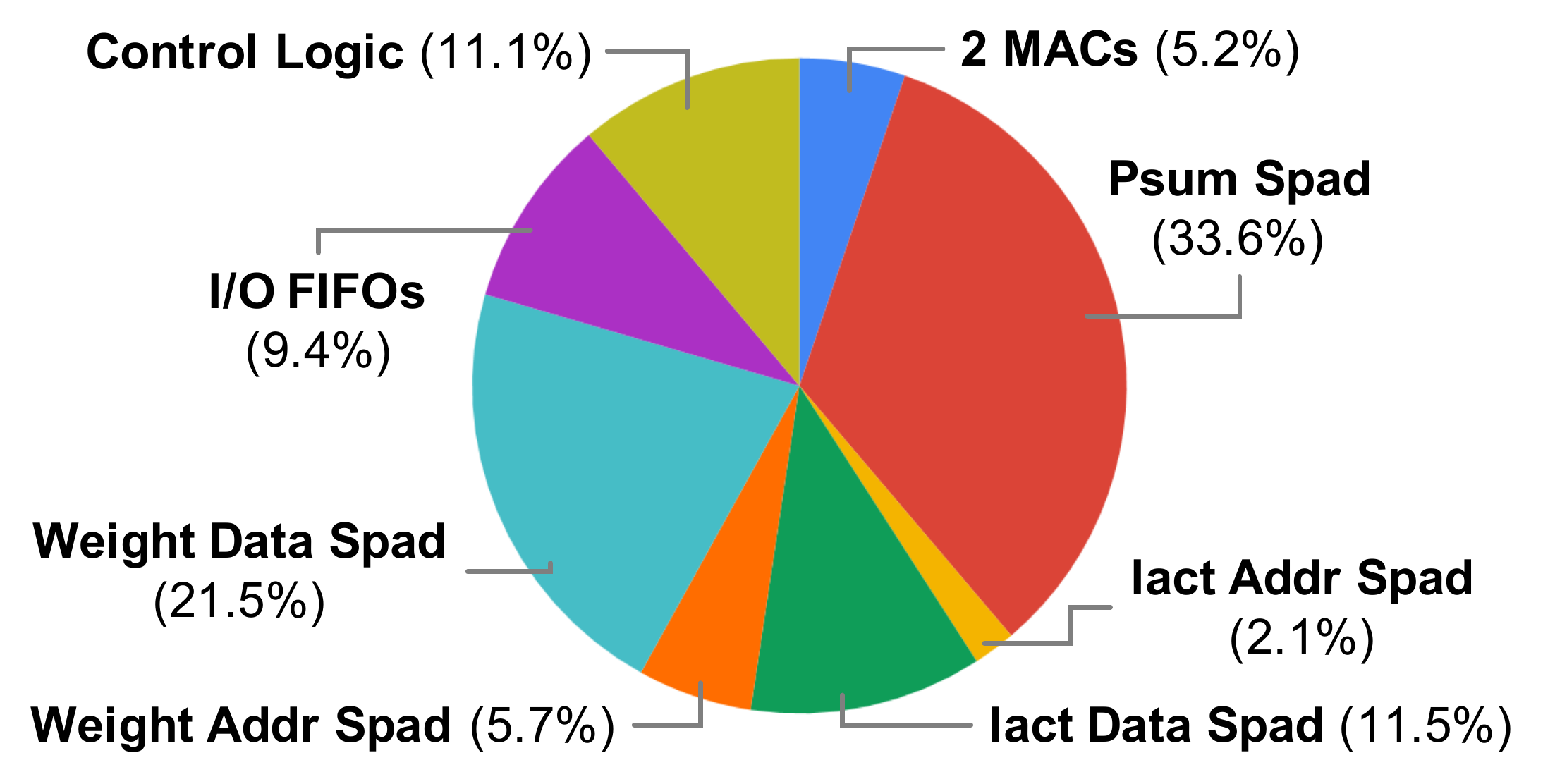}
        \label{fig:eyeriss_v2_pe_area_breakdown}
    }
    \caption{   Eyeriss v2 area breakdown.
            }
    \label{fig:eyeriss_v2_area_breakdown}
\end{figure}

The overall gate count of Eyeriss v2, excluding SRAMs, is approximately 2695k NAND-2 gates. The area breakdown (Fig.~\ref{fig:eyeriss_v2_area_breakdown}) shows that the 192 PEs dominates the area cost, while the area of the hierarchical mesh networks of all data types combined only account for 2.6\% of the area. This result proves that it is possible to build in high flexibility at a low cost. Within each PE, all of the SPads combined account for around 72\% of the area, while the two MAC units only account for 5\%. 

\begin{table}[t]
    \centering
    \footnotesize
    \begin{tabular}{|l|l|}
        \hline
        \textbf{Technology}                             &   TSMC 65nm LP 1P9M                   \\
        \hline
        \textbf{Gate Count (logic only)}                &   2695k (NAND-2)                      \\
        \hline
        \textbf{On-Chip SRAM}                           &   246 KB                              \\
        \hline
        \textbf{Number of PEs}                          &   192                                 \\
        \hline
        \textbf{Global Buffer}                          &   192 KB (SRAM)                       \\
        \hline
        \multirow{4}{*}{\textbf{Scratch Pads}}          &   weight addr: 14B (Reg)              \\
        \multirow{4}{*}{\textbf{(per PE)}}              &   weight data: 288B (SRAM)            \\
                                                        &   iact addr: 4.5B (Reg)               \\
                                                        &   iact data: 24B (Reg)                     \\
                                                        &   psum: 80B (Reg)                     \\
        \hline
        \textbf{Clock Rate}                             &   200 MHz                             \\
        \hline
        \textbf{Peak Throughput}                        &   153.6 GOPS                          \\
        \hline
        \multirow{2}{*}{\textbf{Arithmetic Precision}}  &   weights \& iacts: 8b fixed-point    \\
                                                        &   psums: 20b fixed-point              \\
        \hline
    \end{tabular}
    \vspace{+5pt}
    \caption{   Eyeriss v2 Specifications.
            }
    \label{table:eyeriss_v2_spec}
\end{table}

\subsection{Performance Analysis}
\label{sec:results-perf_analysis}

To demonstrate the throughput and energy efficiency improvements brought on by the hierarchical mesh network and sparse PE architecture, we have implemented three different variants of Eyeriss: v1, v1.5, and v2. Table~\ref{table:eyeriss_diff_summary} lists the key differences between these Eyeriss variants. For the PE architecture, \textit{Dense} means the PE can only clock-gate the cycles with zero data but not skip it, while \textit{Sparse} means the PE can further skip the processing cycles with zero data. Eyeriss v1 is the same design as the original Eyeriss~\cite{jssc2016-chen}, but with the storage capacity, number of PEs and data precision scaled to the same level as v1.5 and v2 for a fair comparison. In short, the comparison between v1 and v1.5 shows the impact of the hierarchical mesh network, while the comparison between v1.5 and v2 shows the impact of the sparse PE architecture along with the support for SIMD processing. These architectures are placed-and-routed and benchmarked with four DNNs that have the same accuracy on the ImageNet dataset: AlexNet~\cite{nips2012-krizhevsky}, MobileNet (with a width multiplier of 0.5 and input size of 128$\times$128)~\cite{arxiv2017-howard}, and the sparse version of them as pruned by the method introduced in~\cite{yang2016}. In this section, unless otherwise specified, AlexNet and MobileNet are referring to the dense model.

The implementation shows that Eyeriss v2 has an area increase of around two times compared to the other versions. The increase in mostly in the PE, which is 73\% larger than the original one. The main reason is due to the need to support sparse processing, which requires deeper pipelining in the control logic and additional SPads to store the CSC compressed data. This contributes to a nearly 50\% increase in area. Supporting SIMD also contributes to an additional 15\% area increase due to the two sets of read and write ports for the psum SPad and the wider bus-width of the PE I/O in addition to the doubling of MAC units.

\begin{table}[t]
    \centering
    \footnotesize
    \begin{tabular}{|l|c|c|c|}
        \hline
                                    &   \textbf{Eyeriss v1} &   \textbf{Eyeriss v1.5}   & \textbf{Eyeriss v2}   \\
        \Xhline{3\arrayrulewidth}
        \textbf{Data Precision}     &   \multicolumn{3}{c|}{activations \& weights: 8b; partial sums: 20b}      \\
        \hline
        \textbf{\# of PEs}          &   \multicolumn{3}{c|}{192}                                                \\
        \hline
        \textbf{\# of MACs}         &   192                 &   192                     &   384                 \\
        \hline
        \textbf{NoC}                &   Multicast           &   Hier. Mesh              &   Hier. Mesh          \\
        \hline
        \textbf{PE Architecture}    &   Dense               &   Dense                   &   Sparse              \\
        \hline
        \textbf{PE SIMD Support}    &   No                  &   No                      &   Yes                 \\
        \hline
        \textbf{Global Buffer Size} &   \multicolumn{3}{c|}{192 kB}                                             \\
        \hline
        \textbf{Area (NAND-2 gates)}&   1394k               &   1354k                   &   2695k               \\
        \hline
    \end{tabular}
    \vspace{+5pt}
    \caption{   Key differences between the three Eyeriss variants. The area is logic only.
            }
    \label{table:eyeriss_diff_summary}
\end{table}

\subsubsection{AlexNet}

\begin{figure*}[t]
    \centering
    \subfloat[Throughput Speedup]
    {
        \includegraphics[width=0.47\linewidth]{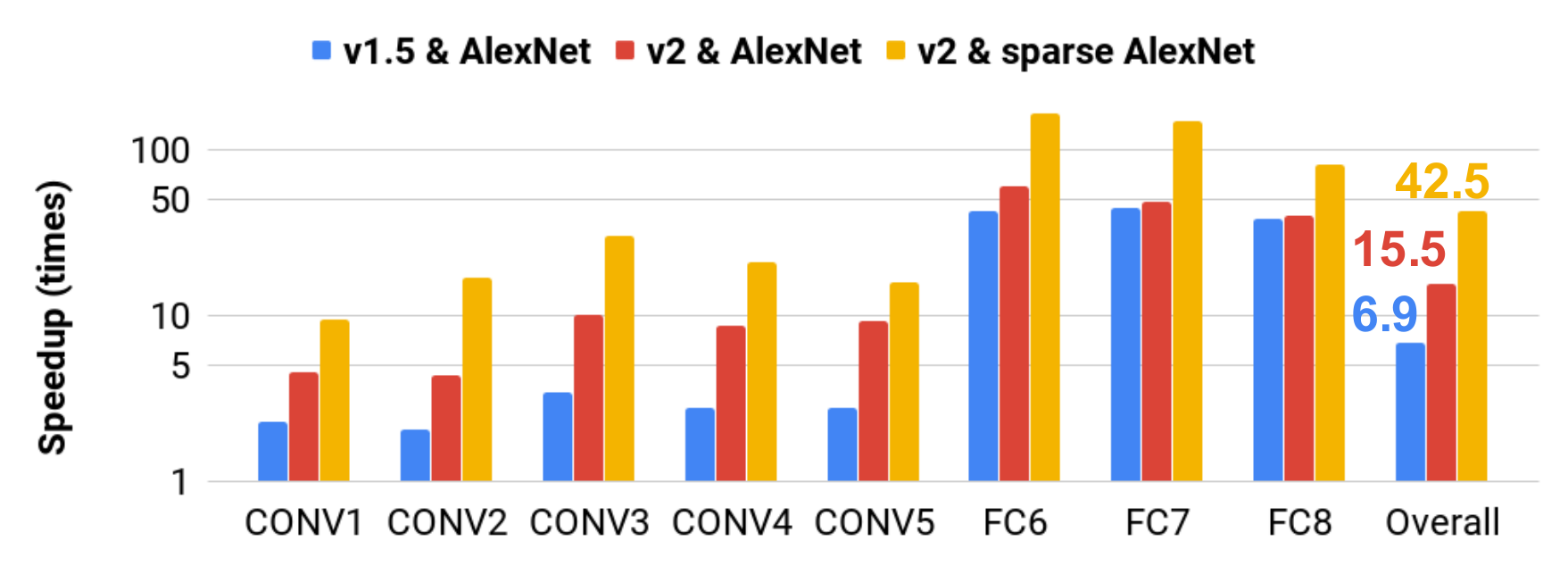}
        \label{fig:eyeriss_versions_comparison_alexnet_speedup}
    }
    \subfloat[Energy Efficiency Improvement]
    {
        \includegraphics[width=0.47\linewidth]{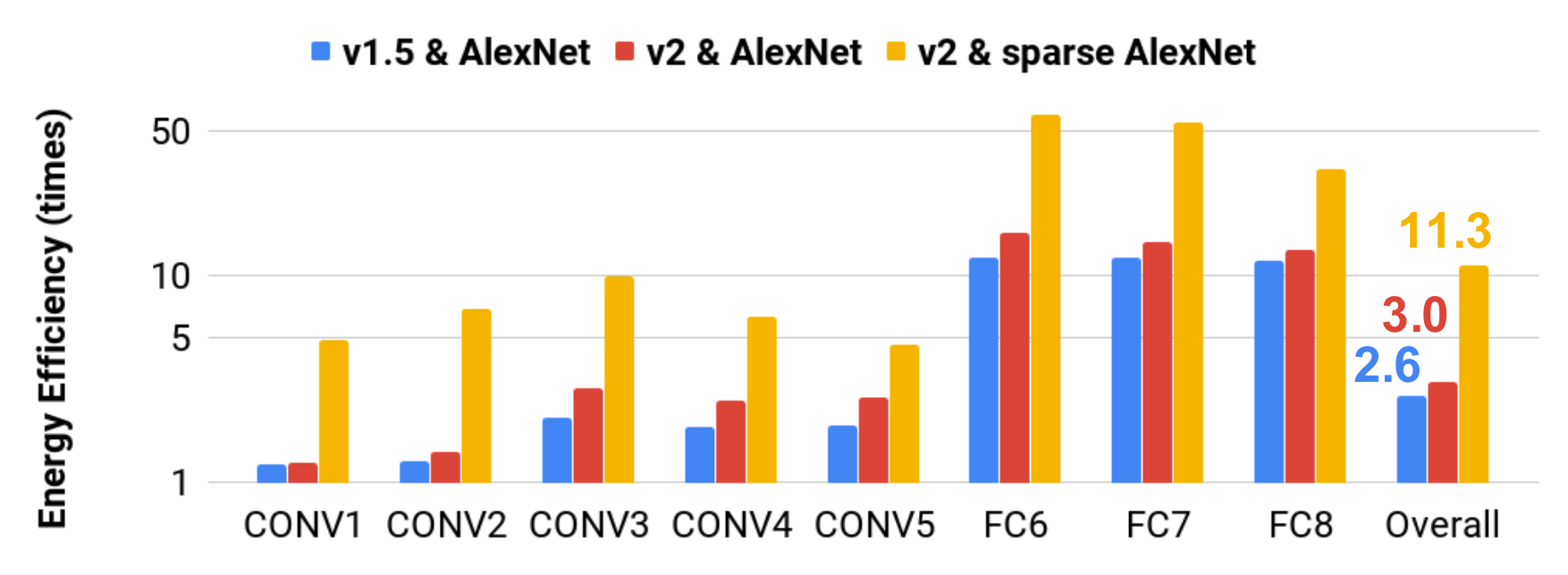}
        \label{fig:eyeriss_versions_comparison_alexnet_efficiency}
    }
    \caption{   (a) Speedup and (b) energy efficiency improvement of different versions of Eyeriss over Eyeriss v1 benchmarked with AlexNet.
            }
    \label{fig:eyeriss_versions_comparison_alexnet}
\end{figure*}

\begin{figure}[t]
    \begin{center}
        \includegraphics[width=0.8\linewidth]{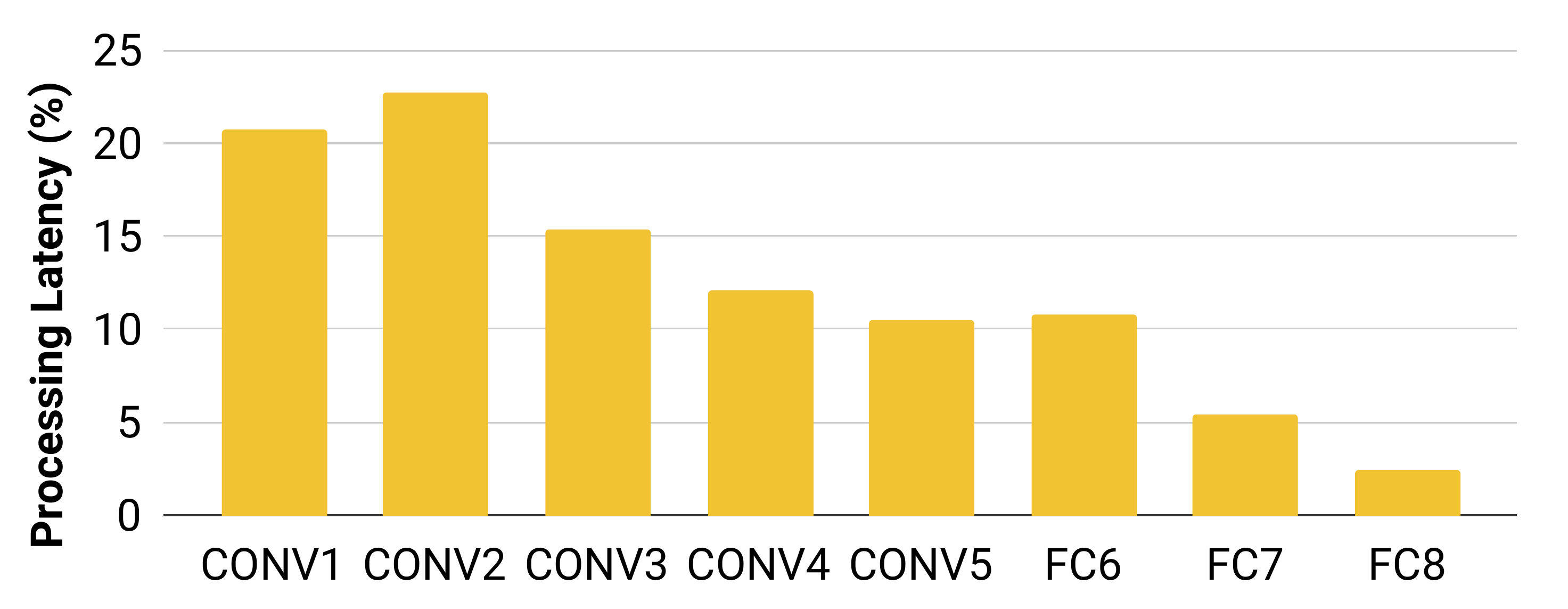}
        \caption{   Breakdown of processing latency across the different layers of sparse AlexNet running on Eyeriss v2.
                }
        \label{fig:eyeriss_v2_sp_alexnet_runtime}
    \end{center}
\end{figure}

Fig.~\ref{fig:eyeriss_versions_comparison_alexnet_speedup} shows the throughput improvements of different versions of Eyeriss on AlexNet over Eyeriss v1. Results on sparse AlexNet are also included (yellow bars) along with a breakdown of the processing latency across the different layers shown in Fig.~\ref{fig:eyeriss_v2_sp_alexnet_runtime}. For AlexNet, the result shows that Eyeriss v1.5 significantly speeds up FC layers. This is because the throughput of FC layers is bandwidth-limited in Eyeriss v1, which is addressed by the hierarchical mesh network in Eyeriss v1.5. Eyeriss v2, on the contrary, significantly speeds up the CONV layers over Eyeriss v1.5 due to the increased number of multipliers and sparsity in the activations.  However, the throughput of the FC layers only shows a marginal improvement because the FC layers are still bandwidth-limited even with the hierarchical mesh network. Therefore, speeding up the processing with sparsity and SIMD does not improve the throughput of FC layers as significantly as in CONV layers. 

The full potential of Eyeriss v2, however, is fully revealed when coupled with sparse AlexNet. The bandwidth requirement of weights is lower in sparse AlexNet since it is very sparse, and the CSC compression can effectively reduce the data traffic. As a result, exploiting sparsity becomes more effective. Overall, Eyeriss v2 achieves 42.5$\times$ speedup with sparse AlexNet over Eyeriss v1 with AlexNet. 

Fig.~\ref{fig:eyeriss_versions_comparison_alexnet_efficiency} shows the improvement on energy efficiency. It largely correlates to the speedup in Fig.~\ref{fig:eyeriss_versions_comparison_alexnet_speedup} since the higher overall utilization of the PEs reduces the proportion of the static power consumption, e.g., clock network. Overall, Eyeriss v2 with sparse AlexNet is 11.3$\times$ more energy efficient than Eyeriss v1 with AlexNet.

\subsubsection{MobileNet (width multiplier of 0.5, input size of 128$\times$128)}

\begin{figure*}[t]
    \centering
    \subfloat[Throughput Speedup]
    {
        \includegraphics[width=0.47\linewidth]{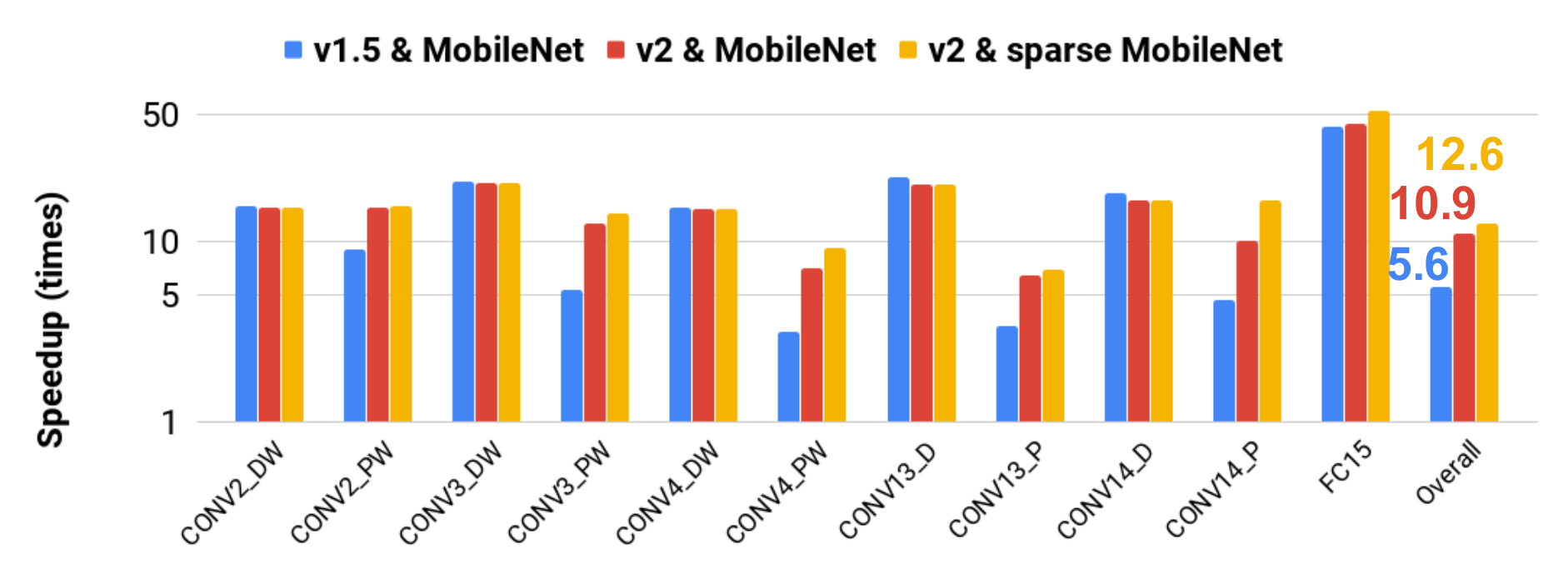}
        \label{fig:eyeriss_versions_comparison_mobilenet_speedup}
    }
    \subfloat[Energy Efficiency Improvement]
    {
        \includegraphics[width=0.47\linewidth]{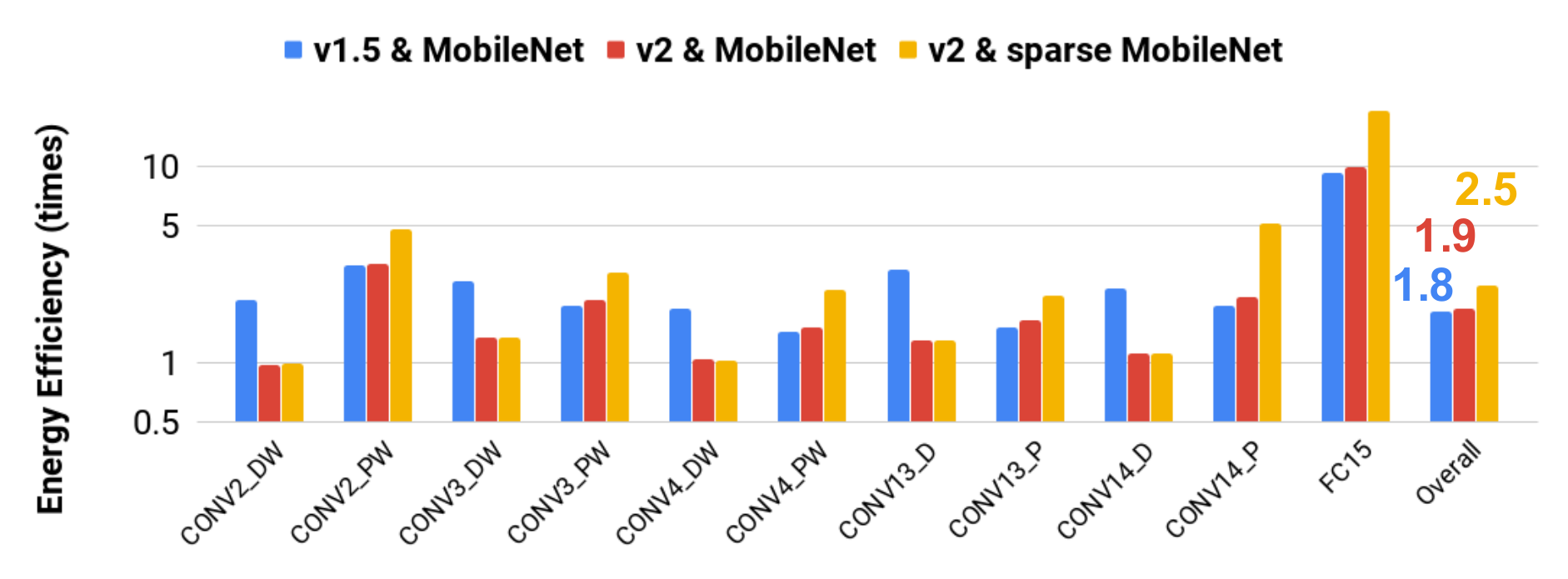}
        \label{fig:eyeriss_versions_comparison_mobilenet_efficiency}
    }
    \caption{   (a) Speedup and (b) energy efficiency improvement of different versions of Eyeriss over Eyeriss v1 benchmarked with MobileNet. Due to the large number of layers, only a few representative layers are presented.
            }
    \label{fig:eyeriss_versions_comparison_mobilenet}
\end{figure*}

Fig.~\ref{fig:eyeriss_versions_comparison_mobilenet_speedup} and~\ref{fig:eyeriss_versions_comparison_mobilenet_efficiency} show the improvement on throughput and energy efficiency, respectively, of different versions of Eyeriss on selected layers of MobileNet over Eyeriss v1. Results on sparse MobileNet are also included (yellow bars). The lack of data reuse in MobileNet results in low throughput on Eyeriss v1 due to the low-bandwidth NoC, which is why Eyeriss v1.5 can achieve a significant speedup over v1. However, the speedup of Eyeriss v2 over v1.5 is a mixed bag. While layers such as CONV1 and the point-wise (PW) layers can still take advantage of the sparsity in input activations to improve the throughput, the throughput of the Depth-wise (DW) CONV layers becomes worse. This is because the CSC compression does not create skippable cycles when the number of input and output channels are both one. Therefore, the sparse PE in Eyeriss v2 does not bring any advantage over the dense PE in Eyeriss v1.5. Furthermore, the deeper pipeline of the sparse PE actually makes the throughput slightly worse in the DW CONV layers. 

Sparse MobileNet brings additional benefits on throughput and energy efficiency on Eyeriss v2; however, the improvement is not as significant as with the sparse AlexNet, since the CSC compression is less effective on sparse MobileNet than on sparse AlexNet due to its small layer sizes.  Overall, Eyeriss v2 with sparse MobileNet is 12.6$\times$ faster and 2.5$\times$ more energy efficient than Eyeriss v1 with MobileNet.

\subsection{Benchmark Results}
\label{sec:results-benchmark}

Table~\ref{table:eyeriss_v2_benchmark} summarizes the throughput and energy efficiency of Eyeriss v2 benchmarked with four DNNs that have comparable accuracy at a batch size of one. Although Eyeriss v2 achieves the highest GOPS/W\footnote{In this paper, we calculate GOPS based on the nominal number of operations in the DNN, i.e., including operations with data values of zero.} with sparse AlexNet, it consumes the least amount of time and energy per inference with the sparse MobileNet. \emph{This result echos the trend of DNN development going toward compact models that are more lightweight but also have less reuse for the hardware to explore, which makes it harder to reduce GOPS/W but can still improve inference/J}. Also, Eyeriss v2 achieves 12.6$\times$ higher inference/sec for MobileNet than AlexNet, which correlates well to the 14.7$\times$ reduction in the nominal number of MACs. This proves that the design has high flexibility to perform well for compact DNN models. 

\begin{table*}[t]
    \centering
    \footnotesize
    \begin{threeparttable}
    \begin{tabular}{|r|c|c|c|c|c|c|c|c|}
        \hline
        \multirow{2}{*}{\textbf{DNN}} & \textbf{ImageNet}           &   \textbf{Nominal Num.}       &   \multirow{2}{*}{\textbf{Inference/sec}} & \multirow{2}{*}{\textbf{Inference/J}} & \multirow{2}{*}{\textbf{GOPS/W}}  &   \textbf{DRAM Acc.}  &   \textbf{PE}                   \\
                                      &   \textbf{Accuracy\tnote{1}}&   \textbf{of MACs}            &                                           &                                       &                                   &   \textbf{(MB)}       &   \textbf{Utilization\tnote{2}} \\
        \Xhline{3\arrayrulewidth}
        \textbf{AlexNet}              &   80.43\%                   &   724.4M                      &   102.1                           &   174.8                           &  253.2            &   71.9                        &   100\%   \\
        \hline
        \textbf{sparse AlexNet}       &   79.56\%                   &   724.4M                      &   278.7                           &   664.6                           &  962.9            &   22.3                        &   100\%   \\
        \hline
        \textbf{MobileNet}            &   79.37\%                   &   49.2M                       &   1282.1                          &   1969.8                          &  193.7            &   4.1                         &   91.5\%  \\
        \hline
        \textbf{sparse MobileNet}     &   79.68\%                   &   49.2M                       &   1470.6                          &   2560.3                          &  251.7            &   3.9                         &   91.5\%  \\
        \hline
    \end{tabular}
    \begin{tablenotes}
        \item[1] top-5 accuracy for the image classification task.
        \item[2] measured in terms of number of utilized MAC datapaths; each PE has 2 MAC datapaths.
    \end{tablenotes}
    \end{threeparttable}
    \vspace{+5pt}
    \caption{   Throughput and energy efficiency of Eyeriss v2 benchmarked with four DNNs that have comparable accuracy and a batch size of 1. Note that the MobileNet used for benchmark has a width multiplier of 0.5 and an input size of 128$\times$128.
            }
            
    \label{table:eyeriss_v2_benchmark}
\end{table*}

\begin{figure}
    \begin{center}
        \includegraphics[width=0.95\linewidth]{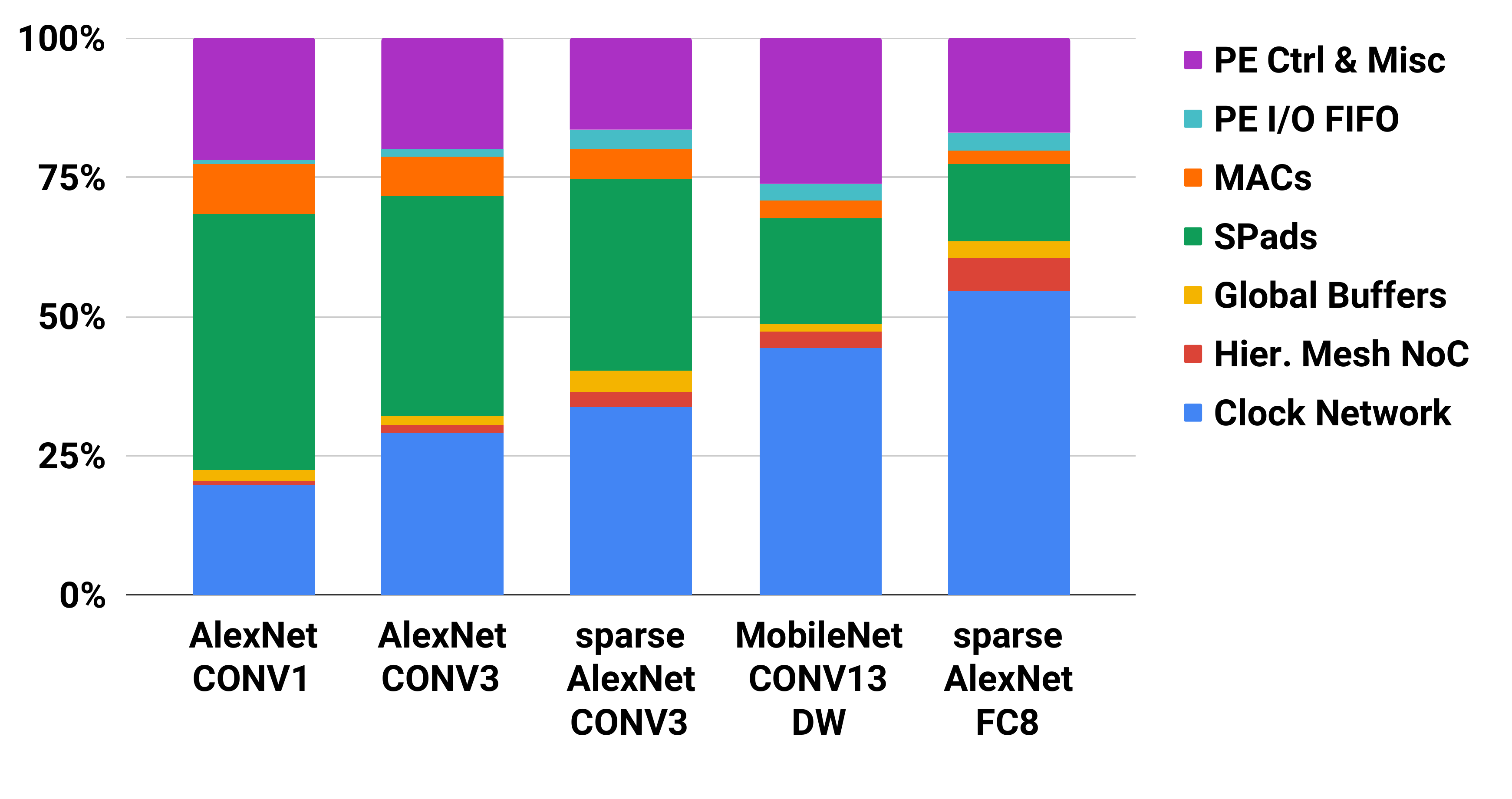}
        \caption{   Eyeriss v2 power breakdown running different DNN layers.
                }
        \label{fig:eyeriss_v2_power_breakdown}
    \end{center}
\end{figure}

Fig.~\ref{fig:eyeriss_v2_power_breakdown} shows the normalized power breakdown of Eyeriss v2 running a variety of DNN layers. We pick a representative set of layers to show how the different characteristics of the DNN layers impact the hardware. Note that these layers have different energy consumption and efficiency. The results are summarized as follows:
\begin{itemize}
    \item CONV1 of AlexNet (148.1 GOPS/W) shows the case of no sparsity in both activations and weights. Compared to other layers, the high utilization of the PEs makes the proportion of the clock network power consumption low. It also has the highest proportion of MAC power consumption.
    \item CONV3 of sparse AlexNet (1423.2 GOPS/W) has the highest amount of sparsity in all layers we have tested. Compared to CONV3 of AlexNet (392.0 GOPS/W), the proportions of the clock network, HM-NoC and GLB power consumption are higher. This is mainly due to the workload imbalance induced by sparsity, which lowers the utilization of the active PEs. However, judging from the large proportion of the SPad and MAC power consumption compared to other components such as PE control logic, the PE is still kept fairly busy and data reuse is effectively exploited by the SPads.
    \item CONV13 DW layer of MobileNet (77.7 GOPS/W) has the lowest GOPS/W among all the layers we have tested. As expected, most of the energy is spent on the clock network. Inside the PE, the lack of reuse and not being able to utilize SIMD also hurt the energy efficiency, which is evident by the fact that most of the energy is spent in the control logic instead of the SPads or MACs.
    \item FC8 of sparse AlexNet (465.1 GOPS/W) shows the case of high sparsity and low data reuse. This combination makes the architecture more bandwidth-limited, and therefore the utilization of active PEs becomes low. That is why this layer has the highest proportion of power consumed by the clock network. The lack of reuse also makes the proportion of the SPad power consumption low and the NoC power consumption high. However, thanks to sparsity, the overall energy efficiency of this layer is still better than CONV1 of AlexNet.
\end{itemize} 

In terms of external DRAM accesses, AlexNet requires much more data than MobileNet as shown in Table~\ref{table:eyeriss_v2_benchmark}, which is mainly due to the large amount of weights in the fully-connected layers. For CONV layers only, the required DRAM accesses are 7.1 MB and 4.6 MB for AlexNet and sparse AlexNet, respectively. Note that Eyeriss v2 does not perform the pooling layers on-chip, and the required DRAM accesses will further decrease if pooling layers are processed on-chip. We have also profiled the impact of a limited peak external bandwidth on the performance. With an aggregated external read and write bandwidth of 25600 MB/s, which is at the level of DDR4-3200, the throughput of Eyeriss v2 running sparse AlexNet and sparse MobileNet will decrease by 16\% and 24\%, respectively, due to the bursty external data access patterns. However, we believe that additional on-chip buffering can alleviate the performance degradation, which we will leave for future endeavors. This result also confirms that, with efficient hardware that can maximize utilization even when data reuse is low, DNNs that do not have enough data reuse to exploit will put more pressure on the external data bandwidth, which should be addressed in the design of future DNN models.

\subsection{Comparison with Prior Art}

Table~\ref{table:sota_comparison} shows the comparison between Eyeriss v2 and the state-of-the-art prior art. Eyeriss v2 is the first one to report benchmark results on both large DNNs, e.g., AlexNet, and compact DNNs, e.g., MobileNet. For AlexNet, Eyeriss v2 still achieves comparable throughput and slightly less energy efficiency compared to other works that are tailored for the large models. This result is achieved with a batch size of one (while other results use larger batch sizes), and the overhead associated with its additional flexibility to handle the drastically different layer shapes in the compact models. We report results for Eyeriss v2 on a sparse network which is a widely used approach for large DNN models, particularly on mobile devices; unfortunately, the available results for the other works are only on AlexNet. We would expect the sparse AlexNet to potentially provide additional energy efficiency improvements on those works, but not throughput improvements.

For MobileNet, Eyeriss v2 achieves 5.3$\times$ throughput improvement and 3.9$\times$ energy improvement over AlexNet, with the same accuracy. Although the other designs do not report results for MobileNet, our understanding of those designs leads us to believe that they would not achieve comparable improvements, similar to the original Eyeriss, due to the NoC limitations as well as additional mapping inefficiencies of the dataflow. However, we conjecture that the NoC limitations can be addressed by the proposed HM-NoC.

\begin{table*}[t]
    \centering
    \footnotesize
    \begin{tabular}{|cl|c|c|c|c|c|c|}
        \hline
                                          &                     &   \textbf{Eyeriss}~\cite{jssc2016-chen} &   \textbf{ENVISION}~\cite{isscc2017-moons}   &   \textbf{Thinker}~\cite{thinker_vlsi_2017} &   \textbf{UNPU}~\cite{isscc2018-unpu}       &   \multicolumn{2}{|c|}{\textbf{This Work}}          \\
        \Xhline{3\arrayrulewidth}
        \multicolumn{2}{|c|}{\textbf{Technology}}               &   65nm                &   28nm                &   65nm                &   65nm                &   \multicolumn{2}{|c|}{65nm}              \\
        \hline
        \multicolumn{2}{|c|}{\multirow{2}{*}{\textbf{Area}}}    &   1176k gates         &   1950k gates         &   2950k gates         &   4.0mm$\times$4.0mm  &   \multicolumn{2}{|c|}{2695k gates}       \\
        \multicolumn{2}{|c|}{}                                  &   (NAND-2)            &   (NAND-2)            &   (NAND-2)            &   (Die Area)          &   \multicolumn{2}{|c|}{(NAND-2)}          \\
        \hline
        \multicolumn{2}{|c|}{\textbf{On-chip SRAM (kB)}}        &   181.5               &   144                 &   348                 &   256                 &   \multicolumn{2}{|c|}{246}               \\
        \hline
        \multicolumn{2}{|c|}{\textbf{Max Core Frequency}}       &   200 MHz             &   200 MHz             &   200 MHz             &   200 MHz             &   \multicolumn{2}{|c|}{200 MHz}           \\
        \hline
        \multicolumn{2}{|c|}{\textbf{Bit Precision}}            &   16b                 &   4b/8b/16b           &   8b/16b              &   1b-16b              &   \multicolumn{2}{|c|}{8b}                \\
        \hline
        \multicolumn{2}{|c|}{\textbf{Num. of MACs}}             &   168 (16b)           &   512 (8b)            &   1024 (8b)           &   13824 (bit-serial)  &   \multicolumn{2}{|c|}{384 (8b)}          \\
        \hline
        \Xhline{3\arrayrulewidth}
        \multicolumn{2}{|c|}{\textbf{DNN Model}}                &   AlexNet             &   AlexNet             &   AlexNet             &   AlexNet             &   sparse AlexNet  & sparse MobileNet      \\
        \hline
        \multicolumn{2}{|c|}{\textbf{Batch Size}}               &   4                   &   N/A                 &   15                  &   N/A                 &   1               &   1                   \\
        \multicolumn{2}{|c|}{\textbf{Core Frequency (MHz)}}     &   200                 &   200                 &   200                 &   200                 &   200             &   200                 \\
        \multicolumn{2}{|c|}{\textbf{Bit Precision}}            &   16b                 &   N/A                 &   adaptive            &   8b                  &   8b              &   8b                  \\
        \hline
        \multirow{2}{*}{\textbf{Inference/sec}} & (CONV only)   &   34.7                &   47                  &   -                   &   346                 &   342.4           &   -                   \\
                                                & (Overall)     &   -                   &   -                   &   254.3               &   -                   &   278.7           &   1470.6              \\
        \hline
        \multirow{2}{*}{\textbf{Inference/J}}   & (CONV only)   &   124.8               &   1068.2              &   -                   &   1097.5              &   743.4           &  -                    \\
                                                & (Overall)     &   -                   &   -                   &   876.6               &   -                   &   664.6           &   2560.3              \\
        \hline
    \end{tabular}
    \vspace{+5pt}
    \caption{   Comparison with State-of-the-art Designs. For Eyeriss v2, the throughput and energy efficiency are benchmarked on the sparse version of AlexNet and MobileNet.
            }
    \label{table:sota_comparison}
\end{table*}

\subsection{Discussion}

Eyeriss v2 focuses its design on improving the throughput and energy efficiency for compact and sparse DNN models, which is very different from the direction taken in many of the state-of-the-art previous works. With a similar amount of resources, i.e., area, Eyeriss v2 has much fewer number of MACs. However, with the flexibility of the on-chip network and the sparse processing logic that effectively improve throughput based on the sparsity of the data, Eyeriss v2 still achieves comparable throughput and energy efficiency for large DNNs against the state-of-the-art that optimizes directly for them. Furthermore, Eyeriss v2 shows a significant throughput and energy efficiency improvement on sparse MobileNet against Eyeriss v1 as shown in Section~\ref{sec:results-perf_analysis}.

Supporting sparse processing is a challenging task from an architecture design point of view. First of all, the PE design complexity and cost becomes much higher due to the additional required logic and storage. This has resulted in a significant increase in area as shown in Table~\ref{table:eyeriss_diff_summary}. In addition, it makes the support for high SIMD width processing difficult because of the workload imbalance and the high cost in the SPad due to the non-deterministic access patterns. Eyeriss v2, however, still demonstrates a design that can effectively translate the sparsity to significant throughput and energy efficiency improvement as compared to Eyeriss v1.

It is worth noting that the flexibility provided by the hierarchical mesh network and the throughput boost from the sparse processing logic can be applied separately. Therefore, if sparse networks are not the target workload, the flexible NoC can still be used in conjunction with other techniques such as lower precision and higher parallelism to achieve higher throughput and energy efficiency.
\section{Conclusion}
\label{sec:conclusion}

DNNs are rapidly evolving due to the significant amount of research in the field; however, the current direction of DNN development also brings new challenges to the design of DNN accelerators due to the widely varying layer shapes in compact DNNs and the varying data sparsity in sparse DNNs. In this work, we propose a new DNN accelerator architecture, called Eyeriss v2, that addresses these challenges. First, the varying layer shapes makes the on-chip network (NoC) the performance bottleneck since conventional NoC design poses strong assumptions on the amount of data reuse and required data bandwidth for each data type, which is too rigid to adapt. We solve this problem by introducing the hierarchical mesh network (HM-NoC). HM-NoC can be configured into different modes that can deliver from high bandwidth to high data reuse. More importantly, its implementation cost is also minimized through the hierarchical design that limit the costly all-to-all communication within local clusters as well as the circuit-switched routing. This helps to bring over an order of magnitude speedup for processing MobileNet compared to the original Eyeriss, i.e., Eyeriss v1, scaled to the same number of multipliers and storage capacity as Eyeriss v2. Furthermore, Eyeriss v2 incorporates a new PE architecture that support processing sparse weights and input activations directly in compressed domain to improve not only energy efficiency but also throughput. It also adds SIMD support so that each PE can process 2 MACs per cycles. Overall, Eyeriss v2 achieves 42.5$\times$ and 11.3$\times$ improvement in throughput and energy efficiency, respectively, with sparse AlexNet compared to Eyeriss v1 running AlexNet; it also achieves 12.6$\times$ and 2.5$\times$ improvement in throughput and energy efficiency, respectively, with sparse MobileNet compared to Eyeriss v1 running MobileNet.
\appendices
\section{Eyexam: Framework for Evaluating Performance}
\label{sec:performance_eval_framework}

Eyexam provides a systematic way of understanding the performance limits for DNN processors as a function of specific characteristics of the workload (i.e., DNN model) and accelerator design (i.e., architecture and microarchitecture); it applies these characteristics as sequential steps to increasingly tighten the bound on the performance limits. Specifically, instead of comparing the overall performance of different designs, which can be affected by many non-architectural factors such as system setup and technology differences, Eyexam provides a step-by-step process that associates a certain amount of performance loss to each architectural design decision (e.g., dataflow, number of PEs, NoC, etc.) as well as the properties of the workload, which for DNNs is dictated by the layer shape and size (e.g., filter shape, feature map size, batch size, etc.). 

Eyexam focuses on two main factors that affect performance: (1) the \emph{number of active PEs} due to the mapping as constrained by the dataflow, (2) the \emph{utilization of active PEs}, i.e. percentage of active cycles for the PE, based on whether the NoC has sufficient bandwidth to deliver data to PEs to keep them active. The product of these two components can be used to compute the \emph{utilization of the PE array} as follows
\begin{equation}
\label{eq:pe_utilization}
    \begin{split}
    &\textrm{{utilization of the PE array}} = \\
    & \textrm{{number of active PEs}} \times \textrm{{utilization of active PEs}}
    \end{split}
\end{equation}
Later in this section, we will see how this approach can use an adapted form of the well-known roofline model~\cite{cacm2009-williams} for the analysis of DNN processors.  

\begin{figure}
    \centering
        \includegraphics[width=0.55\linewidth]{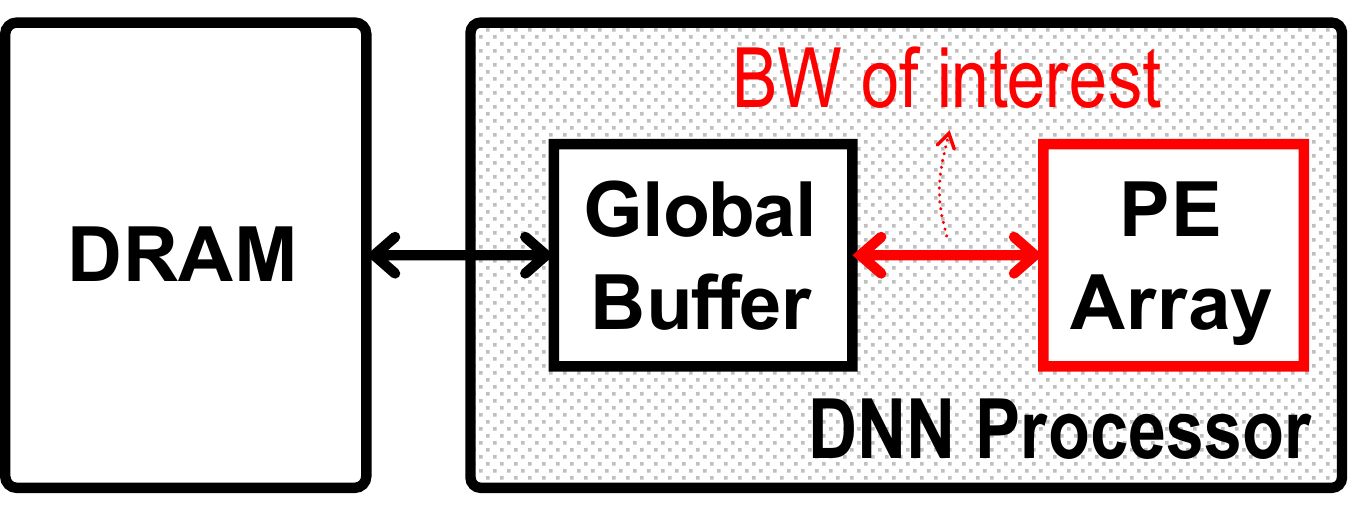}
        \vspace{-5pt}
        \caption{   A generic DNN processor architecture.
                }
        \vspace{-5pt}
        \label{fig:generic_dnn_processor_architecture}
\end{figure}

We will perform this analysis on a generic DNN processor architecture based on a spatial architecture that consists of a global buffer (GLB) and an array of PEs as shown in Figure~\ref{fig:generic_dnn_processor_architecture}. Each PE can have its own scratchpad (SPad) and control logic, and the PE array communicates with the GLB through the NoCs. Separate NoCs are used for the three data types, and Fig.~\ref{fig:common_noc_designs} shows several commonly used NoC designs for different degrees of data reuse and bandwidth requirements. The choice largely depends on how the dataflow exploits spatial data reuse for a specific data type.

The dataflow of a DNN processor is one of the key attributes that define its architecture~\cite{microTP2017-chen}. In this work, we will feature architectures that support the following four popular dataflows~\cite{isca2016-chen, isca2017-parashar}: weight-stationary (WS), output-stationary (OS), input-stationary (IS), and row-stationary (RS). 

To help illustrate the capabilities of Eyexam, we will first describe a simple 1D convolution example in Section~\ref{ssec:simple_example} and walk through the key steps of Eyexam in Section~\ref{ssec:analysis} with the 1D convolution. We will then highlight various insights that Eyexam gives on real DNN workloads and architectures in Section~\ref{ssec:DNN_analysis}.


\subsection{Simple 1D Convolution Example}
\label{ssec:simple_example}

\begin{figure}
    \centering
        \includegraphics[width=0.80\linewidth]{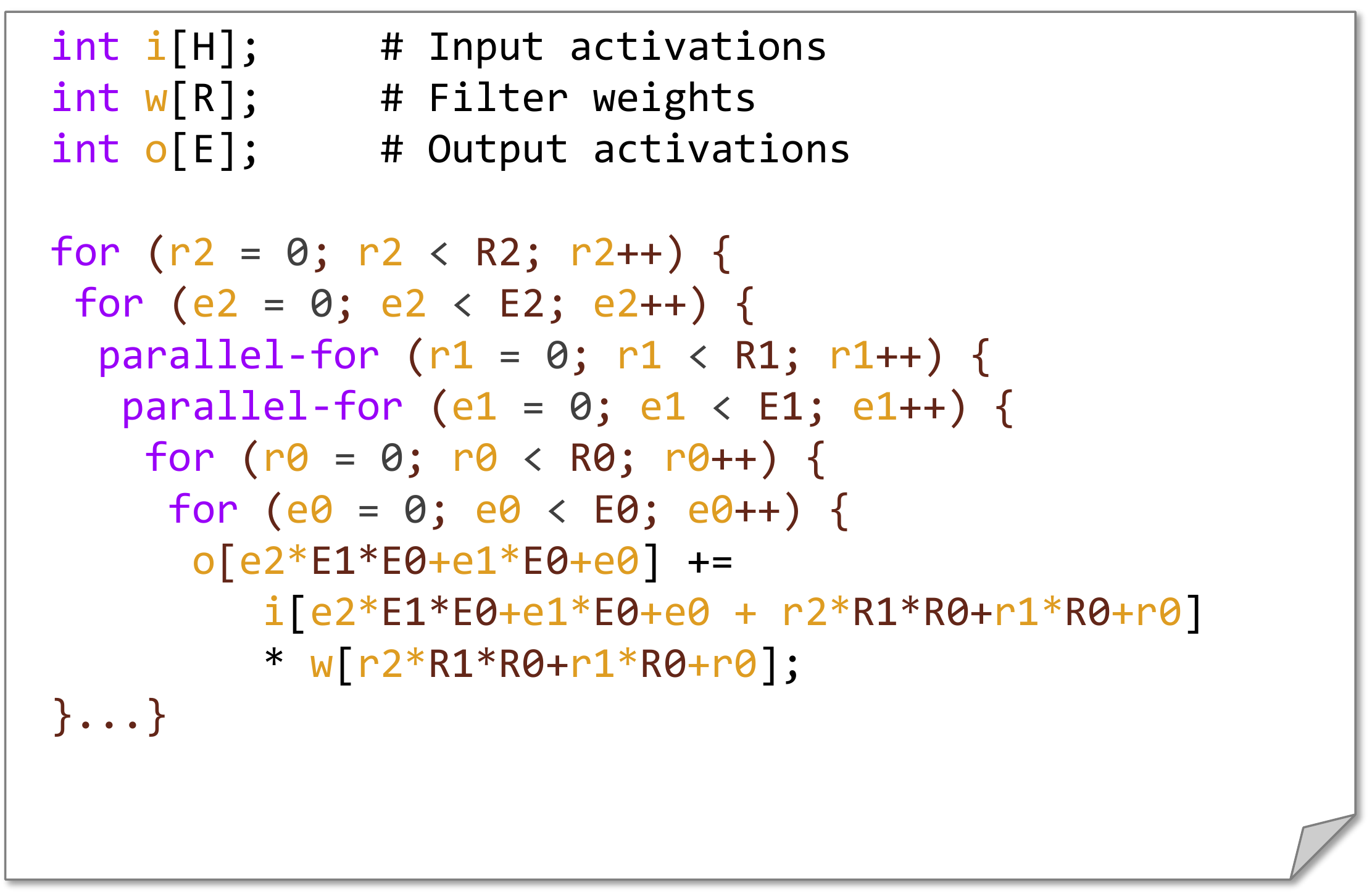}
        \vspace{-5pt}
        \caption{    Example: Loop nest of 1D convolution
                }
        \vspace{-10pt}
        \label{fig:1D_loopnest_example}
\end{figure}

We will start with a simple 1D convolution example. This example illustrates the two components of the problem. The first is the \emph{workload}, which is represented by the shape of the layer for a 1D convolutions. This comprises the filter size $R$ and the input feature map size $H$ and the output feature map size $E$. The second is the \emph{architecture} of the processing unit, for which a key characteristics is the dataflow. The dataflow can be represented by a loop nest as shown in Figure~\ref{fig:1D_loopnest_example}. In this example, the two \texttt{parallel-fors} represent the distribution of computation across multiple PEs (i.e., spatial processing); the inner two \texttt{for loops} represent the temporal processing and SPad accesses within a PE, and the outer two \texttt{for loops} represent the temporal processing of multiple passes across PE array and GLB accesses. For this example, we assume the input activations and weights fit in the GLB.

A mapping assigns specific values to loop limits $E0$, $E1$, $E2$ and $R0$, $R1$, $R2$ to execute a specific workload shape and loop ordering. This assignment of $E0$, $E1$, $E2$ and $R0$, $R1$, $R2$ is constrained by the shape of the workload and the hardware resources. The workload constraints in this example are $E0\times E1\times E2=E$ and $R0\times R1\times R2=R$\footnote{We assume perfect factorization in this example. Imperfect factorization will lead to cycles where no work is done.}.  The architectural constraint in this example is that $E1\times R1$ must be less than the number of PEs (later we will see that the NoC can pose additional restrictions). The size of the SPad allocated to input activations, psums and weights will restrict $E0$ and $R0$, and the space in the GLB allocated to psums restricts $E1$ and $R1$.

While this is a simple 1D example, it can be extended to additional levels of buffering by adding additional levels of loop nest. Furthermore, extending it to support additional dimensionality (e.g., 2D and channels) will also results in additional loops.


\subsection{Apply Performance Analysis Framework to 1D Example}
\label{ssec:analysis}

The goal of Eyexam is to provide a fine-grain performance profile for an architecture. It is a sequential analysis process that involves seven major steps. The process starts with the assumption that the architecture has infinite processing parallelism, storage capacity and data bandwidth. Therefore, it has infinite performance (as measured in MACs/cycle).

For each of the following steps, certain constraints will be added to reflect changes in the assumptions on the architecture or workload. The associated performance loss can therefore be attributed to that change, and the final performance at one step becomes the upper-bound for the next step.

\textbf{Step 1 (Layer Shape and Size):} In this first step, we look at the impact of the workload constraint, specifically the layer shape ($R$, $H$ and $E$), assuming unbounded values for $R1$ and $E1$ since there is no architectural constraints. This allows us to set $R1=R$, $E1=E$, and $E2=E0=1$, $R2=R0=1$, so that there is all spatial (i.e., parallel) processing, and no temporal (i.e., serial) processing. Therefore, the performance upper bound is determined by the finite size of the workload (i.e., the number of MACs in the layer, which is $E\times R$).

\textbf{Step 2 (Dataflow):} In this step, we define the dataflow and examine the impact of this architectural constraint. For example, to configure the example loop nest into a weight-stationary (WS) dataflow, we would set $E1=1$, $E0=E$ and $R1=R$, $R0=1$. This means that each PE stores one weight, that weight is reused $E0$ times within that PE, and the number of PE equals the number of weights. This forces the absolute maximum amount of reuse for weights at the PE. The forced serialization of $E0=E$ reduces the performance upper bound from $E\times R$ to $R$, which is the maximum parallelism of the dataflow.

\textbf{Step 3 (Number of PEs):} In this step, we define a finite number of PEs, and look at the impact of this architectural constraint. For example, in the 1D WS example, where $E1=1$ and $E0=E$, $R1$ is constrained to be less than or equal to the number of PEs, which dictates the theoretical peak performance. There are two scenarios when the actual performance is less than the peak performance. The first scenario is called spatial mapping fragmentation, in which case $R$, and therefore $R1$, is smaller than the number of PEs. In this case, some PEs are completely idle throughout the entire period of processing. The second scenario is called temporal mapping fragmentation, in which case $R$ is larger than the number of PEs but not an integer multiple of it. For example, when the number of PEs is 4, $R = 7$ and $R1 = 4$, it takes two cycles to complete the processing, and none of the PEs are completely idle. However, one of the 4 PEs will only be 50\% active. Therefore, it still does not achieve the theoretical peak performance. In general, however, if the workload does not map into all of the PEs in all cycles, then some PEs will \emph{not} be used at 100\%, which should be taken into account in performance evaluation.

\textbf{Step 4 (Physical dimensions of the PE array):} In this step, we consider the physical dimensions of the PE array (e.g., arranging 12 PEs as 3$\times$4, 2$\times$6 or 4$\times$3, etc.). The spatial partitioning is constrained per dimension which can cause additional performance loss. To explain this step with the simple example, we need to release the WS restriction. Let us assume $E1$ is mapped to the width of the 2D array and $R1$ is mapped to the height of the 2D array. If $E1$ is less than the width of the array or $R1$ is less than the height of the array (spatial mapping fragmentation), not all PEs will be utilized even if without the constraint that $E1\times R1$ is smaller or equal to the number of PE. A similar case can be constructed for the temporal mapping fragmentation as well. This architectural constraint further reduces the number of active PEs.

\textbf{Step 5 (Storage Capacity):} In this step, we consider the impact of making the buffer storage finite. For example, for the WS dataflow example, if the allocated storage for psums in the GLB is limited, it limits the number of weights that can be processed in parallel, which limits the number of PEs that can operate in parallel. Thus an architectural constraint on how many psums can be stored in the GLB restricts $E1$ and $R1$, which again can reduce performance due to less active PEs. 

\textbf{Step 6 (Data Bandwidth):} In this step, we consider the impact of a finite bandwidth for delivering data across the different levels of the loop nest (i.e., memory hierarchy). The amount of data that needs to be transferred between each level of the loop nest and the bandwidth at which we can transmit the data dictate the speed at which the index of the loop can increment (i.e., number of cycles per MAC). For instance, the bandwidth of the SPad in the PE dictates the increment speed of $r0$ and $e0$, the bandwidth of the NoC and GLB dictates the rate of change of $r1$ and $e1$, and the off-chip bandwidth dictates the rate of change of $r2$ and $e2$. In this work, we will focus on the bandwidth between the GLB and the PEs.

\begin{figure}
    \centering
        \includegraphics[width=0.60\linewidth]{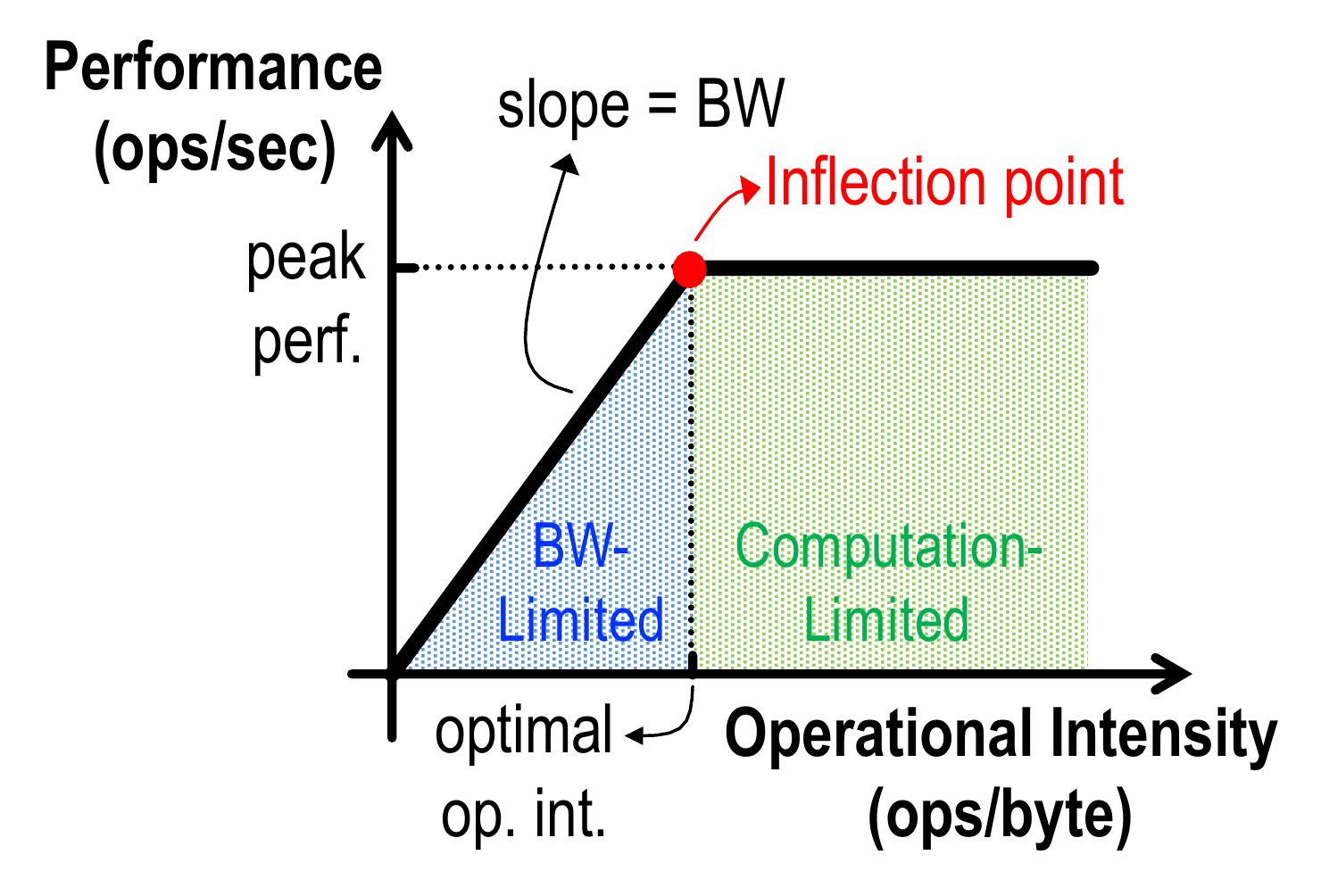}
        \caption{   The roofline model
                }
        \label{fig:roofline_model}
\end{figure}

To quantify the impact on performance from insufficient bandwidth, we can adapt the well-known roofline model~\cite{cacm2009-williams} for the analysis of DNN processors. The roofline model, as shown in Fig.~\ref{fig:roofline_model}, is a tool that visualizes the performance of an architecture under various degrees of operational intensity. It assumes a processing core, e.g., PE array, that has insufficient local memory to fit the entire workload, and therefore its performance can be limited by insufficient bandwidth between the core and the memory, e.g., GLB. When the operational intensity is lower than that at the inflection point, the performance will be bandwidth-limited; otherwise, it is computation-limited. The roofline indicates the performance upper-bound, and the performance of actual workloads sit in the area under the roofline.

\begin{figure}
    \centering
        \includegraphics[width=0.95\linewidth]{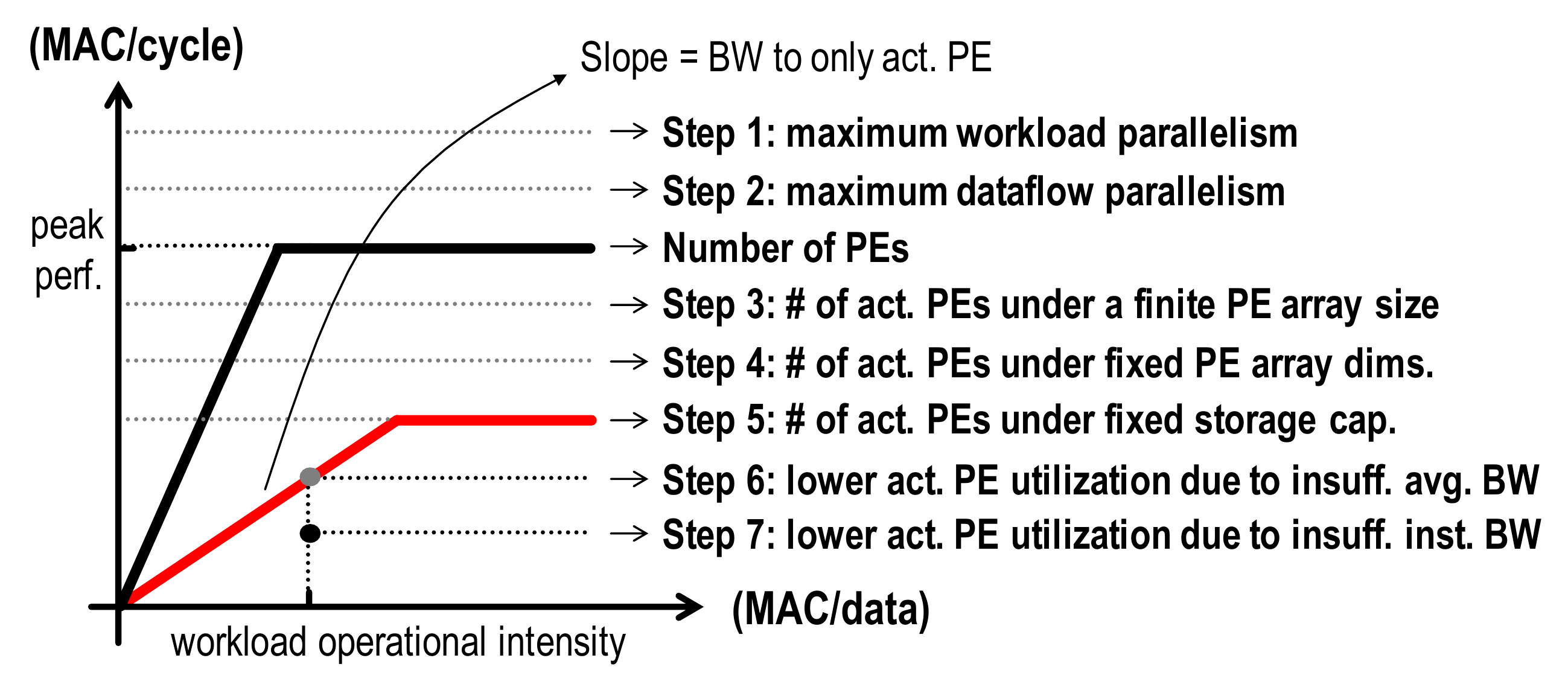}
        \caption{   Impact of steps on the roofline model.
                }
        \label{fig:roofline_model_steps}
\end{figure}

For this analysis, we adapt the roofline model as follows:
\begin{itemize}
    \item We use three separate rooflines for the three data types instead of one with the aggregated bandwidth and operational intensity.\footnote{Ideally, we should draw a \textit{roof-manifold} with the operational intensity of each data type on a separate axis; unfortunately, it will be a 4-D plot that cannot be visualized.} This helps to identify the performance bottleneck and is also a necessary setup since independent NoCs are used for each data type. However, the performance upper-bound will be the worst case of the three rooflines.
    
    \item The roofline is typically drawn with the peak performance of the core and the total bandwidth between the core and memory. However, since we have gone through the first 5 steps in Eyexam, it is possible to get a tighter bound (Fig.~\ref{fig:roofline_model_steps}). The leveled part of the roofline is now at the performance bound from step 5; the slanted part of the roofline should only consider the bandwidth to the active PEs for each data type. Since performance is measured in MACs/cycle, the bandwidth should factor in the clock rate differences between processing and data delivery.
    
    \item For a workload layer, the operational intensity of a data type is the same as its amount of data reuse in the PE array, including both temporal reuse with the SPad and the spatial reuse across PEs. It is measured in MACs per data value (MAC/data) to normalize the differences in bitwidth.
\end{itemize}

\textbf{Step 7 (Varying Data Access Patterns):} In this step, we consider the impact of bandwidth varying across time due to the dynamically changing data access patterns (Step 6 only addresses average bandwidth). For the WS example, during ramp up, the weight NoC will require high bandwidth to load the weights into the SPad of the PEs, but in steady state, the bandwidth requirements of the weight NoC will be low since the weights are reused within the PE. The performance upper bound will be affected by ratio of time spent in ramp up versus steady state, and the ratio of the bandwidth demand versus available bandwidth. This step causes the performance point to fall off the roofline as shown in Fig.~\ref{fig:roofline_model_steps}. There exist many common solutions to address this issue, including using double buffering or increased bus-width for the NoC. Therefore, we will focus less on the performance loss due to this step in this thesis.

Table~\ref{tab:constraints} summarizes the constraints applied at each step. While Eyexam is useful for examining the impact of each step on performance, it can also be used in the architecture design process to iterate through a design. For instance, if one selects a dataflow in step 2 and discovers that the storage capacity in step 5 is not a good match causing a large performance loss, one could return to step 2 to make a different dataflow design choice and then go through the steps again. Another example is that double buffering could be used in step 7 to hide the high bandwidth during ramp up, however, this would require returning to step 5 to change the effective storage capacity constraints. Eyexam can also be applied to consider the trade-off between performance and energy efficiency in combination with the framework for evaluating energy efficiency~\cite{isca2016-chen}, as well as consider the impact of sparsity and workload imbalance on performance. However, this is beyond the scope of this thesis.

\begin{table*}
\scriptsize
\centering
    \begin{tabular}{@{}cp{3.0cm}llp{4.0cm}@{}}
        \toprule
         \textbf{Step} & \textbf{Constraint} & \textbf{Type} & \textbf{New Performance Bound} & \textbf{Reason for Performance Loss} \\\midrule
         \textbf{1} & Layer Size and Shape & Workload & Max workload parallelism & Finite workload size \\ \hline
         \textbf{2} & Dataflow loop nest & Architectural & Max dataflow parallelism & Restricted dataflow mapping space by defined by loop nest \\ \hline
        \textbf{3} & Number of PEs & Architectural & Max PE parallelism & Additional restriction to mapping space due to shape fragmentation \\ \hline
        \textbf{4} & Physical dimensions of PEs array & Architectural & Number of active PEs& Additional restriction to mapping space due to shape fragmentation for \emph{each} dimension \\ \hline  
        \textbf{5} & Fixed Storage Capacity & Architectural & Number of active PEs & Additional restriction to mapping space due to storage of intermediate data (depends on dataflow) \\ \hline  
        \textbf{6} & Fixed Data Bandwidth & Microarchitectural & Max data bandwidth to active PEs & Insufficient average bandwidth to active PEs \\\hline 
        \textbf{7} & Varying Data Access Patterns & Microarchitectural & Actual measured performance & Insufficient instant bandwidth to active PEs \\        
        \bottomrule
    \end{tabular}
    \vspace{+5pt}
    \caption{Summary of steps in Eyexam. }
    \label{tab:constraints}
\end{table*}


\subsection{Performance Analysis Results for DNN Processors and Workloads}
\label{ssec:DNN_analysis}

In this section, we will highlight some of the observations obtained with Eyexam on DNN processors with real DNN workloads (e.g., AlexNet, MobileNet). We will provide results for architectures from all four representative dataflows, including WS, OS, IS, and Row-Stationary~\cite{isca2016-chen}, with different PE array sizes. The dataflows are evaluated on PE arrays where the height and the width are the same, regardless of the number of PEs.

\begin{figure}
    \centering
    \subfloat[AlexNet, 256 PEs]{
		\includegraphics[width=0.95\linewidth]{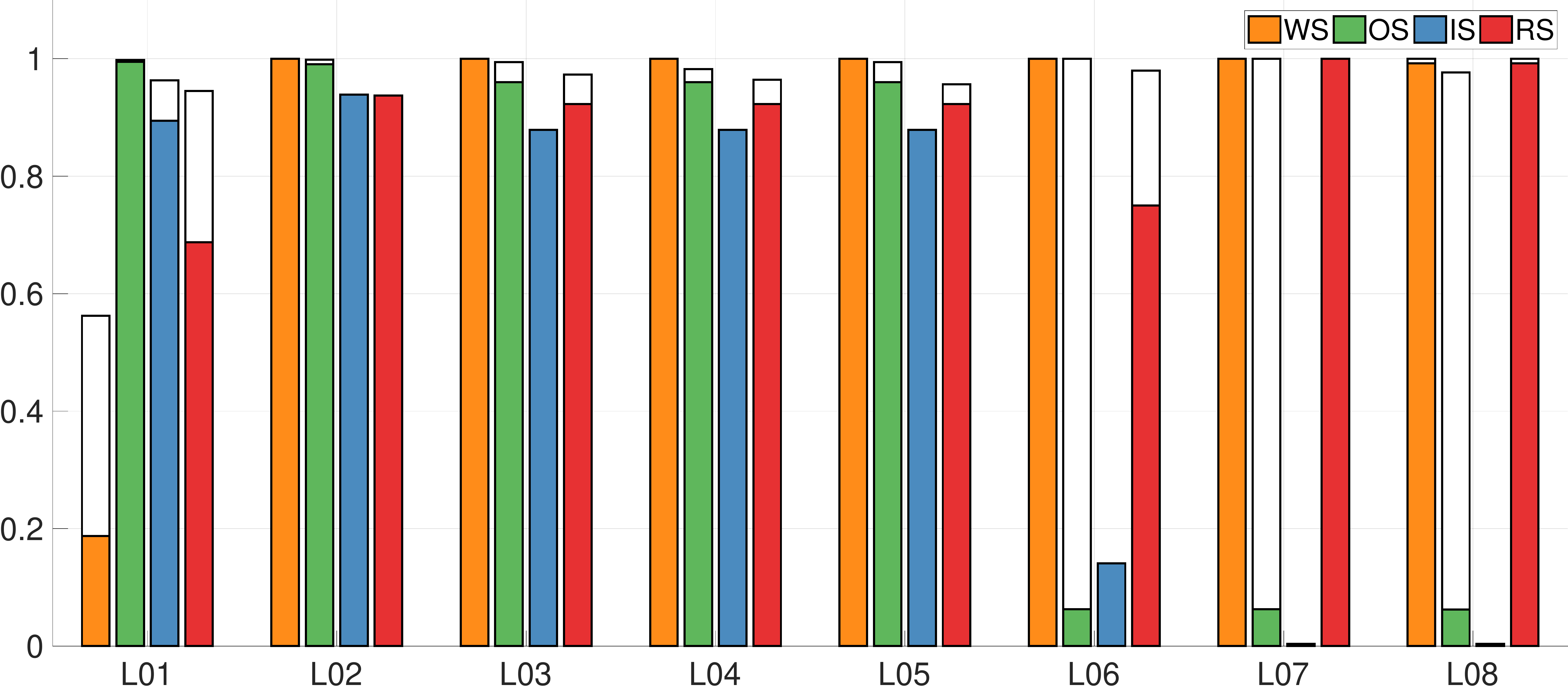}
		\label{fig:step3_4_on_alexnet_256}
	}\hfill
	\subfloat[AlexNet, 16384 PEs]{
		\includegraphics[width=0.95\linewidth]{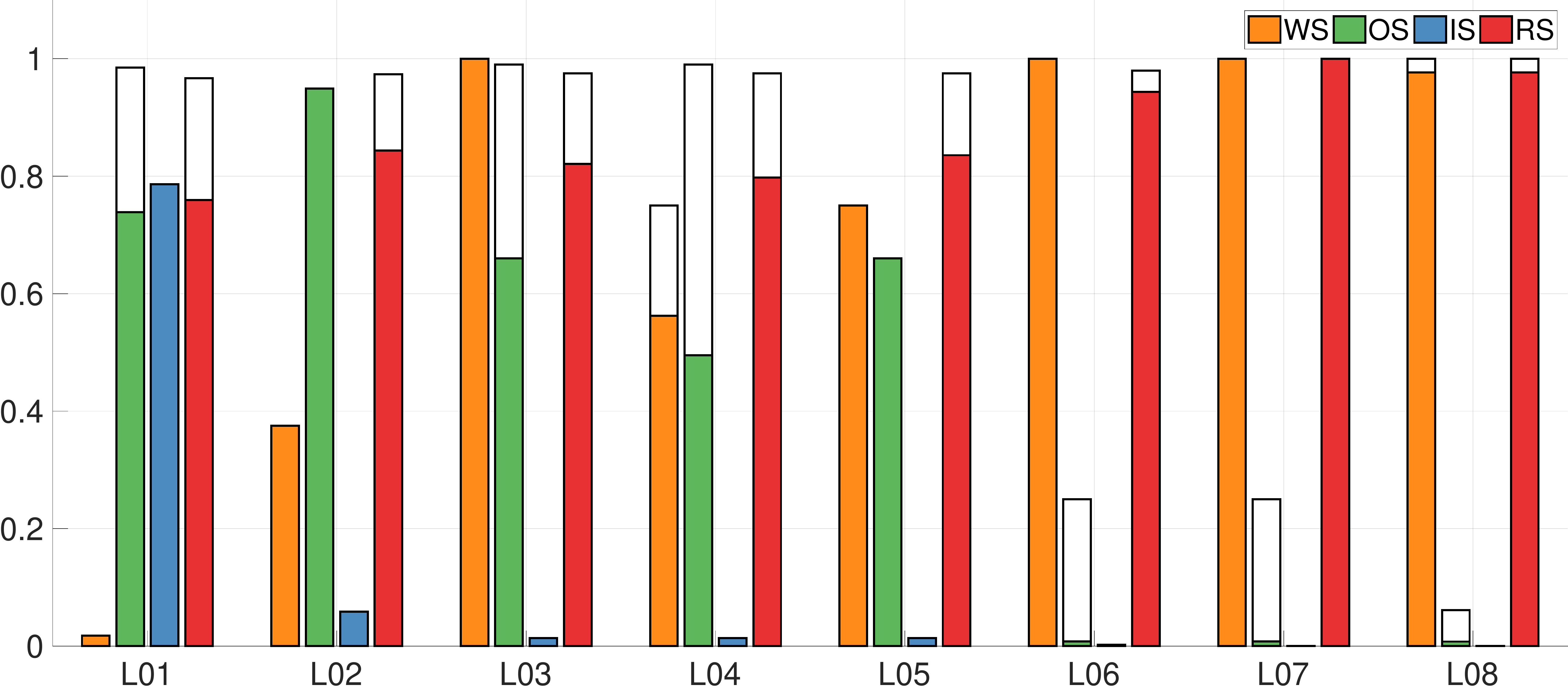}
		\label{fig:step3_4_on_alexnet_16384}
	}\hfill
	\subfloat[MobileNet, 16384 PEs]{
		\includegraphics[width=0.95\linewidth]{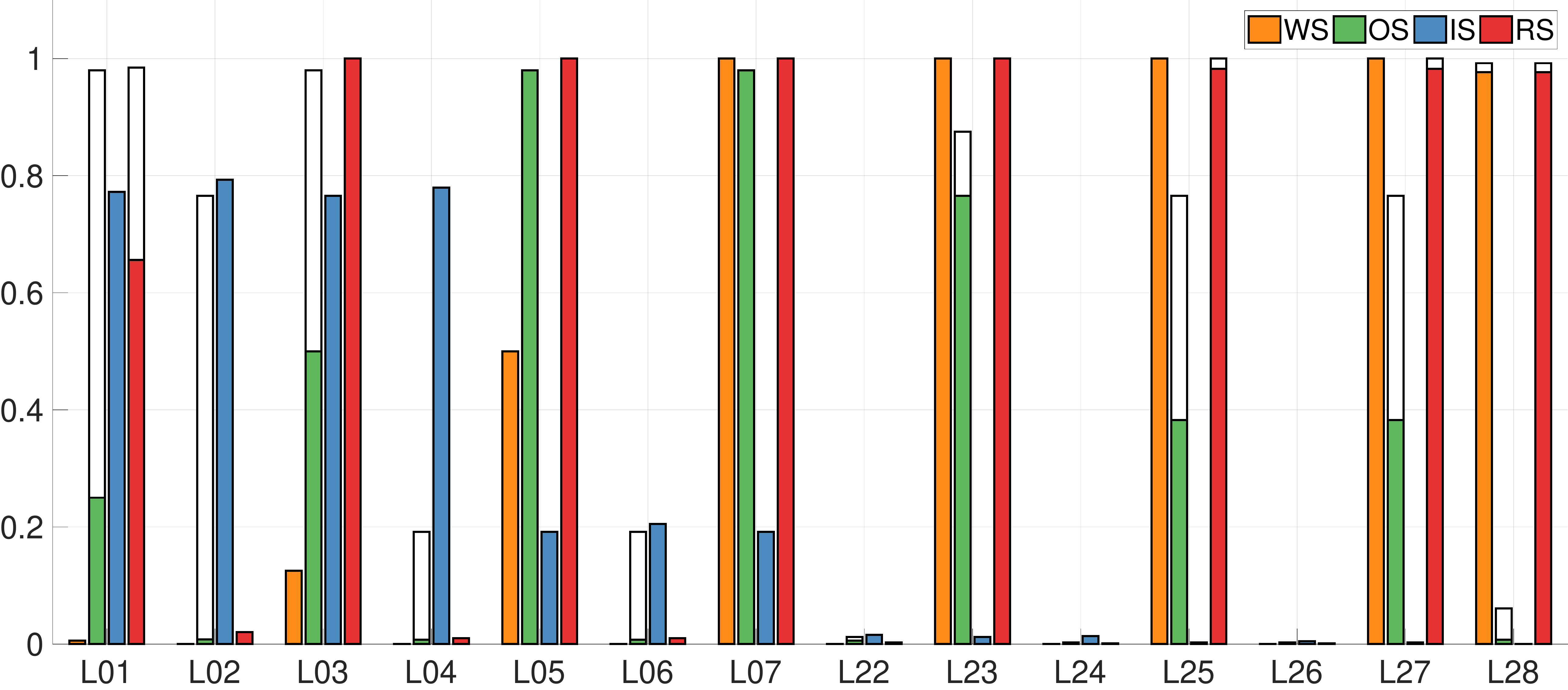}
		\label{fig:step3_4_on_mobilenet_16384}
	}
    \caption{  Impact of the number of PEs and the physical dimensions of the PE array on number of active PEs. The y-axis is the performance normalized to the number of PEs.
            }            
    \label{fig:step3_4_on_dnn}
\end{figure}

Fig.~\ref{fig:step3_4_on_dnn} shows the number of active PEs for the four different architectures in different DNN layers and PE array sizes. It takes into account the mapping of different dataflows in each architecture for different layer shapes under a finite number of PEs. The results are normalized to the total number of PEs in the array. For each bar, the total bar height (white-portion + colored-portion) represent the performance at step 3 of Eyexam, which accounts for the impact of mapping fragmentation due to a finite number of PEs, and the colored-only portion represent the performance at step 4, which further accounts for the impact of the physical dimensions of the PE array. Therefore, the white portion indicates the performance loss from step 3 to 4, which indicates the mapping limitation in the dataflows to adapt to the physical dimensions of the PE array. The results show that 
\begin{itemize}
    \item Fig.~\ref{fig:step3_4_on_alexnet_256} and~\ref{fig:step3_4_on_alexnet_16384} shows the performance impact when scaling the size of PE array. Many of the architectures are not flexible enough to fully utilize the parallelism when it scales up (i.e., increase number of PEs), which indicates that simply increasing hardware resources is not sufficient to achieve a higher performance.
    \item Fig.~\ref{fig:step3_4_on_alexnet_16384} and~\ref{fig:step3_4_on_mobilenet_16384} shows the performance impact when having to support many different layer shapes. Mapping the different layers onto the same architecture according to its dataflow can result in widely varying performance. For example, the featured IS and OS architectures cannot map well in the layers with smaller feature map sizes, while the RS dataflow does not map well in the depth-wise layers due to the lack of channels. The common reasons why each data dimension diminishes is summarized in Table~\ref{table:diminishing_dnn_data_dimensions}. In order to support a wide variety of DNNs, the dataflow has to be flexible enough to deal the diminished reuse available in any data dimensions.
    \item When the PE array size scales up, many of the architectures are not flexible enough to fully utilize the parallelism, which indicates that simply increasing hardware resources is not sufficient to achieve higher performance.
\end{itemize}

In addition to the loss due to the finite number of PEs and the physical PE array shape, there is loss from insufficient bandwidth for data delivery. To avoid performance loss due to insufficient data bandwidth from the GLB, which results in low utilization of the active PEs (step 6), the NoC design should meet the worst-case bandwidth requirement for every data type. In addition, another NoC design objective is to exploit data reuse to minimize the number of GLB accesses, which is usually realized by the multi-cast of broadcast of data from GLB. On the one hand, for an architecture in which the pattern of spatial data reuse is unchanged with mapping, it is straightforward to meet the two requirements at the same time. For example, if a certain type of data is always reused across an entire PE row or column, the systolic or multicast networks will provide sufficient bandwidth and data reuse from GLB. However, this fixed pattern of data delivery can also cause performance loss in step 3 or 4 of Eyexam. On the other hand, if the architecture support very flexible spatial mappings of operations, which potentially can preserve the performance up to step 5 of Eyexam, the pattern of spatial data reuse can vary widely for different layer shapes. While a single broadcast network can exploit data reuse in any spatial reuse patterns, it sacrifices the data bandwidth from GLB. When the amount of data reuse is low, e.g., delivering weights in FC layers with a small batch size, the broadcast network will result in significant performance loss. Therefore, step 6 will become a performance bottleneck.

\bibliographystyle{ieeetr}

\begin{IEEEbiography}
    [{\includegraphics[width=1in,height=1.25in,clip,keepaspectratio]{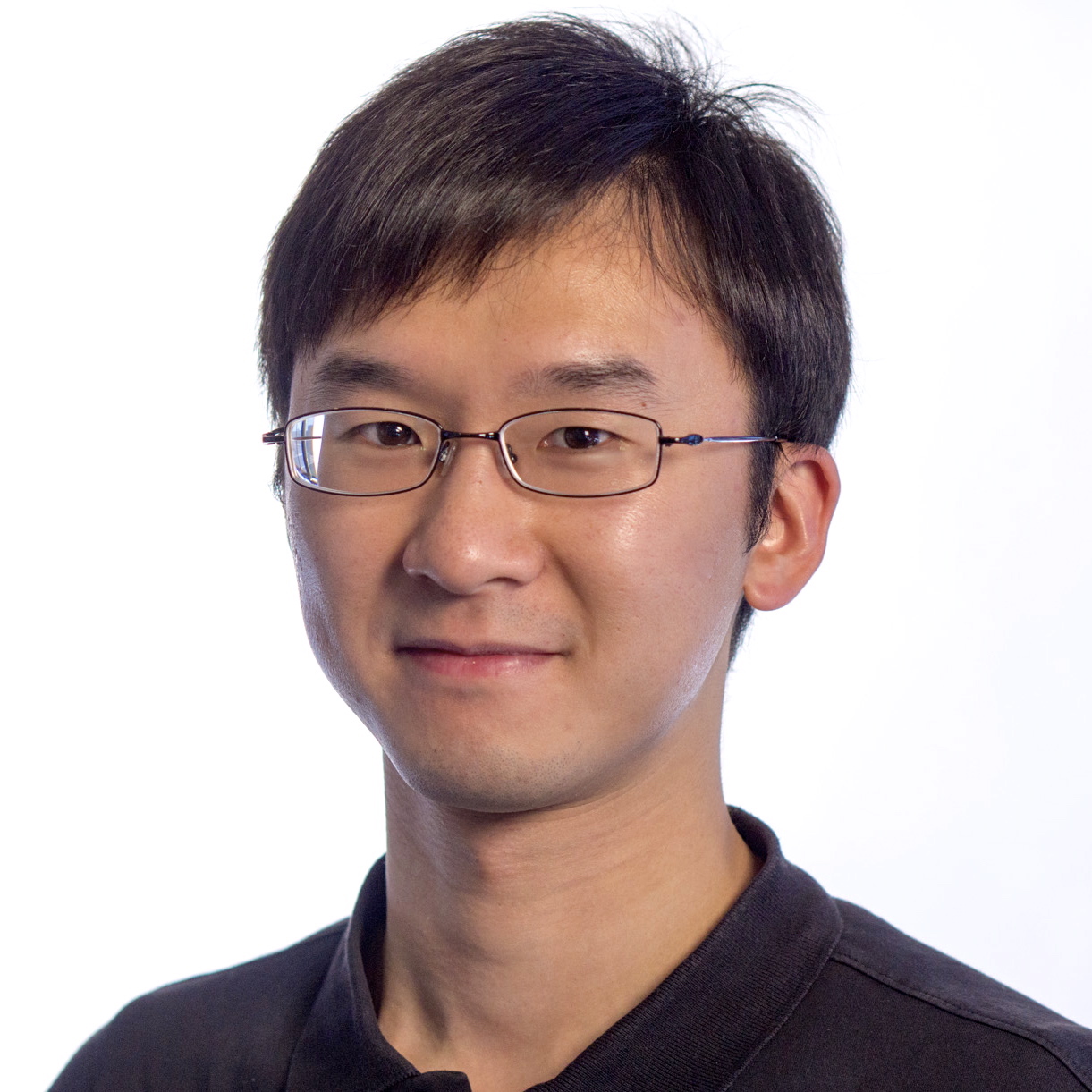}}]
    {Yu-Hsin Chen}
    (S'11) received the B. S. degree in Electrical Engineering from National Taiwan University, Taipei, Taiwan, in 2009, and the M. S. and Ph.D. degrees in Electrical Engineering and Computer Science (EECS) from Massachusetts Institute of Technology (MIT), Cambridge, MA, in 2013 and 2018, respectively. In 2018, he received the Jin-Au Kong Outstanding Doctoral Thesis Prize in Electrical Engineering at MIT. Since September 2018, he has been a Research Scientist in Nvidia's Architecture Research Group in Santa Clara, CA. His current research focuses on the design of computer architectures for machine learning, deep learning, and domain-specific processors. 
    
    He was the recipient of the 2015 Nvidia Graduate Fellowship, 2015 ADI Outstanding Student Designer Award, and 2017 IEEE SSCS Predoctoral Achievement Award. His work on the dataflows for CNN accelerators was selected as one of the Top Picks in Computer Architecture in 2016. He also co-taught a tutorial on ``Hardware Architectures for Deep Neural Networks'' at MICRO-49, ISCA2017, and MICRO-50.
\end{IEEEbiography}

\begin{IEEEbiography}
    [{\includegraphics[width=1in,height=1.25in,clip,keepaspectratio]{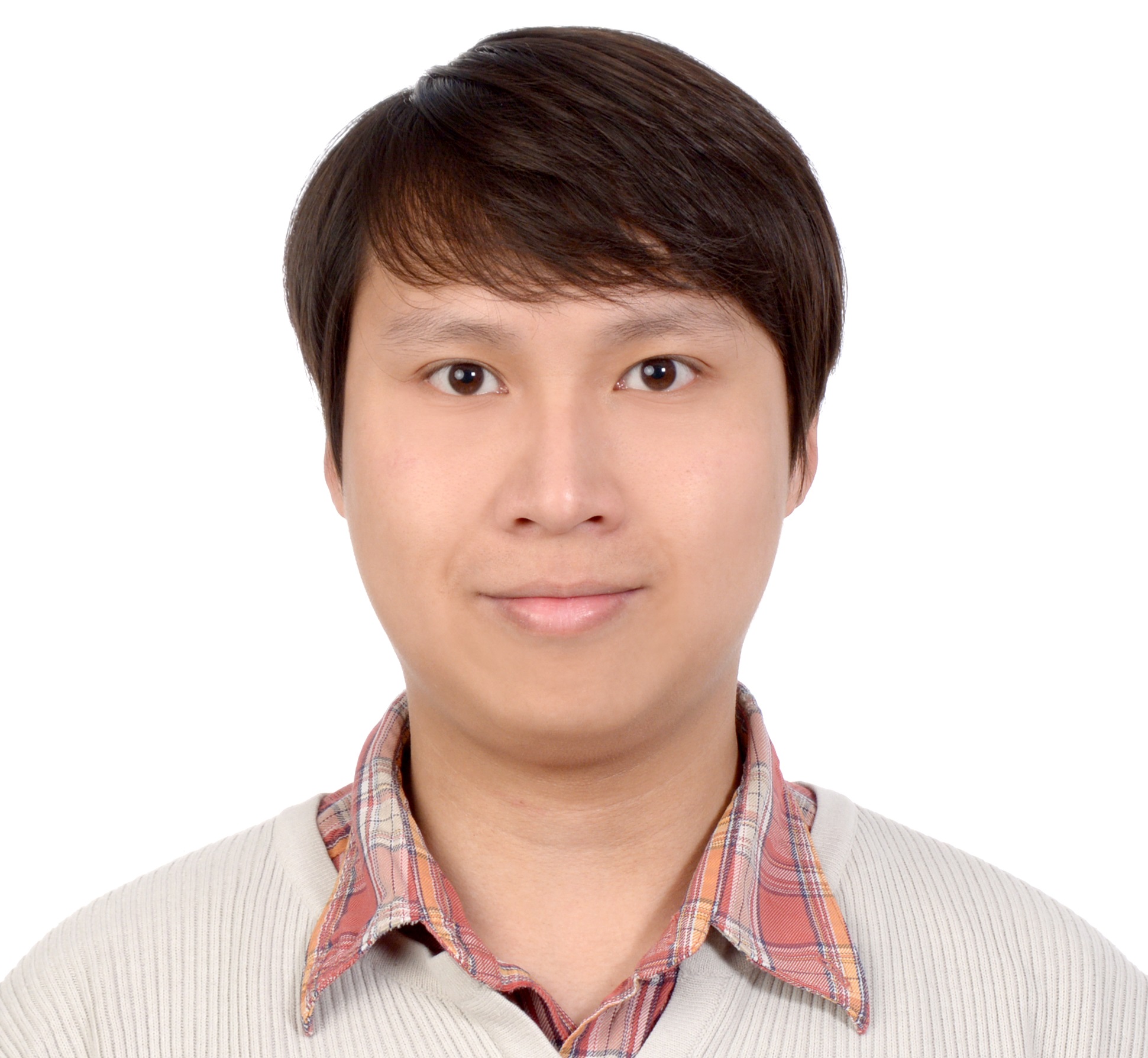}}]
    {Tien-Ju Yang}
    (S'11) received the B. S. degree in Electrical Engineering from National Taiwan University (NTU), Taipei, Taiwan, in 2010, and the M. S. degree in Electronics Engineering from NTU in 2012. Between 2012 and 2015, he worked in the Intelligent Vision Processing Group, MediaTek Inc., Hsinchu, Taiwan, as an engineer. He is currently a Ph.D. candidate in Electrical Engineering and Computer Science at Massachusetts Institute of Technology, Cambridge, MA, working on energy-efficient deep neural network design. His research interest spans the area of computer vision, machine learning, image/video processing, and VLSI system design. He won the first place of the 2011 National Taiwan University Innovation Contest.
\end{IEEEbiography}

\begin{IEEEbiography}
    [{\includegraphics[width=1in,height=1.25in,clip,keepaspectratio]{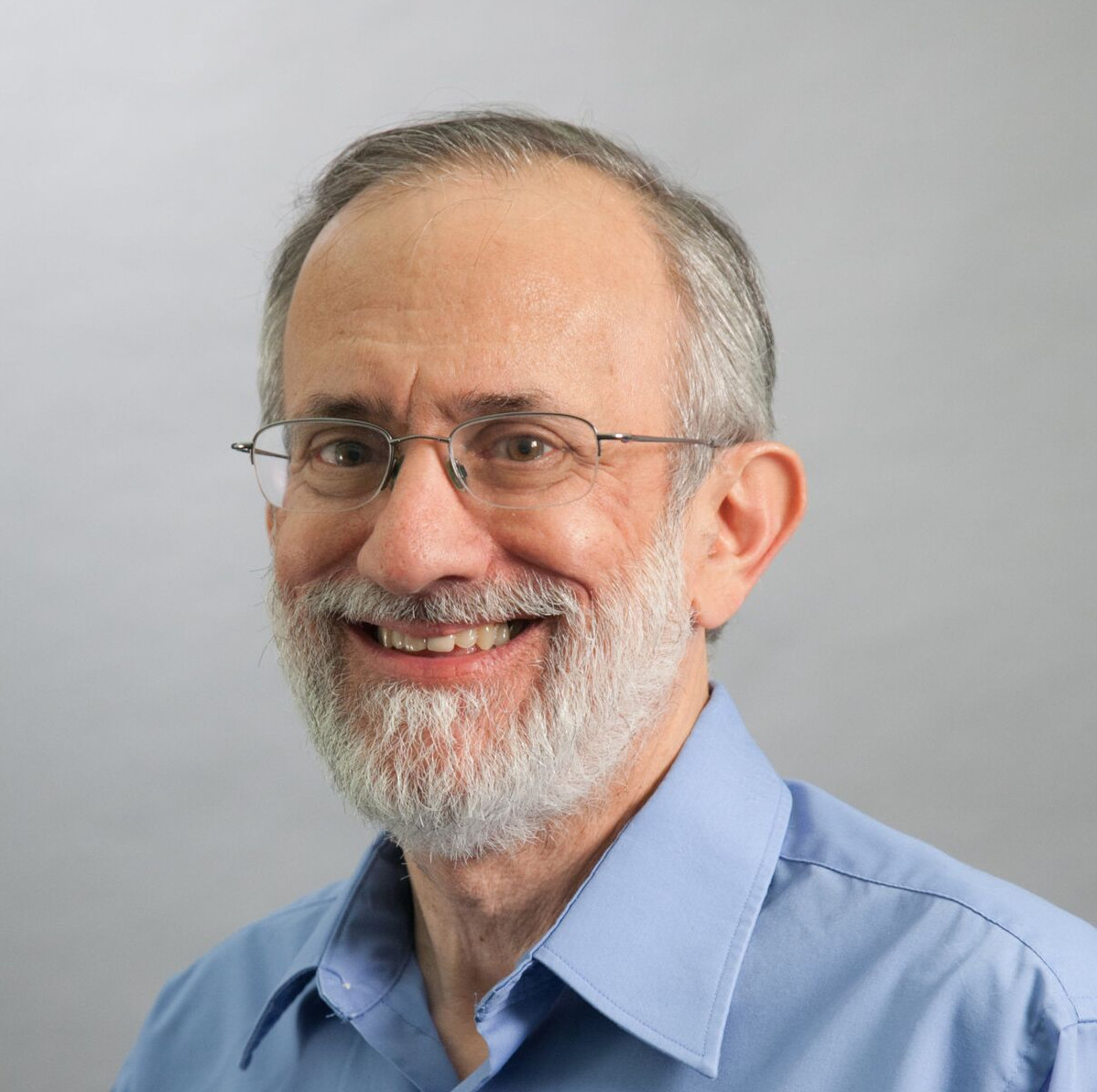}}]
    {Joel S. Emer}
    (M'73–-SM'03–-F'04) received the B.S. (Hons.) and M.S. degrees in electrical engineering from Purdue University, West Lafayette, IN, USA, in 1974 and 1975, respectively, and the Ph.D. degree in electrical engineering from the University of Illinois at Urbana–Champaign, Champaign, IL, USA, in 1979.
    
    He was with Intel, where he was an Intel Fellow and the Director of Microarchitecture Research. At Intel, he led the VSSAD Group, which he had previously been a member of at Compaq and Digital Equipment Corporation. He is currently a Senior Distinguished Research Scientist with the Nvidia's Architecture Research Group, Westford, MA, USA, where he is responsible for exploration of future architectures and modeling and analysis methodologies. He is also a Professor of the Practice at the Massachusetts Institute of Technology, Cambridge, MA, USA, where he teaches computer architecture and supervises graduate students. He has held various research and advanced development positions investigating processor microarchitecture and developing performance modeling and evaluation techniques. He has made architectural contributions to a number of VAX, Alpha, and X86 processors and is recognized as one of the developers of the widely employed quantitative approach to processor performance evaluation. He has been recognized for his contributions in the advancement of simultaneous multithreading technology, processor reliability analysis, cache organization, and spatial architectures for deep learning.
    
    Dr. Emer is a Fellow of the ACM. He has been a recipient of numerous public recognitions. In 2009, he received the Eckert-Mauchly Award for lifetime contributions in computer architecture, the Purdue University Outstanding Electrical and Computer Engineer Alumni Award, and the University of Illinois Electrical and Computer Engineering Distinguished Alumni Award in 2010 and 2011, respectively. His 1996 paper on simultaneous multithreading received the ACM/SIGARCH-IEEE-CS/TCCA: Most Influential Paper Award in 2011. He was named to the ISCA and Micro Halls of Fame in 2005 and 2015, respectively. He has had six papers selected for the IEEE Micro’s Top Picks in Computer Architecture, in 2003, 2004, 2007, 2013, 2015, and 2016. He was the Program Chair of ISCA in 2000 and Micro in 2017.
\end{IEEEbiography}

\begin{IEEEbiography}
    [{\includegraphics[width=1in,height=1.25in,clip,keepaspectratio]{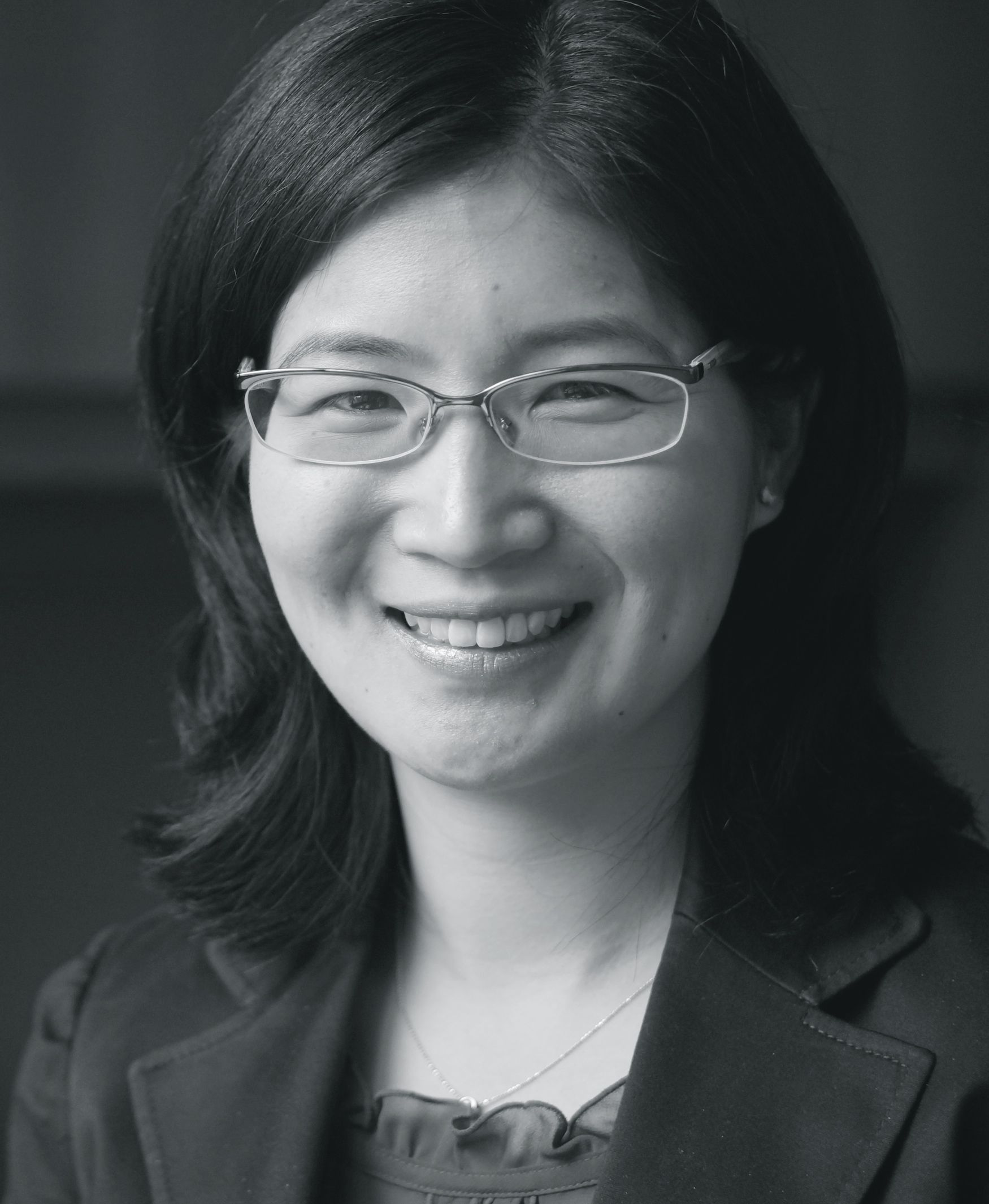}}]
    {Vivienne Sze}
    (S'04--M'10--SM'16) received the B.A.Sc. (Hons) degree in electrical engineering from the University of Toronto, Toronto, ON, Canada, in 2004, and the S.M. and Ph.D. degree in electrical engineering from the Massachusetts Institute of Technology (MIT), Cambridge, MA, in 2006 and 2010 respectively.  In 2011, she received the Jin-Au Kong Outstanding Doctoral Thesis Prize in Electrical Engineering at MIT.
    
    She is an Associate Professor at MIT in the Electrical Engineering and Computer Science Department.  Her research interests include energy-aware signal processing algorithms, and low-power circuit and system design for portable multimedia applications including computer vision, deep learning, autonomous navigation, image processing, and video coding. Prior to joining MIT, she was a Member of Technical Staff in the Systems and Applications R\&D Center at Texas Instruments (TI), Dallas, TX, where she designed low-power algorithms and architectures for video coding.  She also represented TI in the JCT-VC committee of ITU-T and ISO/IEC standards body during the development of High Efficiency Video Coding (HEVC), which received a Primetime Engineering Emmy Award.  Within the committee, she was the primary coordinator of the core experiment on coefficient scanning and coding, and has chaired/vice-chaired several ad hoc groups on entropy coding.  She is a co-editor of “High Efficiency Video Coding (HEVC): Algorithms and Architectures” (Springer, 2014).
    
    Prof. Sze is a recipient of the 2019 Edgerton Faculty Achievement Award at MIT, 2018 Facebook Faculty Award, 2018 \& 2017 Qualcomm Faculty Award, 2018 \& 2016 Google Faculty Research Award, 2016 AFOSR Young Investigator Research Program (YIP) Award, 2016 3M Non-Tenured Faculty Award, 2014 DARPA Young Faculty Award, 2007 DAC/ISSCC Student Design Contest Award and a co-recipient of the 2018 VLSI Best Student Paper Award, 2017 CICC Outstanding Invited Paper Award, 2016 IEEE Micro Top Picks Award and the 2008 A-SSCC Outstanding Design Award.  She is a Distinguished Lecturer of the IEEE Solid-State Circuits Society (SSCS), and currently serves on the technical program committee for the International Solid-State Circuits Conference (ISSCC) and the SSCS Advisory Committee (AdCom). She has also served on the technical program committees for VLSI Circuits Symposium, Micro and the Conference on Systems and Machine Learning (SysML), and as a guest editor for the IEEE Transactions on Circuits and Systems for Video Technology (TCSVT).  Prof. Sze will be the Systems Program Chair of SysML in 2020.
\end{IEEEbiography}
\end{document}